\documentclass[a4paper,11pt,twoside,openright]{book}

\usepackage{tipx}
\usepackage[named]{libtex/algo}
\usepackage{algorithm}
\usepackage[noend]{algorithmic}
\usepackage{amssymb}
\usepackage{graphicx}
\usepackage{amsfonts}
\usepackage{epsfig}
\usepackage{multirow}
\usepackage{amsfonts,amsmath}
\usepackage{srcltx}
\usepackage{multirow}
\usepackage{dsfont}
\usepackage{booktabs}
\usepackage{subfloat}
\usepackage[table]{xcolor}
\usepackage{tabularx}
\usepackage{subfig}
\usepackage{url}
\usepackage{makeidx}
\usepackage{libtex/thesisStyle}

\usepackage[utf8]{inputenc} % necessary for greek letters

\usepackage[hidelinks]{hyperref} % hidelinks removes boxes around links
\usepackage[autostyle]{csquotes}
\usepackage[labelfont=bf,margin=0.2in,labelsep=endash,justification=raggedright]{caption}
\usepackage{enumitem}

\usepackage{listings} % Replacement for minted

\usepackage{fancyhdr} % create footer for every page

\usepackage{todonotes}
\usepackage{layouts}
\usepackage{colortbl}
\usepackage[T1]{fontenc} % Fix textmu error - loaded AFTER colortbl to avoid rowcolor bug
\usepackage{textcomp} % Fix textmu error
\usepackage{amsmath} % For \floor command in Chapter2
\usepackage[numbers,sort&compress]{natbib}

\definecolor{codegreen}{rgb}{0,0.6,0}
\definecolor{codegray}{rgb}{0.5,0.5,0.5}
\definecolor{codepurple}{rgb}{0.58,0,0.82}
\definecolor{backcolour}{rgb}{0,0,0}
\definecolor{whitecolour}{rgb}{1,1,1}
\usepackage{listings}
\lstdefinestyle{mystyle}{
	backgroundcolor=\color{backcolour},
	basicstyle=\ttfamily\color{white},
	commentstyle=\color{codegreen},
	keywordstyle=\color{magenta},
	numberstyle=\tiny\color{codegray},
	stringstyle=\color{codepurple},
	basicstyle=\footnotesize\color{white},
	breakatwhitespace=false,
	breaklines=true,
	captionpos=b,
	numbers=left,
	numbersep=5pt,
	showspaces=false,
	showstringspaces=false,
	showtabs=false,
	tabsize=2,
	language=bash,
	xleftmargin=0.025\textwidth,
	aboveskip=2mm,
	belowskip=2mm,
	framexleftmargin=0mm
}

\long\def\comment#1{}

\newlength{\figtblfootnotemargin}
\newlength{\figtblfootnotewidth}

\newcommand{\SquareOfSize}[2]{
 \fbox{\hsize #1cm \hbox to #1cm{\vbox{#2}}}
}

\newsavebox{\wholeWidthLine}
\sbox{\wholeWidthLine} {\rule[0.1in]{\textwidth}{.01in}}

\newcommand{\bq}{\begin{quote}}
\newcommand{\eq}{\end{quote}}
\newcommand{\be}{\begin{enumerate}}
\newcommand{\ee}{\end{enumerate}}
\newcommand{\bi}{\begin{itemize}}
\newcommand{\ei}{\end{itemize}}
\newcommand{\bie}{\begin{itemize}\begin{enumerate}}
\newcommand{\eie}{\end{enumerate}\end{itemize}}
\newcommand{\ba}{\begin{array}}
\newcommand{\ea}{\end{array}}
\newcommand{\btbl}{\begin{tabular}}
\newcommand{\etbl}{\end{tabular}}
\newcommand{\bequ}{\begin{displaymath}}
\newcommand{\eequ}{\end{displaymath}}
\newcommand{\bequa}{\begin{eqnarray*}}
\newcommand{\eequa}{\end{eqnarray*}}
\newcommand{\bc}{\begin{center}}
\newcommand{\ec}{\end{center}}
\newcommand{\btab}{\begin{tabbing}}
\newcommand{\etab}{\end{tabbing}}

\def\mpr#1{\ifmmode #1 \else #1 \fi}

\newcommand{\DPATTRNAME}{{\large{{\bf SC}$_{attr}$}}}
\newcommand{\DPISANAME}{{\large{{\bf SC}$_{isA}$}}}
\newcommand{\DPINNAME}{{\large{{\bf SC}$_{in}$}}}

\usepackage[greek,american]{babel}
\newcommand{\selg}[0]{\selectlanguage{greek}}
\newcommand{\selam}[0]{\selectlanguage{american}}

\small\normalsize
\makeindex

\long\def\symbolfootnote[#1]#2
    {\begingroup
    \def\thefootnote{\fnsymbol{footnote}}\footnote[#1]{#2}
    \endgroup}

\newcommand{\worksupportedby}{\symbolfootnote[0]{This work has been
    performed at and supported by the Computer Architecture and VLSI Systems (CARV) Laboratory, Institute of Computer Science (ICS), Foundation for Research and Technology - Hellas (FORTH), within the ExaNeSt project, funded by the European Commission under the Horizon 2020 Framework Programme (Grant Agreement 671553).
    }}

\newcommand{\thesisdate}{July 2019}
\newcolumntype {+}{>{\global \let \currentrowstyle \relax}}
\newcolumntype {^}{>{\currentrowstyle}}
\newcommand {\rowstyle}[1]{\gdef \currentrowstyle {#1} %
#1\ignorespaces
}

\graphicspath{{Figures/}{./}} % Specifies where to look for included images

\fancypagestyle{headings}{%
    
\fancyhead[RO]{\thepage}
\fancyhead[LE]{\thepage}
\fancyfoot[L]{CARV, ICS, FORTH}
 \fancyfoot[C]{M.Sc. Thesis by A. Psistakis}
\fancyfoot[R]{CSD, Univ. Of Crete}
 
}

\newcommand{\thesistitle}{Handling of Memory Page Faults during Virtual-Address RDMA}
\newcommand{\owner}{Antonis Psistakis}
\newcommand{\firstprof}{Manolis GH Katevenis}
\newcommand{\secondprof}{Angelos Bilas}
\newcommand{\thirdprof}{Evangelos Markatos}
\newcommand{\csdchair}{Antonios Argyros}

\begin{document}
\newcommand{\source}[1]{\caption*{Source: {#1}} }
% Compact spacing to match original page count
\setlength{\parskip}{0pt}
\setlength{\textfloatsep}{8pt plus 2pt minus 2pt}
\setlength{\floatsep}{8pt plus 2pt minus 2pt}
\setlength{\intextsep}{8pt plus 2pt minus 2pt}
\setlength{\abovecaptionskip}{6pt}
\setlength{\belowcaptionskip}{0pt}

\begin{titlepage}
\thispagestyle{empty}
\begin{center}

\Large \textbf{UNIVERSITY OF CRETE}\\%[0.25cm]
\Large \textbf{SCHOOL OF SCIENCES AND ENGINEERING}\\%[0.25cm]
\Large \textbf{COMPUTER SCIENCE DEPARTMENT}\\%[0.25cm]
%\LARGE \textit{Antonis Psistakis}\\[0.5cm]

\vfill

\Large{\MakeUppercase{\thesistitle}}

\vfill

\Large{\MakeUppercase{\owner}}\\[0.5cm]

\vfill

\Large{\MakeUppercase{MASTER THESIS}}\\[0.5cm]

\vfill

\Large{\MakeUppercase{Heraklion, \thesisdate{}}}

\end{center}
\end{titlepage}

\cleardoublepage

\begin{titlepage}
\thispagestyle{empty}
\begin{center}

\LARGE \textbf{Handling of Memory Page Faults during Virtual-Address RDMA}\\[0.5cm]
\LARGE \textit{Antonis Psistakis}\\[0.5cm]

\vfill

\normalsize{
Thesis submitted in partial fulfillment of the requirements for the\\[0.30cm]

\textit{Masters' of Science degree in Computer Science and Engineering}}\\[0.30cm]

University of Crete\\
School of Sciences and Engineering\\
Computer Science Department\\
Voutes University Campus, 700 13 Heraklion, Crete, Greece\\[0.5cm]

\vfill

\Large{Thesis Advisor: Prof. \emph{\firstprof}}\\[3cm]

\large{Thesis Defense: Heraklion, \thesisdate}\\
\large{Thesis Publication: Heraklion, October 2019}\\
% \large{Heraklion, October 2019 (Publication)}\\

\vfill

\end{center}

%\workperformedat{} 
\worksupportedby{}
\end{titlepage}

\cleardoublepage

%\listoftodos

\thispagestyle{empty}

\begin{titlepage}

\begin{center}
\textsc{University of Crete}\\
\textsc{Computer Science Department}\\
\vspace{0.4cm}
\noindent {\textbf{\thesistitle{}}}\\
\vspace{0.4cm}
\noindent Thesis submitted by\\
\textbf{\owner{}}\\
in partial fulfillment of the requirements for the\\
Masters' of Science degree in Computer Science\\
\vspace{0.4cm} THESIS APPROVAL

\vspace{0.4cm}

\begin{tabular}{rl}
\\
Author: & \underline{\phantom{123456789012345678901234567890123456789012}}\\
    & \owner{}\\
    \\
    \\
    \\
Committee approvals: & \underline{\phantom{123456789012345678901234567890123456789012}}\\
    & \firstprof{}\\
    & {\small Professor, Thesis Supervisor}\\
    \vspace{0.2cm}
    \\
    \\
& \underline{\phantom{123456789012345678901234567890123456789012}}\\
    & \secondprof{}\\
    & {\small Professor, Committee Member}\\
    \vspace{0.2cm}
    \\
    \\
& \underline{\phantom{123456789012345678901234567890123456789012}}\\
    & \thirdprof{}\\
    & {\small  Professor, Committee Member}\\
    \vspace{0.2cm}
    \\
    \\
\hspace{1.4ex}Departmental approval: & \underline{\phantom{123456789012345678901234567890123456789012}}\\
    & \csdchair{}\\
    & {\small Professor, Director of Graduate Studies}\\
\end{tabular}
\\

\vfill Heraklion, \thesisdate{}
\end{center}

\thispagestyle{empty}

\end{titlepage}

\cleardoublepage

\thispagestyle{empty}
\begin{titlepage}
\begin{center}
%\author{}
%\bc \Large{ \textbf{\thesistitle{}}} \ec \bc \owner \ec \bc {\bf\large Abstract}\\\ec
\bc \Large{ \textbf{\thesistitle{}}} \ec \bc {\bf\large Abstract}\\\ec

\end{center}

Nowadays, avoiding system calls %comes as a necessity 
during cluster communication (e.g. in Data Centers, High Performance Computing etc) in modern high-speed interconnection networks comes as a necessity, due to the high overhead of the multiple copies (kernel-to-user and user-to-kernel). User-level zero-copy Remote Direct Memory Access (RDMA) technologies overcome this problem and, as a result, increase the performance and reduce the energy consumption of the system. Common RDMA Engines like these cannot tolerate page faults caused by them and choose different ways to circumvent them.

The state-of-the-art RDMA techniques usually include pinning address spaces or multiple pages per application. This approach has some disadvantages in the long run, as a consequence of the complexity induced in the programming model (pinning/unpinning buffers), the limit of bytes that an application is allowed to pin and the overall memory utilization. Furthermore, pinning does not guarantee that someone will not experience any page-faults, due to internal optimization mechanisms, such as Transparent Huge Pages (THP), which is enabled by default in modern Linux operating systems. 

This thesis implements a page fault handling mechanism in association with the DMA Engine of the ExaNeSt project. First, the fault is detected by the fault handler of the ARM System Memory Management Unit (SMMU). Then, our hardware-software solution resolves the fault. Finally, a retransmission %of the transfer 
is requested by the mechanism, if needed. In our system, this mechanism required modifications to the Linux driver of the SMMU, a new library in software, alterations to the hardware of the DMA engine and adjustments to the scheduler of the DMA transfers. Our tests were run on the Quad-FPGA Daughter Board (QFDB) of ExaNeSt, which contains Xilinx Zynq UltraScale+ MPSoCs.

We evaluate our mechanism and we compare against alternatives such as pinning or \enquote{pre-faulting} pages, and we discuss the merits of our approach.

\vfill
\end{titlepage}

\cleardoublepage

\thispagestyle{empty}

\selg

\begin{titlepage}
\begin{center}
\bc \Large{\textbf{Χειρισμός των Σφαλμάτων Σελίδας Μνήμης στη Διάρκεια Άμεσων Απομακρυσμένων Προσβάσεων Μνήμης με Εικονικές Διευθύνσεις}} \ec \bc {\bf\large Περίληψη}\\\ec
\end{center}

Στις ημέρες μας, η αποφυγή των κλήσεων συστήματος κατά την διάρκεια επικοινωνίας συστάδων υπολογιστών (π.χ. Κέντρα Δεδομένων, Υπολογισμοί Υψηλής Επίδοσης κ.ά.) στα μοντέρνα δίκτυα διασύνδεσης υψηλής-ταχήτητας προκύπτει ως ανάγκη, λόγω του υψηλού κόστους των πολλαπλών αντιγράφων (πυρήνα-σε-χρήστη και χρήστη-σε-πυρήνα). Οι τεχνολογίες Άμεσων Απομακρυσμένων Προσβάσεων Μνήμης (\textlatin{Remote Direct Memory Accesses -- RDMA}), οι οποίες είναι επιπέδου-χρήστη και μηδενικών{-}αντιγράφων, ξεπερνούν αυτό το πρόβλημα και ως αποτέλεσμα αυξάνουν την επίδοση και μειώνουν την κατανάλωση ενέργειας του συστήματος. Κοινές μηχανές \textlatin{RDMA} όπως αυτές, δεν ανέχονται τα σφάλματα σελίδας μνήμης που προκύπτουν από εκείνες.

Οι τελευταίας τεχνολογίας τεχνικές συνήθως περιλαμβάνουν \enquote{καρφίτσωμα} \textlatin{(pinning)} των χώρων διευθύνσεων ή πολλαπλών σελίδων μνήμης για κάθε εφαρμογή. Αυτή η προσέγγιση έχει κάποια μειονεκτήματα μακροπρόθεσμα, ως συνέπεια της πολυπλοκότητας, η οποία προκαλείται στο προγραμματιστικό μοντέλο (\enquote{καρφίτσωμα}/\enquote{ξε-καρφίτσωμα} ενταμιευτών), του ορίου από \textlatin{bytes} τα οποία μία εφαρμογή επιτρέπεται να κάνει \enquote{\textlatin{pin}} και της συνολικής χρήσης της μνήμης. Επιπλέον, το \enquote{καρφίτσωμα} σελίδων μνήμης δεν εξασφαλίζει ότι κάποιος δεν θα αντιμετωπίσει κανένα απολύτως σφάλμα σελίδας, εξαιτίας των εσωτερικών μηχανισμών βελτιστοποίησης, όπως ο \textlatin{Transparent Huge Pages (THP)}, ο οποίος είναι ενεργοποιημένος ως προεπιλογή στα μοντέρνα λειτουργικά συστήματα \textlatin{Linux}.

Αυτή η εργασία υλοποιεί έναν μηχανισμό διαχείρισης των σφαλμάτων σελίδας μνήμης σε συνεργασία με την μηχανή \textlatin{DMA} του έργου \textlatin{ExaNeSt}. Πρώτα, ανιχνεύεται το λάθος από τον διαχειριστή σφαλμάτων του προγράμματος οδήγησης \textlatin{(driver)} της Μονάδας Διαχείρισης Μνήμης Εισόδων/Εξόδων \textlatin{(IOMMU)} της \textlatin{ARM}, η οποία ονομάζεται \textlatin{SMMU}. Έπειτα, η λύση υλικού-λογισμικού μας επιλύει το σφάλμα. Τέλος, στέλνεται ένα αίτημα ώστε να γίνει επαν-αποστολή, %της πληροφορίας
όταν χρειάζεται. Στο σύστημα μας, ο μηχανισμός αυτός χρειάστηκε τροποποιήσεις στο πρόγραμμα οδήγησης \textlatin{(driver)} \textlatin{Linux} για την \textlatin{SMMU}, την υλοποίηση μίας νέας βιβλιοθήκης λογισμικού, αλλαγές στο υλικό της μηχανής Άμεσων Απομακρυσμένων Προσβάσεων Μνήμης και τροποποιήσεις στον χρονοπρογραμματιστή των μεταφορών Απομακρυσμένων Προσβάσεων Μνήμης. %(λογισμικό επεξεργαστή \textlatin{R5}). 
Οι δοκιμές μας έγιναν επάνω στο \textlatin{Quad-FPGA Daughter Board (QFDB)} του \textlatin{ExaNeSt}, το οποίο εμπεριέχει τα \textlatin{Xilinx Zynq UltraScale+ MPSoCs}.

Εκτιμούμε το κόστος του μηχανισμού μας και κάνουμε σύγκριση με τις εναλλακτικές επιλογές όπως το \enquote{καρφίτσωμα} ή την πρώιμη πρόκληση σφαλμάτων σελίδας μνήμης, και συζητάμε τα οφέλη της δικής μας προσέγγισης.

\vfill

\end{titlepage}

\cleardoublepage

\selam %new...
\thispagestyle{empty}
%\bc \large{\textbf{Ευχαριστίες}} \ec
\bc \large{\textbf{Acknowledgements}} \ec

The work for this thesis was performed at the CARV Laboratory of ICS-FORTH, from June 2017 to June 2019 and was supported by the ExaNeSt project, which was funded by the European Commission under the Horizon 2020 Framework Programme (Grant Agreement 671553).

There are many people I would like to thank and give credit to because of their generous help in many ways throughout my M.Sc. thesis.

First of all, I would like to thank my supervisor Prof.~\firstprof, who supported, guided and trusted me during my thesis. We knew from the beginning that the topic of my M.Sc. thesis is challenging, yet we did not hesitate to choose and work on it.

Secondly, I would like to take this opportunity to thank Prof.~\secondprof~and Prof.~\thirdprof, for being members of my M.Sc.~Committee and their feedback.

I would like to express my gratitude to Dr.~Fabien Chaix. He was always supportive from the beginning of my M.Sc. thesis. I am grateful for his guidance and interest in my work and progress. Dr.~Chaix provided to me constructive feedback during the design, implementation and verification of the FIFO that was implemented, as part of this thesis.%; I believe it is fundamental for the progress of any work to have someone to regularly ask questions, challenge you and give you constructive feedback --I am lucky Dr.~Chaix did this for me throughout my M.Sc. thesis. 

Also, I would like to thank Dr.~Nikolaos Chrysos. We had brief discussions in the beginning of my M.Sc.~thesis that became more regular and extended in the latter part of my work, when we spent more time both on the translation-fault path of the FORTH PLDMA as well as the mechanism implemented as part of this thesis. I am thankful for Dr.~Chrysos' patience, guidance and support. 

I would like to give credits to Marios Asiminakis and Vasilis Flouris. Their advice, support and mindset helped me overcome many obstacles in the implementation, the evaluation and even the way of presenting the results. %The rewarding system Vassilis introduced, that mostly he and Marios used, helped me to stay focused, be more precise and somehow braked my random reckless character.

Also, I would like to thank Dr.~Vassilis Papaefstathiou for his feedback in the design of the mechanism, especially during the beginning of the thesis. 

I would like to thank Michalis Giannioudis, Pantelis Xirouchakis, Leandros Tzanakis, and Dr.~Nikolaos Chrysos, who have worked on and implemented the FORTH PLDMA. I collaborated with Giannioudis, Xirouchakis, and Dr.~Chrysos on the debugging of the translation-fault path of the PLDMA. I mostly worked with Giannioudis, since he is the designer of the receive path of the FORTH PLDMA, where the FIFO part of this thesis resides.

I would also like to thank Dr.~Manolis Marazakis, Nikolaos Dimou, Nikolaos Kossifidis and Panagiotis Peristerakis, for their support, whenever it was needed.

I greatly appreciate the feedback that Dr.~Fabien Chaix, Sotiris Totomis, Vasilis Flouris, Nikolaos Dimou, Pantelis Xirouchakis gave me for the document of my thesis before I submitted it.

% Last but not least and while I am hoping I did not forget to thank personally anyone that I should have, I would like to thank everyone in the CARV Laboratory of the Institute of Computer Science, FORTH, for all the support throughout my thesis. I could not be more grateful.

Last but not least and while I am hoping I did not forget to thank personally anyone that I should have, I would like to thank everyone in CARV, for all the support throughout my thesis. I could not be more grateful.
	
\cleardoublepage

%\selg %new..
\begin{flushright}
\emph{
\newline \newline \newline \newline \newline \newline \newline \newline
To my father George, mother Rania and sister Marianna} \\
\end{flushright}

\cleardoublepage

\selam

%\thispagestyle{empty}

% \AtBeginDocument{%
%  \addtocontents{toc}{\protect\thispagestyle{plain}}%
%  % \addtocontents{lof}{\protect\thispagestyle{plain}}%
% }
% \pagestyle{plain}
% \fancyhf{}                    % clear all fields
% \fancyfoot[C]{\thepage}       % center footer

\selam

\pagestyle{plain}
\pagenumbering{roman}
\setcounter{page}{1}
\setcounter{tocdepth}{3}

\tableofcontents
    % \addtocontents{toc}{\protect\contentsline {chapter}{Table of Contents}{i}}

\listoftables
    % \addtocontents{toc}{\protect\contentsline {chapter}{List of Tables}{iii}}

\listoffigures
    % \addtocontents{toc}{\protect\contentsline {chapter}{List of Figures}{v}}

\cleardoublepage

\pagenumbering{arabic}

\pagestyle{headings}
%\pagestyle{IHA-fancy-style}

% Chapter 1
\chapter{Introduction}\label{Chapter1}

In this chapter we will describe the motivation behind this thesis, the contributions and the background information, that was needed for the purposes of this thesis, hardware- and software-wise.

\section{Motivation}
\label{unimem}

\begin{figure}[h]
        \centering
        \captionbox[A system with many nodes that are connected]{A system with many nodes (N in total) that are connected \label{fig:unimem}}{%
        \fontsize{7}{11}\selectfont\includegraphics[width=1\textwidth]{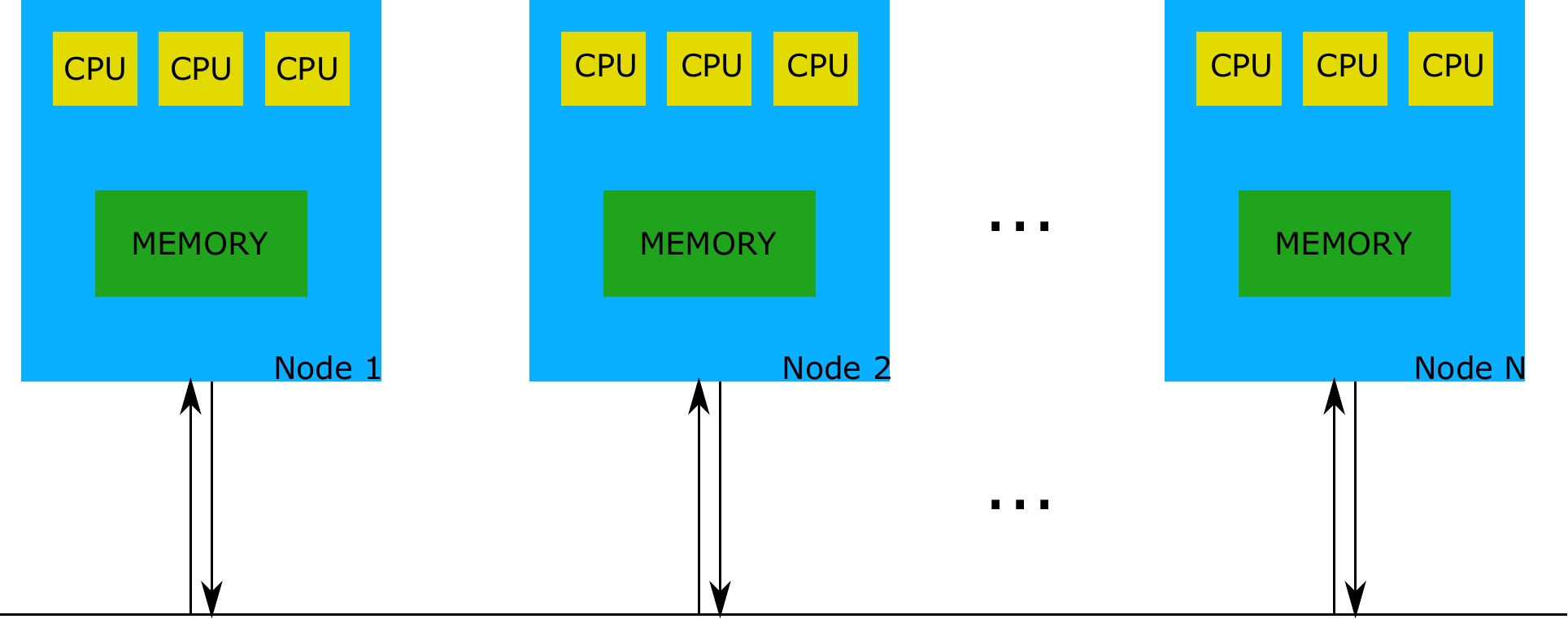}
        }
        %\caption{SVG in superior directory}
\end{figure}

As the years go by, when it comes to Computer Science and the industry, it seems there is a direction in creating more complex sets of computers that will be able to solve difficult problems. An example would be many computing nodes, or \enquote{Coherent Islands}, as they are described in \textit{Unimem} \cite{exanest}, trying to communicate with each other (e.g. transfer data). Figure~\ref{fig:unimem} provides a very simplistic view of that system.  

Coherence among all cores (CPUs) in one node is somehow granted. The problem occurs when we have many nodes embedding many coherent cores and we want to achieve coherence among them -- the more nodes the bigger the problem. 

The ExaNeSt project \cite{exanest} tries to address this issue by allowing remote memory accesses through a Global Virtual Address Space (GVAS).

% OLD
% \begin{figure}[h]
% 	\centering
% 	%\caption{lalala}
% 	\captionbox[A system with many nodes that are connected]{A system with many nodes that are connected \label{fig:unimem}}{%
% 		\includegraphics[width=0.7\textwidth]{unimem}  
% 	}
% \end{figure}

% % NOT WORKING: 
% \begin{figure}[h]
%         \centering
%         \includesvg[width=1\textwidth]{Figures/computing_nodes.svg}
%         \caption{SVG in superior directory}
% \end{figure}
% % NOT WORKING:
% % \begin{figure}[h]
% %         \centering
% %         \escapeus{\includesvg[width=1\textwidth]{Figures/drawing}}
% %         \caption{SVG in superior directory}
% % \end{figure}

In order for such system to be coherent, each node that needs to access information, that belongs to the memory of another node, should request the information from the other node, without keeping a copy to its memory. In order to achieve this goal, Unimem proposed a virtualized system that distinguishes all nodes by utilizing a virtual global address space. This has as a result the necessity of an \enquote{arbiter} in each node, that will handle all incoming transactions in order to access safely their memory. 
% \vfill
% \footnoterule
% \pagebreak
This arbiter is called I/O Memory Management Unit (IOMMU) and is necessary in each node.

We prototype this approach by using one Xilinx Zynq UltraScale+ \cite{zynq_usplus} per node. This chip consists of many components including four (4) ARM 64-bit A53 cores and one IOMMU (or in other words, ARM SMMU - System Memory Management Unit).
%In general, SMMU enables the translation of virtual addresses to their corresponding physical addresses.

Kernel-level transactions, as the name reveals, are initiated by the kernel (e.g. Linux) of a node. An alternative is the user-level initiation of remote DMA (RDMA) transactions, which eliminate the involvement of kernel (e.g. system calls) during the transfers, by reducing unwanted overheads such as the initiation latency~\cite{DBLP:conf/hpca/MarkatosK97}. User-level zero-copy RDMA transfers allow page migration, simplify multi-programming and also improve security, just as virtualization did in the past and well-known single-node systems \cite{virtualization}. 

Common user-level RDMA approaches cannot tolerate page faults during the RDMA transfer, thus avoid them by pinning multiple pages or whole address spaces of the processes. Pinning can hinder the memory utilization \cite{DBLP:conf/ipps/BellB03} and might not suffice due to optimizations of the Operating System (e.g. Transparent Huge Pages) or lack of privileges for a user/application. We also argue that pinning complicates the programming model because a kind of synchronization might be needed before/after the DMA and some extra actions from the application/user side are expected (e.g. someone has to unpin the memory after the DMA). %\cite{DBLP:conf/asplos/AmitTS14}, \cite{DBLP:journals/cacm/GuptaLVSSVVV10},\cite{DBLP:journals/sigops/WoodTSDCC09}
A non-pinned memory-pages design results to page faults that require handling, which is the main work of this thesis.

% \vfill
% \footnoterule
% \pagebreak

%%%
%%% NEW SECTION
%%% TOPIC: Contributions
%%%
%%%
%%%

\section{Contributions}

In cluster-communication systems we prefer to eliminate the kernel involvement and the multiple copies (kernel-to-user and user-to-kernel) when communicating. User-level zero-copy Remote DMA transfers answer this problem. This thesis addresses the need of supporting page faults that might occur during these RDMA transfers. This work was supported by the ExaNeSt project.

%This thesis is trying to address the fundamental need of ExaNeSt and any system really that wants to support virtual-address remote Direct Memory Accesses (DMA) -- which is handling the page faults, that might occur.%% as described in Section \ref{unimem}, by understanding how the SMMU (ARM's IOMMU) works and how we could use it in order to achieve virtual-address remote Direct Memory Access (DMA) transactions in one or between multiple nodes -- for the purposes of this thesis only one node was considered in the experiments, but the information and results generated by this thesis could also support future implementations for multiple nodes. 

Initially, the author of this thesis implemented and tested the main components of the page fault handling mechanism using the Low-Power-Domain (LPD) DMA Engine embedded in the Processing System of the Zynq UltraScale+. When the prototype of ExaNest became more mature, this author worked on handling the page faults occurred during RDMA transfers through the custom DMA engine residing in the Programmable Logic (FORTH PLDMA). The PLDMA Engine --FORTH PLDMA (designed by other members of the Lab) could provide useful and necessary information for the page fault handling mechanism, that required a different approach to be used than the approach considered when using the LPD DMA Engine in the Processing System. For the rest of this thesis, the author describes only the mechanism developed when using the FORTH PLDMA.

In order to achieve the goals of this thesis, the work consisted of three different parts. 

First, this author worked on the background/theory of a page fault -- what are the main reasons that it can be caused and the extensive research needed in kernel code in order to handle the fault properly. %how deep someone should go in kernel code to handle it properly etc. 
This includes all the information and knowledge that was collected and used in order to proceed with the implementation of the necessary modifications in the Linux kernel driver of ARM's IOMMU (SMMU) and the library for any application/user. Chapter~\ref{Chapter1} and Chapter~\ref{Chapter2} provide a detailed description of all this information.

Second, this author worked on the implementation: all modifications in the Linux kernel driver of ARM's IOMMU and the user library necessary for the mechanism, that will be activated when a page fault is caused during a RDMA from a user application. Also, as part of the implementation, the author modified the receiver hardware block of the FORTH PLDMA Engine and added extra functionality on the scheduler of the RDMA transfers (firmware), which is the R5 processor. Chapter~\ref{Chapter3} has a detailed description for the implementation.

Third, this author worked on evaluating the mechanism. This was essential in order to see the good and bad aspects of the current mechanism and propose %where a reader should focus in the future, when trying to do optimizations. 
possible future optimizations.
Detailed information about the measurements and how they were conducted, can be found in Chapter~\ref{chap:conclusion}.

% \vfill
% \footnoterule
% \pagebreak

%%%
%%% NEW SECTION
%%% TOPIC: Context/Environment
%%%
%%%
%%%

\section{Environment}
\label{sec:environment}
\subsection{Hardware}

\subsubsection{Trenz board}\label{trenz}

The hardware that was initially used was a \href{https://wiki.trenz-electronic.de/pages/viewpage.action?pageId=24153560}{TET0808 Trenz board} with the FPGA: XCZU9EG-FFVC900-1 \cite{te0808trm}. It is an MPSoC module integrating a Xilinx Zynq UltraScale+, that includes a 2 Giga Byte DDR4 SDRAM with 64-Bit width, 64 MByte Flash memory for configuration and operation, 20 Gigabit transceivers, and switch-mode power supplies for all on-board voltages.\par

\subsubsection{Quad-FPGA Daughter Board}
%// NEED TO CHECK ALL DETAILS\n

%The ExaNet Network IP (ExaNIP) is the component in charge of managing the communications (i) among the FPGAs inside a QFDB (Intra-Node Network), (ii) QFDBs within the same mezzanine (Intra-Mezzanine communication) and (iii) mezzanines within a Chassis (Inter-Mezzanine communication).

The hardware that was later and finally used in order to achieve our goals was based on Quad-FPGA Daughter Boards (QFDB), each one of them embedding four (4) Xilinx Zynq UltraScale+ MPSoCs: XCZU9EG-FFVC900 \cite{zynq_usplus}, with 64 Giga-Byte of DDR4 SDRAM (16GB/FPGA at 160Gb/s), 512 Giga-Byte SSD/NVMe (4x PCIe v2 (8 GBytes/s)) and 10 High Speed Serial (HSS) links (10 Gb/s per link). 

According to ExaNeSt project~\cite{exanest}, the QFDB FPGAs are specialized for different tasks: 
\begin{itemize}
    \item two of them are pure computing nodes
    \item the \enquote{Storage FPGA}, among other things, manages an SSD interface
    \item the \enquote{Network FPGA} is the QFDB network peer
\end{itemize}

\begin{figure}[h!]
	\centering
	\captionbox[Schematic of the QFDB module]{Schematic of the QFDB module \label{fig:qfdb_schematic}\\{\small Source: ExaNeSt \cite{exanest}}}{%
		\includegraphics[width=0.7\textwidth]{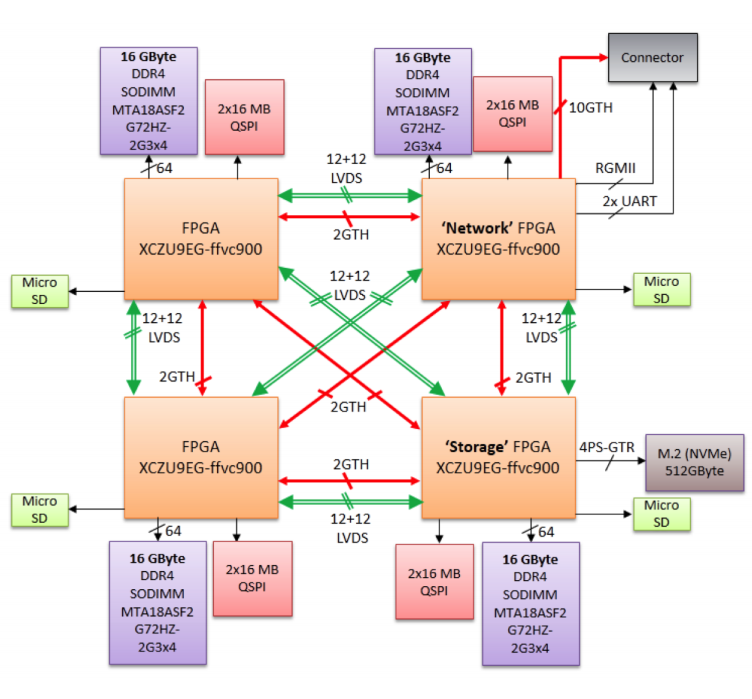}
	}
\end{figure}

The first experiments were conducted on QFDBs on mini feeders, and the final target was QFDBs embedded on a Mezzanine board. FPGAs on one QFDB (one hop) are connected (intra-QFDB) and QFDBs on a Mezzanine are also connected, consisting a network of nodes. The routing between nodes occurs using coordinates that differentiate them. In Figure~\ref{fig:qfdb_schematic} from \cite{exanest}, we can see a schematic of the QFDB module. \par

%\todo[inline]{Find figure of QFDB schematic in .svg?!}

% \vfill
% \footnoterule
% \pagebreak

\subsubsection{Zynq UltraScale+}\label{r5_ref}

%\noindent

Zynq UltraScale+ MPSoC is the Xilinx second-generation Zynq platform, combining a Processing System (PS) and user-Programmable Logic (PL) into the same device. 
The Zynq UltraScale+ MPSoC has four (4) different power domains.

\begin{itemize}
	\item Low-power domain (LPD)
	\item Full-power domain (FPD)
\end{itemize}

It also has the PL power domain (PLPD) and the Battery power domain (BPD), that are outside of the scope of this thesis. %The FPD was also not used.
The Zynq UltraScale+ MPSoC PS block has three major processing units:

\begin{itemize}
	\item Cortex-A53 application processing unit (APU)—ARM v8 architecture-based 64-bit quad-core multiprocessing CPU
	\item Cortex-R5 real-time processing unit (RPU)—ARM v7 architecture-based 32-bit dual real-time processing unit with dedicated tightly coupled memory (TCM)
\end{itemize}

The third processing unit is the Mali-400 graphics processing unit (GPU) with a pixel and geometry processor and a 64KB L2 cache. For the purposes of this thesis we will not use it.

\begin{figure}[h!]
	\centering
	\captionbox[Zynq UltraScale+ MPSoC Top-Level Block Diagram]{Zynq UltraScale+ MPSoC Top-Level Block Diagram \label{fig:zynqusplus_toplvldiagram}\\{\small Source: Xilinx \cite{zynq_usplus}}}{%
		\includegraphics[width=0.9\textwidth]{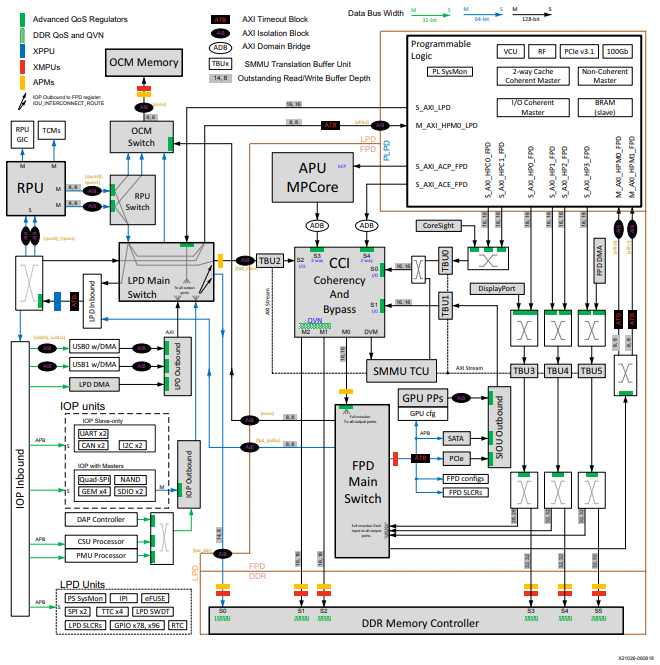}  
	}
\end{figure}

As we can see in Figure \ref{fig:zynqusplus_toplvldiagram}, in the top-level block diagram of the Zynq UltraScale+ MPSoC, part of the Processing System (PS) is the APU that consists of four (4) Cortex-A53 processors that will run the system and more specifically they will run Linux that we will use in order to achieve our goals.
%, using the IOMMU. 
Also, part of the PS is the RPU that consists of two (2) Cortex-R5 Real-Time processors, which is utilized by FORTH as the scheduler of ExaNeT, a network developed fully by FORTH as part of ExaNeSt.

As we can see, the SMMU and the CCI (Cache Coherent Interconnect) are in the same block in the top-level block diagram -- that of course does not mean that they indeed are part of the same block, but that they collaborate in order to have coherent accesses to the memory. %For the purposes of the thesis and because of the limited time, we will not focus on coherence issues. Also, as we will see later, we will use both the LPD-DMA (Low-Power Domain DMA) and the FPD-DMA (FPD-DMA) that are parts of the PS, in order to trigger transactions that we want to go through the SMMU, when we have virtual addresses. 
Last but not least, part of the top-level diagram is the Programmable Logic (PL). A user, among other things, can program the FPGA and add blocks in PL that can access the PS. This is where the custom PLDMA developed at FORTH (FORTH PLDMA) resides in, and we will embed part of our mechanism there.

\subsubsection{Memory Management Unit}

Serving the purposes of this thesis, we utilized one of Zynq UltraScale+ MPSoC's major components, which is the System Memory Management Unit (SMMU). Before we continue with the remaining of the thesis, we should explain the basic idea of the Memory Management Unit (MMU).

\paragraph{The Memory Management Unit (MMU)}\mbox{}

\begin{figure}[h!]
        \centering
        \captionbox[A simple use of the MMU]{A simple use of the MMU \label{fig:mmu}}{%
        \fontsize{8}{11}\selectfont\includegraphics[width=0.6\textwidth]{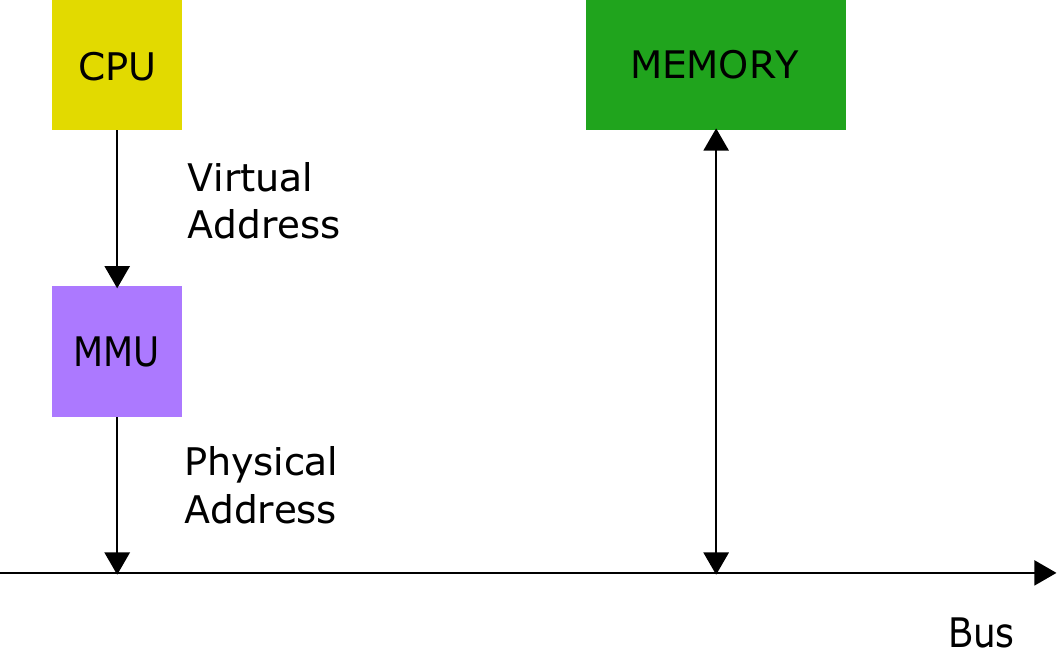}
        }
        %\caption{SVG in superior directory}
\end{figure}

The Memory Management Unit (MMU), in general, is a computer hardware unit having all memory references passed through itself, primarily performing the translation of virtual memory addresses to physical addresses. It is usually implemented as part of the CPU, but it also can be in the form of a separate integrated circuit. We can see a simplified use of the MMU in Figure~\ref{fig:mmu}.

An MMU can effectively perform virtual memory management, handling at the same time memory protection, cache control, bus arbitration and, in simpler computer architectures (especially 8-bit systems), bank switching. 

% \begin{figure}[h!]
% 	\centering
% 	\captionbox[A simple use of the MMU]{A simple use of the MMU \label{fig:3.3}}{%
% 		\includegraphics[width=0.6\textwidth]{mmu}  
% 	}
% \end{figure}

% \vfill
% \footnoterule
% \pagebreak

\paragraph{The I/O Memory Management Unit (IOMMU)}\mbox{}

\begin{figure}[h!]
	
	\centering
	\captionbox[Comparison of the IOMMU to the MMU]{Comparison of the IOMMU to the MMU \label{fig:iommuvsmmu}\\{\small Source: \href{https://en.wikipedia.org/wiki/Input-output_memory_management_unit}{Wikipedia}}}{
		\includegraphics[width=0.4\textwidth]{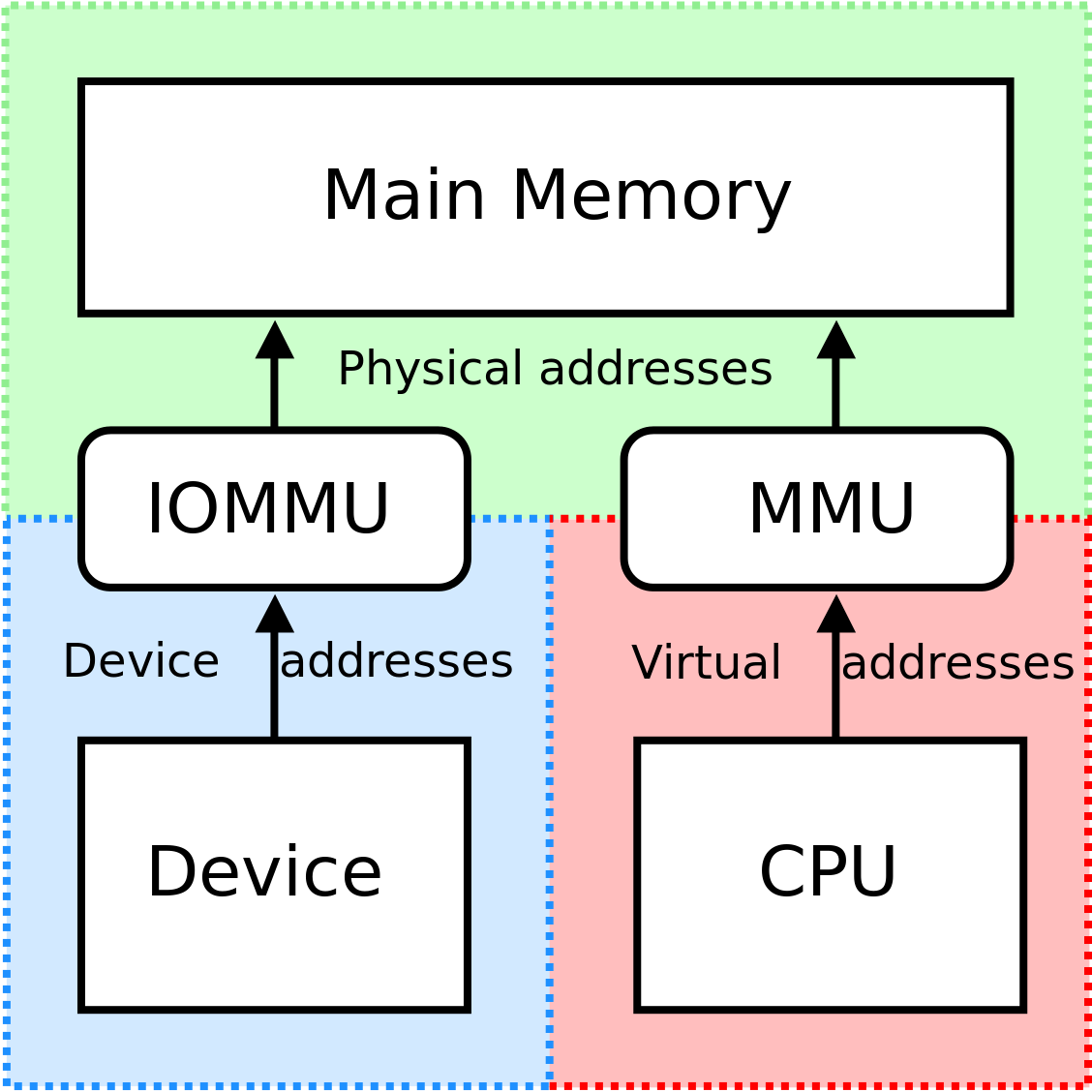}  
	}
\end{figure}

I/O Memory Management Unit (IOMMU) is an MMU that connects a Direct Memory Access-capable (DMA-capable) I/O bus to the main memory. IOMMU is basically responsible to map device-visible virtual addresses (also called device addresses or I/O addresses in this context) to physical addresses. In Figure~\ref{fig:iommuvsmmu} we can see a comparison of the IOMMU to the MMU.

% \vfill
% \footnoterule
% \pagebreak
\paragraph{The System Memory Management Unit (SMMU)}\mbox{}

In order to address some issues (e.g.: memory fragmentation, multiple DMA-capable masters, system integrity guarantee) and limitations, the ARM Architecture Virtualization Extensions introduced the System Memory Management Unit (SMMU) concept to the ARM Architecture. In other words, as we mentioned before, ARM's IOMMU is called SMMU.

% The SMMU performs address translation of an incoming AXI (Advanced eXtensible Interface) \cite{axiguide} address and AXI ID (mapped to context) to an outgoing address (physical address - PA), based on address mapping and memory attribute information held in translation tables.

The SMMU performs address translation of an incoming AXI (Advanced eX-tensible  Interface)~\cite{axiguide} address (virtual address - VA)  and  AXI  ID  (mapped  to  a context) to an outgoing address (physical address - PA), based on address mapping and memory attribute information held in translation tables~\cite{zynq_usplus}.

\begin{figure}[h!]
	\centering
	\captionbox[Examples of where a SMMU could be located in a system]{Examples of where a SMMU could be located in a system. \label{fig:arm_smmu}\\{\small Source: \href{https://www.arm.com/files/pdf/System-MMU-Whitepaper-v8.0.pdf}{ARM} }}{%
		\includegraphics[width=0.7\textwidth]{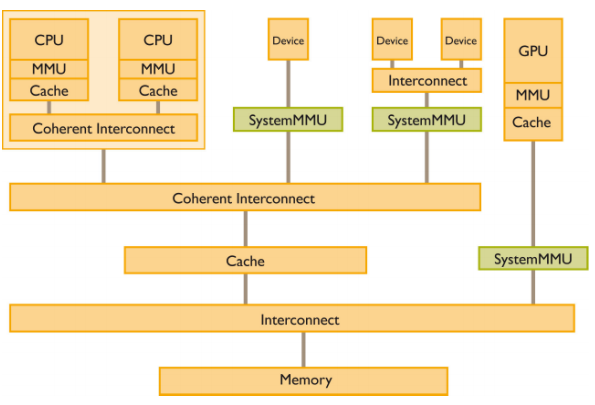}  
	}
\end{figure}

In Figure~\ref{fig:arm_smmu}, we can see examples of where a SMMU block could be located in a system -- coherent interconnects ensure cache coherence between masters.

\paragraph{Translation Context}\mbox{}\label{translation_cb}

The translation context (as known as the context bank) provides information and resources required by the SMMU to process a transaction. SMMU can process multiple transaction streams from different threads of execution and supports multiple live translation contexts.\\

A translation context bank includes:
\begin{itemize}
	\item state for configuring the translation process 
	\item capturing fault status, and operations for maintaining cached translations
\end{itemize}

In implementations that support stage 1 followed by stage 2 translations: 
\begin{itemize}
	\item one translation context bank is specified for single-stage translation
	\item two translation context banks are specified for two-stage translation
\end{itemize}

A translation context bank is arranged as a table in the SMMU configuration address map. Each entry in the table occupies a 4 KB or 64 KB address space.

In theory, SMMU architecture provides space for up to 128 translation context banks. In our Zynq UltraScale+ MPSoC, we have only 16 translation context banks.

Each context bank of the SMMU can be considered as one page table -- to be more accurate, each context bank has a field that points to a unique page table for this context. Each context bank has the \textbf{SMMU{\_}CB$n${\_}TTBR$m$}, where $m$ can be 0 or 1. TTBR0, as known as the Translation Table Base Register 0 holds the base address of translation table 0. Respectively, the TTBR1 holds the base address of translation table 0.

It is recommended by ARM that TTBR0 should be used to store the offset of the page tables used by user processes and TTBR1 should be used to store the offset of the page tables used by the kernel. It seems that most Linux implementations for ARM have decided to eliminate the use of TTBR1 and stick to using TTBR0 for everything -- this is also happening with the ARM SMMU driver (arm-smmu.c), we worked with.

For each context bank there is also the Translation Control Register, called \textbf{SMMU{\_}CB$n${\_}TCR}, that determines translation properties, including which one of the Translation Table Base Registers,
SMMU{\_}CB$n${\_}TTBR$m$, defines the base address for the translation table walk required when an input address is not found in the TLB. An extension of the SMMU{\_}CB$n${\_}TCR exists with the name \textbf{SMMU{\_}CB$n${\_}TCR2}, that basically extends the SMMU{\_}CB$n${\_}TCR by adding control information about the translation granule size and the size of the intermediate physical address.

\subsection{Firmware}\label{custom_PLDMA}

Page Fault Handling mechanism handles page faults caused during Remote Direct Memory Accesses (RDMAs). The mechanism, as we mentioned before, initially was built to support page faults caused when using the Direct Memory Access Engine, called ZDMA, embedded in the Processing System of the Zynq UltraScale+ MPSoC, which is a fundamental component of the QFDB. ZDMA was the first Engine that we used in order to support Virtual-Address Remote DMA, as part of the ExaNeSt. ZDMA has some fundamental limitations that would constrain the ultimate goal of ExaNeSt. Some of them include limited number of channels (8 for the low-power domain DMA engine), limited use of the address space (with ZDMA we could have access to a window of $\sim$500GB/s remotely -- which is equal to the physical address space from PS to PL, meaning we could only have 500~GB memory space per node), one acknowledgement per transfer (not the best option for a resilient environment), \enquote{end-to-end} retransmission, which is not possible with ZDMA (page faults or packet transmission errors), not clear if multi-pathing is actually supported by ZDMA, fast notifications (ZDMA would incur one (1) extra round-trip-time (RTT) latency). Also, low performance was detectable with ZDMA, probably because of the small packet size of 64 Bytes and the small number of outstanding transactions, which was six (6). These are some of the reasons that led ExaNeSt to design and implement a new custom-made programmable logic Direct Memory Access (PLDMA) Engine. This is a work designed, implemented and supported mainly by our colleagues at FORTH \cite{TzanakisMScThesis2019, XirouchakisMScThesis2019}. %\todo[inline]{ADD PROPER CITATION: PROPER CITATION for Michalis Giannioudis is missing.} 
The new PLDMA among other things supports low latency transfers, resilience (fast retransmissions) and multi-pathing.

When ExaNeSt’s own custom PLDMA was more mature, we moved our testing environment and implementation efforts for the Page Fault Handling mechanism to it. From this point on, we will focus only on the work implemented and used in order to perform the transfers that will be completed (successful) after deliberate failures due to one or more page faults. %If we need to describe components during the time we used ZDMA, we will state it explicitly.

Although the design and implementation of the FORTH PLDMA Engine was not part of this thesis, it was a necessary component during the testing process of our Page Fault Handling mechanism, which is the reason why we will briefly describe its fundamental parts and the basic idea of it.

%\todo[inline]{The new PLDMA supports (ADD INFO)…..}

%We will now try to describe briefly the flow of an RDMA transfer. 
In Section~\ref{r5_ref} we mentioned the Real Time co-processor R5 embedded in the Processing System (PS) of the Zynq UltraScale+ MPSoC. Co-processor’s main task is to segment, prioritize, initiate and monitor the DMA transfers. %Each transfer can use 1 of the 1024 virtualized channels, that might consist of many transactions. 
Each process, which runs under a Protection Domain can use 64 virtualized channels of the PLDMA. Our system supports up to 16 Protection Domains (since our SMMU has 16 context banks), thus the PLDMA can support up to $64x16=1024$ outstanding transfers.

%, that the R5 will make sure to split into 256 Byte blocks. 
Each transaction can be up to 16~KB. Real-time processor (R5) segments transfers into 16~KB blocks, and makes sure that they are 16~KB aligned (as a result some blocks might be $<16$~KB). Each transfer can have a parameterized number of outstanding transactions (currently the number is two (2)). The hardware segments 16~KB transactions into 256~Byte blocks. The main reason the transfers are splitted into packets with Maximum Transfer Unit (MTU) of 256~Bytes is that this size has shown to be more efficient at doing network congestion work. Furthermore small MTU can guarantee network buffers to be small, which saves utilization space and therefore cost, as mentioned in \cite{XirouchakisMScThesis2019}. It is really important that R5 can monitor the state of each block, by using acknowledgments and if necessary fast retransmissions. 

\subsubsection{Remote Write — Initiator/Sender part}

\begin{figure}[h]
	\centering
	\captionbox[Remote write flow path between two nodes]{Remote write flow path between two nodes \label{fig:remwr_schematic}}{%
		\includegraphics[width=0.9\textwidth]{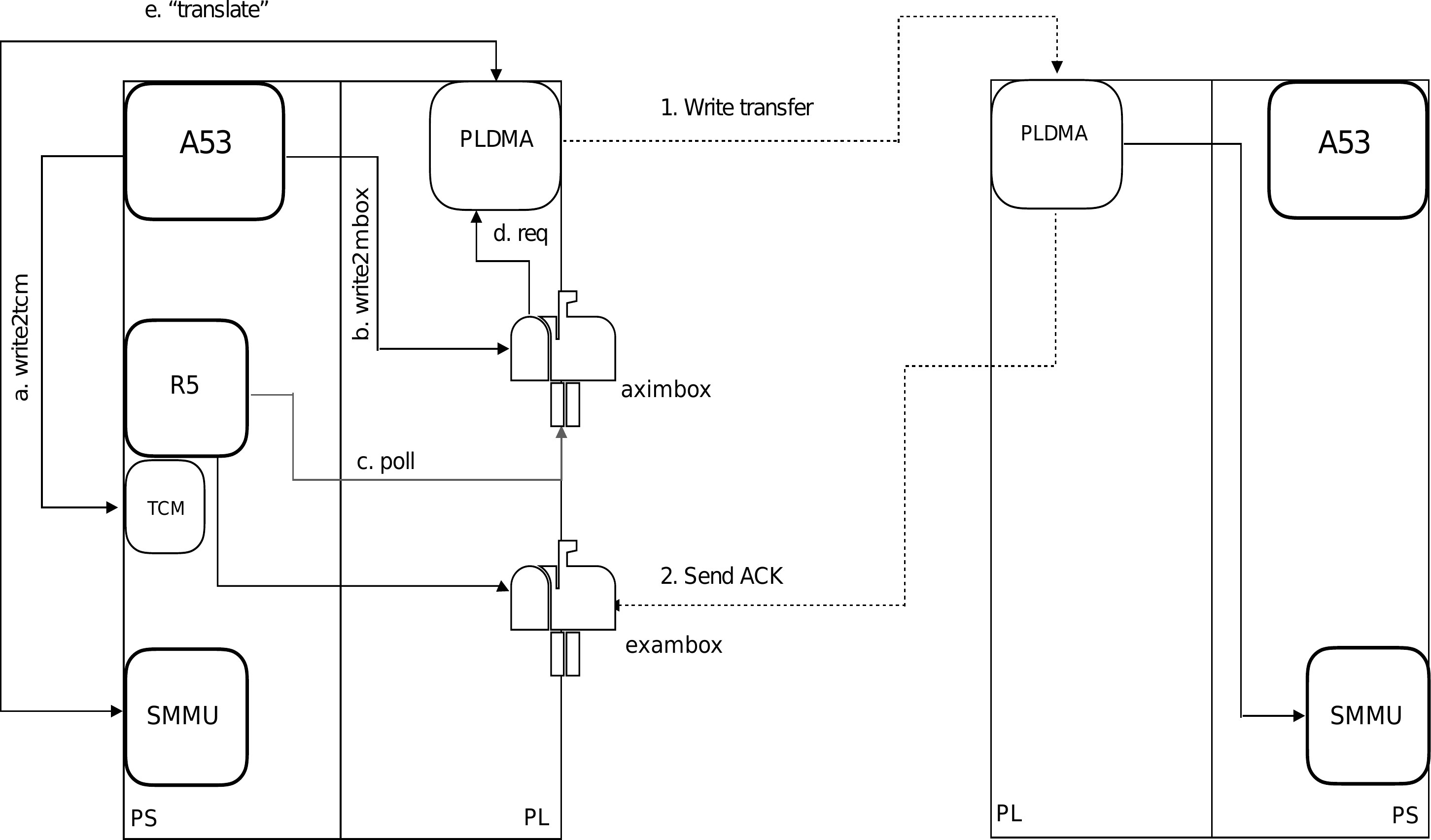}  
	}
\end{figure}

First, ARM’s A53 cores and the user-space library provided to support RDMA, initialize both the TCM (scratchpad) of R5 and the mailbox dedicated to an R5 core. When the R5 core detects new information in its dedicated mailbox, it initiates an RDMA transfer, by also initializing accordingly the necessary components of the hardware. The custom hardware of PLDMA will be advised by the System Memory Management Unit (SMMU) to translate the remote virtual-address based on the protection domain of the process that triggered the DMA \cite{PsistakisBScThesis2017, PsistakisACACES2018}. Using the SMMU we can achieve transfers to/from pages that reside in DRAMs of different computing nodes. 
If this translation is successful we have a remote DMA write transfer to another (or the same) computing node. If the translation at the other node using, once again, the SMMU is successful, then we probably (not certainly, because it might trigger a permission fault) have the memory access, that eventually generates a positive acknowledgment that is sent to another mailbox dedicated to the R5 of the initiator node. This is how the node that initiated the transfer is becoming aware of the completion of the transfer.

\subsubsection{Remote Read — Target/Receiver part}

For a remote read transaction, the process, running on an ARM A53 core, triggers the transfers using a user-space library that sets a packetizer responsible to forward the read request to the corresponding (probably remote) node. Then, a mailbox dedicated to the R5 of the target node receives the request and transforms, in a way, this request to a write transfer. This way we make sure that we have the same mechanism as a write request, which makes easier the programming model. In simple words, this allows us to reuse same blocks of hardware that can perfectly fit our purposes. The target node will write to the memory of the initiator node, which effectively acts like a read request.

% \vfill
% \footnoterule
% \pagebreak

\subsection{Software}

Most of the effort for this thesis was related to Software aspects. From the environment we used to test our prototype-mechanism to the user-space library of the mechanism and the generation of the needed image files (kernel, real-time, etc). Part of the software that was used for our mechanism includes the Netlink sockets, which we will briefly describe later in this sub-section.

\subsubsection{Linux}

We chose to use the Linux OS as the Operating System for the purposes of this thesis - the Linux version was 4.9.0.
%\todo[inline]{UPDATE Linux part...}

At first, while working on Trenz boards (Section~\ref{trenz}), we were using SD cards loaded with the necessary image files, such as the image file of the kernel (Image), the device tree, the FSBL etc. Before migrating to our brand new prototype, the QFDB, a tool called \enquote{yat} was implemented at FORTH to automatically generate all necessary image files according to the requested and given as input details of the platform. After setting it up, by giving the Xilinx Software Development Kit (SDK) version and the name of the profile (trenz, QFDB, etc..), we could generate our images and finally our BOOT.bin.

Since most part of this work was mainly kernel development, when it comes to Linux we had to generate a new Linux kernel image after every new modification to the ARM SMMU Linux driver. After the build was done, we could \enquote{kexec} the new kernel Image on top of the Linux environment on QFDBs, that was already boot-ed. The process of booting our own kernel image can be found in Appendix~\ref{Appendix_Environment}.

% At first, we used the pre-built images provided by Trenz in order to flash the QSPI (Queued Serial Peripheral Interface -- a type of SPI controller that uses a data queue to transfer data across the SPI bus) memory with Linux. Although the pre-built images worked, it was not enough for our purposes because those pre-built images had a very minimal version of Linux. When we tried to find another way to create our own Linux images, the Petalinux method was the suggested one by Trenz and Xilinx. Although, again, there was a document with the process of creating images via Petalinux environment, the steps and the details were not clear, and after spending much time trying to make them work, we decided to use part of the working files (or images) like the device tree (.dtb) and the first-stage-boot-loader (FSBL) and the rest of the images, that we needed, from other projects that were working already.

% As mentioned before, we used to boot from the QSPI (that we were flashing using the JTAG), but currently, since we now have more extensive base boards that support the SD card, we are booting from there. Appendix \ref{AppendixA} describes the process of booting from the SD Card. 

\subsubsection{Xilinx Environment}

We used Vivado, a software suite produced by Xilinx for synthesis and analysis of HDL designs, and SDK (Software Development Kit), that is the Integrated Design Environment for creating embedded applications on any of Xilinx's microprocessors like the Zynq UltraScale+ MPSoC. 

Although it was not part of this work, Xilinx Vivado was used to export the bitstream (.bit) and the hardware design/description file (.hdf), and the Xilinx SDK to create the device tree (.dtb) and the first-stage-boot-loader (.fsbl) from the exported hardware design/description file (.hdf). As we mentioned before, prior to our migration to the QFDB prototype, we had in our hands a tool, called \enquote{yat}, that could generate all necessary files for us, such as the First Stage Boot Loader (FSBL), the Power Management Unit Firmware (PMUFW), the Flatened Device Tree image for the board (DTB), the EL3 Secure Monitor (BL31), the Second Stage Boot Loader (U-Boot), the Linux Kernel Image (Kernel), the Ramfs image (Initramfs) and finally the BOOT.bin, that included all the necessary image files.

Part of this thesis was to build a modified, by the author, version of Real-Time R5 co-processor, in order to produce a firmware that could support our mechanism. In order to achieve this we used Xilinx SDK tool.

The version of SDK mainly used during this work was: 2017.2 at first, and then 2017.4.

\subsubsection{Modules/Drivers}\label{driversvsmodules}

At this point, it is good to define \enquote{modules} and \enquote{drivers}, that were a big part of this thesis.

A device driver (commonly referred to simply as a driver) is a computer program that operates or controls a particular type of device that is attached to a computer. A driver provides a software interface to hardware devices, enabling operating systems and other computer programs to access hardware functions without needing to know precise details of the hardware being used.

A module is a piece of code that can be loaded and unloaded into the kernel upon demand. Modules extend the functionality of the kernel without the need to reboot the system. For example, one type of module is the device driver, which allows the kernel to access hardware connected to the system. Without modules, we would have to build monolithic kernels and add new functionality directly into the kernel image. Besides having larger kernels, this has the disadvantage of requiring us to rebuild and reboot the kernel every time we want to add a new functionality. 

In order to test and evaluate the mechanism of this thesis, the author developed and modified some drivers and modules, which will be described in Chapter~\ref{Chapter3}.

\begin{figure}[h!]
	\centering
	\captionbox[The Kernel, The Processes And The Hardware]{The Kernel, The Processes And The Hardware.\label{fig:kernel_process_hw}\\{\small Source: \href{http://haifux.org/lectures/86-sil/kernel-modules-drivers/kernel-modules-drivers.html}{haifux.org}}}{%
		\includegraphics[width=0.5\textwidth]{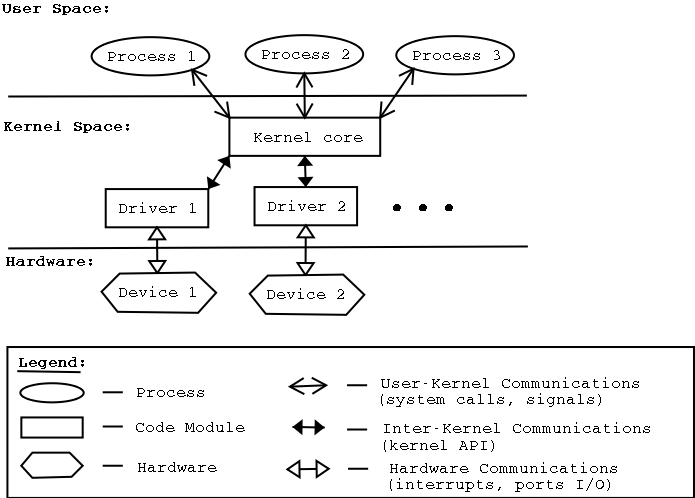}  
	}
\end{figure}

In Figure~\ref{fig:kernel_process_hw}, we can see how a system can use drivers and modules.

\subsubsection{Netlink Sockets}

Netlink sockets first appeared in Linux kernel 2.2 mainly as a flexible alternative to the IOCTL communication method, that can be used between userspace processes and the kernel. One of the main disadvantages of IOCTL method, apart from the complexity to use, is that the IOCTL handlers cannot send asynchronous messages to the userspace process from the kernel. When using the IOCTL method it is required from the programmer to define IOCTL numbers, that of course by itself increases the complexity considerably. The Linux header file\textit{ /usr/include/asm/iocth.h} defines macros that must be used to create the IOCTL command number for each command used through IOCTL. This number should be unique for the whole system, and picking it arbitrarily is a bad idea, that could lead to bad situations, including damage to hardware \cite{ioctl_lab6}. The userspace processes that use the Netlink library \enquote{open} and then \enquote{register} a Netlink socket, in order to handle bidirectional communication (send and receive messages) from the kernel.

It should be mentioned that in theory Netlink sockets can also be used for the communication between two (2) userspace processes, but this is not very common and certainly not the original goal of Netlink sockets.

Some advantages using Netlink sockets over other ways of communication:
\begin{itemize}
    \item No need for polling
    \item Kernel can be the initiator of sending asynchronous messages to userspace. For alternatives such as IOCTL or sysfs entry, an action from userspace is required.
    \item Supports multicast transmission
\end{itemize}

According to the manual, Netlink protocol is not considered reliable and may drop messages when an out-of-memory condition or other errors occur. When a message from kernel to userspace cannot be sent, the message will be dropped, which means the application and the kernel will no longer be able to have the same view of the kernel state. It is the responsibility of the application to detect it when this happens.

\subsubsection{Tasklets}\label{tasklet}

Several tasks among those executed by the kernel are not critical, which means that they can be delayed for a long period of time, if it is necessary.

In general, interrupt service routines are not preempt-able, until the corresponding interrupt handler has terminated. Tasklets overcome this critical restriction, by being preempt-able, which means they can execute while having all interrupts enabled. By having this functionality, we can keep the kernel response time relatively small: something very important for time-critical applications, whose interrupt requests should be serviced in a few milliseconds. Another name for tasklets is deferrable functions. 
A given tasklet will run on only one CPU, the CPU on which the tasklet was scheduled. The same tasklet will never run on more than one CPU of a given processor at the same time. Different tasklets can run on different CPUs simultaneously.

In order to use a tasklet in a driver someone has to declare it first e.g DECLARE\_TASKLET($<$name of tasklet$>$, $<$name of function/handler$>$, $<$input$>$), where name of $<$function/handler$>$ points to the code to be executed and the $<$input$>$ is the data (input parameter) passed to the tasklet. When using this call, the tasklet is enabled by default, which means it will be executed as soon as possible after we schedule it. The second part is to schedule the tasklet, calling the tasklet\_schedule($<$name of tasklet$>$) method. The scheduling of a tasklet can happen anytime as part of the main interrupt handler, but the tasklet will run when it is safe to run. When the driver is about to exit, we can remove the tasklet by calling tasklet\_kill($<$name of tasklet$>$). This function ensures that the tasklet will not run again.%, and if the tasklet is currently running, it will be terminated (and not run again) after its completion.

%\cleardoublepage

%----------------------------------------------------------------------------------------
%\noindent

%----------------------------------------------------------------------------------------

% \vfill
% \footnoterule
% Chapter 2
\chapter{Related Work}\label{Chapter2}

\section{Page Fault Support in NICs}\label{npf_asplos17_paper}
While working on our topic, that is based on Page Fault support necessary when using RDMA technology, we found a similar topic-wise work, that was presented in ASPLOS 2017. The title of this work is \enquote{Page Fault Support for Network Controllers}, which was conducted by I.~Lesokhin~et.~al.~\cite{DBLP:conf/asplos/LesokhinERSGLBA17}. This seems to be the most relevant work to ours, which is why we are going to cover it in this Chapter.

Isolation of the address spaces between different applications (or virtual machines) is one key benefit that comes with virtual memory. Also there is a simplicity that comes with it for the programmers; they do not bother to properly manage the memory or the storage. Last but not least, virtual memory allows optimizations that are quite important performance- and memory-utilization-wise, such as demand paging. According to the authors, programmers who write software that initiates DMAs do not enjoy all the benefits coming with virtual memory, because DMAs cannot tolerate page faults.

% \begin{tabular}{c|c}
%      Term & Description  \\
%      IOchannel & hardware-provided virtual NIC instance 
% \end{tabular}

\begin{table}[th!]
\caption{Direct network I/O terminology}
\centering %
    \begin{tabular}{+ l^ c}
    \toprule \rowstyle {\bfseries}%
        Term & Description \\\toprule %
        IOchannel & hardware-provided virtual NIC instance \\
        IOuser & untrusted process or VM assigned with IOchannel \\
        IOprovider & trusted operating system (OS) or hypervisor
    \end{tabular}
    \label{tab:pfsfornic_terms}
\end{table}

The authors of the paper use the terminology that can be found in Table~\ref{tab:pfsfornic_terms}. NICs allow IOusers to bypass the IOprovider, which is something that led to the sudden increase of NICs usage. According to the authors, a lot of research focused on benefits and improvements that can happen in such ecosystem, though mainly ignoring one thing: losing virtual memory benefits due to lack of DMA page fault support. There are currently two ways to avoid DMA page faults: static and dynamic pinning. Static pinning of the whole address space of an application enjoys simple programming model, but loses on canonical optimizations, such as demand-paging. 
Alternatively, buffers can be pinned and unpinned before and after they are DMAed. Although this latter approach enjoys the canonical optimizations of virtual memory, it has two drawbacks: it complicates the programming model and when frequently used it hampers the performance.

% There are basically two existing DMA mechanisms that support page faults: PCI-SIG and AMD's HSA. Both seem to cover only page fault cases that the data is local. They authors, go beyond that claiming it is necessary to support remote network page faults (rNPFs) as well.

% The proposed solution for rNPFs should have these properties:
% \begin{itemize}
%     \item No additional hardware costs (adding buffers would needlessly increase the NIC price)
%     \item Stream isolation: solution should not affect unrelated to page faults traffic
%     \item No IOuser pinning: authors state that it is wrong to believe that only a small portion of buffers will be pinned on NICs. They also mention that in many multi-tenant setups, tenants are not allowed to pin as much space as they please (e.g. by default there is a 64KB limitation on mlock() method of Linux 4.9)
%     \item No IOuser TCP changes: although TCP/IP running on CPU can be beneficial, this is not an option for NICs.
%     \item No IOuser NPF handling: Having a buffer per IOuser that the NIC could directly use would not work for many different reasons such as, increased complexity exposed to each IOuser, can be wasteful, hinders portability, can be riskier with interrupt handlers of IOusers not getting scheduled immediately (could lead to buffers overflow) etc. 
% \end{itemize}

% Authors state their solution is complete, since it entirely eliminates packet loss (Ethernet) and alternatively allows quick recovery of data (InfiniBand).

\subsection{InfiniBand Page Fault Support}\label{npf_asplos17_ib}

The mechanism of InfiniBand page fault support consists of two different flows.
\begin{itemize}
    \item Network Page Fault (NPF)
        \begin{enumerate}
            \item When a new request is received, NIC consults IOMMU page tables. The NIC finds and marks the page involved that is not present.
            \item The modified firmware detects the fault and raises NPF interrupt.
            \item The driver catches the interrupt.
            \item The driver's NPF interrupt handler queries the OS regarding the physical address of the faulting IOVA (I/O virtual address). If necessary, the OS allocates the pages, possibly retrieving their content from secondary storage.
            \item The driver updates the IOMMU page table with the physical address and informs the firmware that the NPF has been resolved.
        \end{enumerate}
    \item Invalidation
        \begin{enumerate}
            \item The operating system requests from the driver to remove the old IOVA and stop the device from using it.
            \item The driver updates the IOMMU page tables accordingly and issues the invalidation.
            \item The NIC acknowledges the invalidation.
            \item And then, the driver notifies the operating system that the relevant pages can safely be reused.
        \end{enumerate}
\end{itemize}

InfiniBand supports Reliable Connection (RC), which means that the mechanism can let the sender know to stop sending upon dealing with a page fault and retransmit when ready -- this works because the data is local when a sender encounters a page fault. Receiving can be more tricky, but still doable using RNR (receiver-not-ready) messages to suspend sender upon NPFs. In addition to send/receive, RC supports RDMA operations, but in some cases RC does not permit RNR negative acknowledgements. For instance, there is no way to stop the sender during remote read operations, which means packets will be dropped. The only way for the sender to retransmit is to rewind the transfer after the page fault is resolved.

\begin{figure}[h!]
	\centering
	\captionbox[Execution breakdown of NPF and invalidation]{Execution breakdown of NPF and invalidation \label{fig:npfsupport_exec_breakdown}\\{\small Source: \cite{DBLP:conf/asplos/LesokhinERSGLBA17}}}{%
		\includegraphics[width=0.8\textwidth]{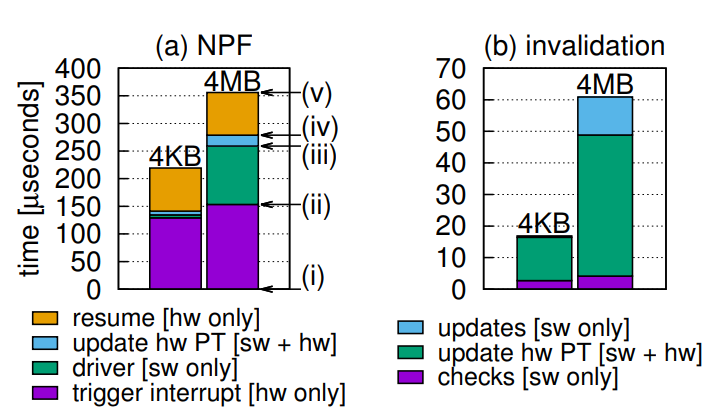}
	}
\end{figure}

% \begin{figure}[h!]
% 	\centering
% 	\captionbox[High level design of the backup ring]{High level design of the backup ring \label{fig:npf_aslplos17_exec_backupring}\addtocounter{figure}{-1}}{%
% 	\includegraphics[width=0.8\textwidth]{npf_aslplos17_exec_backupring}  
% 	}
% \end{figure}

In Figure~\ref{fig:npfsupport_exec_breakdown}~(a), we can see the average overhead breakdown of minor NPFs, which means no disk access, when sending 4KB and 4MB messages. We see a breakdown of the following events:
\begin{enumerate}[label=(\roman*)]
    \item an NPF that occurs is observed by the IOMMU and an interrupt is triggered
    \item invocation of the driver's NPF handler
    \item the mapping from the OS (the physical address that corresponds to the IOVA) is \enquote{recovered} and later sent to the driver
    \item the driver finishes updating the IOMMU page table accordingly (due to coherence issues the driver needs to communicate with the IOMMU that is on the NIC -- page tables normally reside in DRAM)
    \item the NIC identifies the update and resumes the transmission
\end{enumerate}

As noted in the paper and can also be seen on Figure~\ref{fig:npfsupport_exec_breakdown}~(a), a minor NPF takes 220~μsec for a 4KB message, 90\% of which is due to hardware (firmware). As it is explained later, this is a typical duration for Mellanox NIC firmware activity (not only for NPFs), \enquote{as the goal of the NIC circuitry that runs the firmware is usually to handle error paths}, which is why it is considered acceptable to be slow. When the message is 4MB, the duration increases to 350~μsec, due to software overheads such as more translations and allocations by the OS.

In Figure~\ref{fig:npfsupport_exec_breakdown}~(b) we can see the overhead caused by the invalidation flow. First, the driver identifies the memory region that needs to be invalidated and checks if the mapping exists in the IOMMU on the NIC. If the mapping does not exist, no additional overhead is incurred. Secondly, the driver needs to update the IOMMU page tables and its own internal state. 

\subsection{Ethernet Page Fault Support}\label{npf_asplos17_ethernet}

Ethernet is considered more mainstream as a scenario with NICs. IOuser (application or process) utilizes a direct network channel through a regular Ethernet NIC. An IOuser probably uses TCP/IP protocol to drive its direct channel. This approach does not have the benefits of InfiniBand (Section~\ref{npf_asplos17_ib}) technology, because this is a very different ecosystem. There was hope that they could be benefited from TCP reliable communication (which includes retransmissions when a packet is lost), but dropping of packets does not seem a viable solution.

Dropping packets upon rNPFs seems to be an important problem due to the cold ring problem. At start, no buffers are pinned and as a result page faults are triggered one after the other. Meanwhile, packets get dropped, triggering TCP retransmissions and congestion avoidance that nearly deadlock the communication, or as the authors claim, completely halt it in the worst case. Cold ring problem does not occur only on startup situations, other examples would be when a virtual machine is resumed, brought back from swap memory, due to NUMA migration, copy-on-write (COW) semantics etc. Their proposed solution for rNPFs includes a Backup ring.

\begin{figure}[h!]
	\centering
	\captionbox[High level design of the backup ring]{High level design of the backup ring \label{fig:npf_aslplos17_exec_backupring}}{%
	\includegraphics[width=0.8\textwidth]{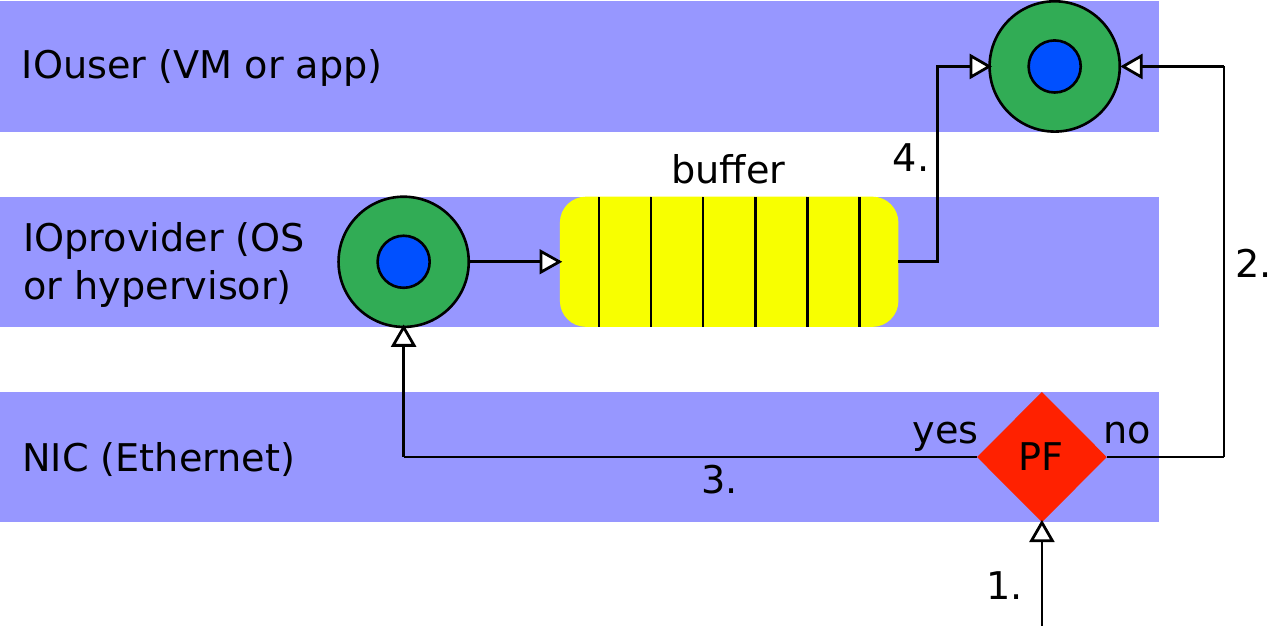}
	}
\end{figure}

As we can see in the high level of the design of the backup ring (Figure~\ref{fig:npf_aslplos17_exec_backupring}), there is a communication between the NIC, the IOprovider and the IOuser in order for the mechaninsm to work. The backup ring is denoted as \enquote{buffer} in the schematic.

Below we can see a brief description of all steps followed in accordance with Figure~\ref{fig:npf_aslplos17_exec_backupring}.

\begin{enumerate}
    \item Traffic is received from the network.
    \item For each incoming packet, NIC inspects the target receive buffer of IOuser. If the buffer is available, data is written directly into it.
    \item For each incoming packet, NIC inspects the target receive buffer of IOuser. If a page fault is encountered, packet is written to a small pinned backup ring, owned by the IOprovider.
    \item After the IOprovider resolves the rNPFs, it copies (or merges) the packet into the original receive buffer of the IOuser.
\end{enumerate}

In order to maintain the ordering, the NIC does not report reception of new packets to the IOuser until all previous page faults have been handled.

% \begin{figure}[h!]
% 	\centering
% 	\captionbox[High level design of the backup ring]{High level design of the backup ring %\label{fig:npf_aslplos17_exec_backupring.png}\addtocounter{figure}{-1}}{%
% 	\label{fig:npf_aslplos17_exec_backupring}}{
% 	\includegraphics[width=0.8\textwidth]{npf_aslplos17_exec_backupring}  
% 	}
% \end{figure}

As the authors reveal, in practice it was not possible to implement the backup ring in firmware. Instead, they prototyped the driver within the IOprovider. All incoming packets were duplicated by the NIC into two receive rings: a primary $p$, where page faults could occur, and a secondary $s$, which was populated with pinned buffers. If no rNPFs on $p$, the duplicate packets in $s$ were discarded. If rNPF on $p$, the driver would utilize $s$ as the backup ring, copying faulty packets from $s$ to an intermediate queue $q$. After resolving the fault, the driver would copy the packets from $q$ to $p$. Unfortunately, the lack of true hardware support leads to halved throughput, due to the duplication of packets. %that was necessary.

Their evaluation was splitted in three different categories: memory utilization, performance overheads and code complexity. In terms of memory utilization, their experiments show that NPF dynamic working set is more advantageous than static pinning. As for the performance overheads, they showed that there is an advantage of RDMA (zero-copying) over copying, especially for larger messages ($\geq 32KBytes$) and that the NPF configuration has similar performance as pin-down cache (a technique that unpins after exceeding an upper limit). In some HPC workloads, the NPF support can provide benefits without the need to pin. As for the code complexity, although as they say it is difficult to quantitatively measure the benefit of NPFs in reducing code complexity, they propose considering lines of code (LOC) involved. The authors state that very few lines (a few tens) of code needed to be modified or added to the design in order to work, compared to implementation requiring pinning (a few thousands LOC).

\section{Pinning-Based Networks}

For the approach of pinning-based techniques, one of the fundamental works we read was \enquote{New DMA Registration Strategy for Pinning-Based High Performance Networks} \cite{DBLP:conf/ipps/BellB03}. 

In this work, the authors propose a new memory registration strategy for supporting RDMA operations over pinning-based networks. One of the motivations was that existing approaches at that time were not efficient when implementing GAS (Global Address Space) languages. In fact, existing approaches, although they could often maximize the bandwidth, required a level of synchronization that discouraged one-side communication and caused significant latency costs for small messages.

Their proposed memory registration strategy is described by an algorithm. The \enquote{Firehose} algorithm exposes one-sided zero-copy communication in the common case and at the same time minimizes the number of synchronization messages that is required in order to support remote memory operations.

Algorithm description:
\begin{itemize}
    \item Determine largest amount of application memory that can be registered (upper bound on total number of physical page frames that can be pinned simultaneously -- this is limited to some reasonable fraction of existing physical memory). If this amount is in total M bytes, using P byte pages, then only a total of M/P pages can be pinned at any time during execution. This mechanism supports many nodes, which is why available space is equally divided and $F= \lfloor{\frac{M}{P*(nodes-1)}}\rfloor$ physical pages can assigned to each remote node.
    \item Conceptually, a firehose is a handle to a remote page. Each node owns F of these firehoses. The authors mention that a functionality of freeing firehoses to establish new mappings to remote pages in order to pin and then serve the pending remote memory operations is supported.
    \item A round-trip synchronization message is required in order for a node to situate one of its firehoses, by mapping it to a region in the remote virtual memory. This way it is guaranteed by the remote node that the virtual page will be pinned for the duration of the mapping.
\end{itemize}

In general, the Firehose algorithm includes some tunable parameters, such as the maximum amount of the physical memory used for remote firehoses (M), the maximum size of bucket victim FIFO queue (MAXVICTIM), which will be used to unpin only when necessary. Also, the bucket size, which is the basic unit of physical and virtual memory of the Firehose algorithm, is configurable (equal to page size by default).

The main benefit of this approach is that %they do not pay the price of 
the handshake per transfer is avoided, which in general case is necessary in Rendezvous approaches, because it is the only way to advertise and make sure that the related virtual pages are resident (pinned) in memory. In the Firehose approach, pinning happens only once at the beginning in the common case and the handshake is required for the non common case, which comes with a cost that is negligible. This was the main focus of this work; to reduce the frequency of the registration operations.

Existing pinning-based strategies:
\begin{itemize}
    \item \enquote{Pin Everything}: This method is not on-demand-based. A single segment of memory is pinned at startup and kept pinned until the program terminates. This method seems rational when the total memory requirements are known and constrained to a relatively reasonable size within the physical memory limits of the host. In this case, it is preferable to pin the entire remotely-accessible region of memory at startup. It does not require additional synchronization and can complete as one-sided.
    \item \enquote{Bounce Buffer}: This method, uses temporary buffers residing in pinned memory to hold data for outgoing and incoming DMA operations. In case of a \enquote{put}, the DMA operation completes, the target processor is informed of its delivery and must copy the data to the final remote destination. When a \enquote{get} request is received, targeted code copies the data into a bounce buffer and executes a \enquote{put} operation to the requesting node. The main advantage is that the cost of registration is being paid only at startup and no more pinning is required. It has some disadvantages as well. It is two-sided in a strict way (latency for remote transfer operations is likely to increase), copying costs may be significant (even for small messages, because of the interrupts and the different kind of CPU and TLB invalidations) and %, can introduce 
    and complexity and handshake overheads may appear, arising questions about the scalability of the mechanism.
    \item \enquote{Rendezvous}: For large transfers, cost of pinning on-demand can be amortized over more data and provide performance improvements over the use of many bounce buffers. Rendezvous includes two (2) main steps:
    \begin{enumerate}
        \item Send a message to the remote node indicating the region to be pinned 
        \begin{itemize}
            \item for \enquote{puts}, remote node processes the message and pins the relevant memory region and then sends a reply, indicating that the DMA transfer can be initiated.
            \item for \enquote{gets}, similar approach with \enquote{puts}: as an optimization, reply may coalesce acknowledgement and payload.
        \end{itemize}
        \item Optionally, there may be some final handshaking to unpin the relevant regions once the DMA transfer is complete.
    \end{enumerate}
    The cost of registration is paid on every operation, which is prohibitive %costly 
    for small messages and debatable for larger messages.
\end{itemize}

\begin{figure}[h!]
	\centering
	\captionbox[Firehose: 8-byte put latency over increasing working set memory size (M = 400MB)]{8-byte put latency over increasing working set memory size (M = 400MB) 
	\label{fig:firehose}\\{\small Source: \cite{DBLP:conf/ipps/BellB03}}}{%
	\includegraphics[width=0.8\textwidth]{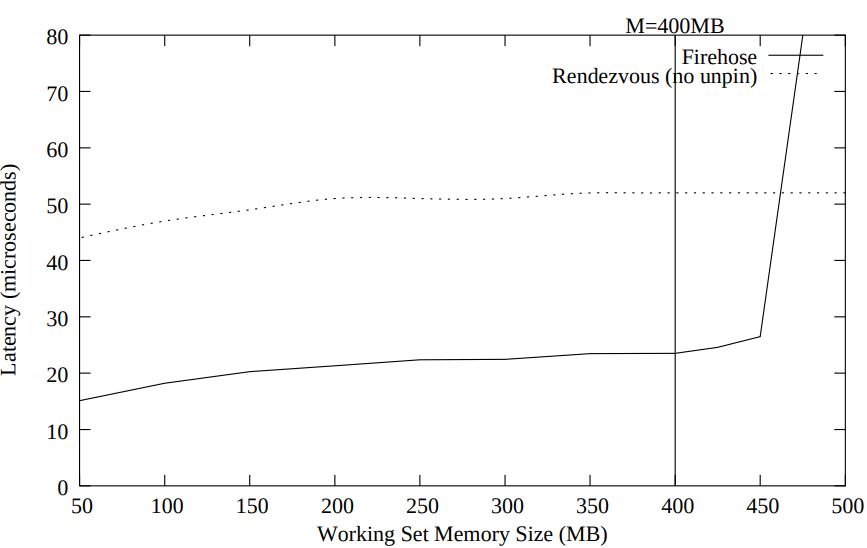} 
	}
\end{figure}

% \begin{figure}[h!]
% 	\centering
% 	\captionbox[Execution breakdown of NPF and invalidation]{Execution breakdown of NPF and invalidation \label{fig:npfsupport_exec_breakdown}\\\subcaption{Source: \cite{DBLP:conf/asplos/LesokhinERSGLBA17}}}{%
% 		\includegraphics[width=0.8\textwidth]{npf_aslplos17_exec_breakdown}  
% 	}
% \end{figure}

For the evaluation, Firehose assumed M=400MB of pinnable memory, MAXVICTIM=50MB, and total pinned memory: M+MAXVICTIM=450MB. Tests are run long enough to reach a steady state. Bucket size is set to single-page buckets to provide an upper bound on the overhead in managing Firehose data.

The authors used two parallel applications implemented in Titanium (GAS language) as their benchmarks: Cannon’s Matrix Multiplication and a Bitonic Sort. 

Their results show that Firehose is twice as good as the Rendezvous average put latency, with no-unpin. However, the results of the synthetic microbenchmark (Figure~\ref{fig:firehose}), show that the latency of Firehose past the M+MAXVICTIM point (450MB) increases sharply, approaching the Rendezvous with no-unpin performance. Also, although not depicted in Figure~\ref{fig:firehose}, their results for the large-message bandwidth show that when the working set exceeds M, the Firehose hit rate decreases (more \enquote{handshaking} is required).

% \begin{figure}[h!]
% 	\centering
% 	\captionbox[Firehose: 8-byte put latency over increasing working set memory size (M = 400MB)]{8-byte put latency over increasing working set memory size (M = 400MB) \label{fig:firehose}\addtocounter{figure}{-1}}{%
% 	\includegraphics[width=0.8\textwidth]{Figures/firehose.png}  
% 	}
% \end{figure}

The above results show that there is a high overhead of pinning, which can become non-tolerable after the point of pinnable memory (M) -- which by itself can become a limiting factor. %(what is a good enough value? complex programming model for a programmable)
Also, we argue that pinning the whole address-space of the process of each node at startup has to be possible size-wise and it cannot be the case all the times. Furthermore, it does not look like a viable solution, since a user application might need to allocate (request) more virtual pages, which either we will have to pin (costly) or we will need a mechanism to page them in when a page fault occurs. %We need to mention here that there are ways for someone to pin even new pages added in address space
Besides that, Linux Operating System comes with some optimization techniques like Transparent Huge Pages --THP (see Section~\ref{transparenthugepages}) enabled by default, that will eventually cause (minor) page faults that require handling. If we disable THP we will not take advantage of the possible performance benefits provided by it. 

%\printinunitsof{in}\prntlen{\textwidth}

%\cleardoublepage

\cleardoublepage 
% Chapter 3
\chapter{Remote Page Fault Handling}\label{Chapter3}

\section{Theory}

\subsection{Definition}

According to a definition provided by Daniel P. Bovet and Marco Cesati~\cite{under_the_lnx_kernel} a page fault occurs when \enquote{the addressed page is not present in memory, the corresponding Page Table entry is null, or a violation of the paging protection mechanism has occurred}. Before attempting to find a solution for page faults, it is important to try to break down this definition in a way to find what is the actual problem.

The first part of the definition states that the addressed page is not present in memory, which is probably the most obvious assumption that everyone has in mind when they hear about a page fault. For different reasons, some of them will be covered later (see Section~\ref{pf_reasons}), it is possible that a page does not reside in RAM (memory). This means, that when a legitimate user (process) tries to have access to a page (whether it is a read or write request), they will have to wait until kernel has provided the related physical page (frame) to them by bringing (or allocating) a page frame. While doing this work, kernel context-switches to a different process that can utilize the core, until the page fault has been resolved, which is when the core is ready to switch back to the process that initially triggered the page fault.

The second part says that there is no mapping (yet), meaning a mapping of the virtual address to the physical address of the page. This can happen for many reasons. One reason is that the virtual address is illegal, in a sense that it should not be allowed to be translated and eventually have access to the memory. Operating Systems such as Windows report invalid memory references in this way. Another cause is that the process has not shown interest in accessing it yet, so the kernel \enquote{lazily} decided not to have a corresponding page frame for it, since it might not use it (see Demand Paging, Section~\ref{demand_paging}).

The third part describes a violation of the paging protection mechanism. % has occurred, is not easy to explicitly define. 
Memory protection in paging is achieved by having protection bits for each page. These bits are associated with each Page Table Entry (also known as PTE) and specify the protection on the corresponding page. A valid/invalid (\enquote{v}) bit guards against a process trying to access a page that does not belong to its address space. Read (\enquote{r}), write (\enquote{w}), and execute (\enquote{x}) bits are used to allow accesses of the corresponding type. Illegal attempts institute a memory-protection violation, that causes a hardware trap to the Operating System \cite{pgf_protection_violation}.

Page faults are mainly categorized in two types: minor and major page faults. Minor page faults are those which can be handled by just reclaiming (or allocating) a page frame. Major page faults are the faults, whose handling requires I/O, e.g. disk access. Having this in mind, it is safe to say that major page faults induce greater overheads than minor page faults, because more time is required in order to recover from them.

%One way to approach it is the permissions of a page. For example, a page that has read privileges cannot be written, which means when someone tries to write to it, will cause a Page Fault that the kernel will decide how to treat it.

\subsection{Page Fault Causes}\label{pf_reasons}

In general, two of the main reasons page faults are caused in CPUs are due to some techniques in Operating Systems (e.g. Linux). In our system, we examine and handle page faults triggered in the System Memory Management Unit (SMMU). Another reason for a page fault to occur is due to the Transparent Huge Pages (THP) mechanism.

\subsubsection{Demand Paging}\label{demand_paging}

Demand paging is a dynamic memory allocation technique that defers page frame allocation until the last possible moment. This moment is when a process attempts to address a page that does not reside in RAM (memory), thus causing a page fault.

The reason why this mechanism makes sense is that processes do not utilize all the pages that their addresses are mapped, as part of their address space, right from the beginning. In fact, some of these addresses may never be used by the process. %-- so one question is, why have them residing in the memory? 
The principle of program locality ensures that at each stage of execution only a small subset of addresses will be used \cite{under_the_lnx_kernel}.

This technique increases the average number of free page frames in the system and therefore allows better utilization of memory. Also, as mentioned in Section~\cite{under_the_lnx_kernel}, it allows the system as a whole to get better throughput with the same amount of memory.

\subsubsection{Copy On Write}\label{cow}

The creation of processes from the first-generation Unix systems was in a way \enquote{graceless}. During the fork() system call, kernel was responsible to duplicate the whole parent process address space and assign the copy to the child process. This included many steps such as: allocating page frames for the page tables of the child process, for the pages of the child process, initializing the page table of the child process and copying the pages of the parent into the corresponding pages of the child process. Because of this procedure, we had many memory accesses and a great consumption in CPU cycles, which means that this activity was time consuming.

Modern Unix kernels, including Linux, utilize a new approach called Copy on Write (COW). Initially, the page frames are shared between the parent and the child process, instead of having a duplicate. Neither the parent nor the child process are allowed to modify any content on the page frames (read-only). If any of them want to modify any page frame, an exception occurs. This is the moment that the kernel duplicates the page into a new page frame and marks it as writable, while the original page frame remains read-only. When the other process tries to write into it, the kernel checks whether it is the only owner and in such case makes the page frame writable for the process.

\subsubsection{Transparent Huge Pages}\label{transparenthugepages}

Most of the architectures supported by Linux are able to work with pages larger than 4KB, such as 2MB or even 1GB pages. These are considered \enquote{huge pages} compared to what is considered normal 4KB page size. Huge pages are in common case beneficial performance-wise, since they can mostly offload the Table Look-aside Buffers (TLBs), making at the same time TLB misses less expensive. According to \cite{trasparent_huge_pages}, the mechanism of Transparent Huge Pages (THP) works quietly, substituting huge pages into a process' address space, when these physically contiguous pages are available and it appears that the process would benefit from this mechanism. This feature was first added in 2.6.38 kernel.

One important part of this mechanism is the \textit{khugepaged} kernel thread, that occasionally attempts to substitute smaller pages being used currently with a hugepage allocation, thus maximizing transparent huge page usage. The reason it is called transparent, is because the user does not need to modify the applications in order to work with the new page size. This kernel thread will automatically start when transparent\_hugepage/enabled option is set to \enquote{always} or \enquote{madvise}, and it will be automatically shutdown if it is set to \enquote{never}. This option exists in the path: /sys/kernel/mm/transparent\_hugepage/enabled and can have two values:
\begin{itemize}
    \item always - always use THP
    \item never - disable THP
\end{itemize}
It might be the case that the kernel thread khugepaged is not successful in converting \enquote{small}-sized pages (e.g. 4KB) to huge pages (e.g. 2MB). Even in that case, it may still be taking processor time to search for candidate pages \cite{thp_redhat}.

This %was a \enquote{hidden gem}, that 
triggers random page faults during RDMA transfers even when the buffers are touched and there is no swap device for the kernel to move pages. The reason we could actually witness page faults due to this mechanism was because it seems that while kernel was trying to merge some pages, the previous mappings of these pages were invalid the moment a transfer would try to be translated from the local SMMU. Since someone might want to keep the optimization that the THP mechanism offers in performance, it works as a great motivation for us to have a mechanism that supports page faults during an RDMA transfer. 

\section{Approach}

In Section~\ref{custom_PLDMA}, we briefly described the DMA engine and the environment that we worked on in order to implement a hardware-software co-design that would support the page faults caused during virtual-address RDMAs.

During an RDMA that is based on virtual-addresses, a page fault might occur in:
\begin{itemize}
    \item the source buffer (address)
    \item the destination buffer (address)
    \item both buffers (addresses)
\end{itemize}

We believe that the most common case for a page fault to occur during an RDMA is the destination buffer. We expect that when an application triggers an RDMA transfer, the source buffer will be \enquote{touched} prior to the transfer. We will still cover this scenario, which is possible to happen for many reasons, e.g internal optimizations in Linux, such as Transparent Huge Pages (as described in Section~\ref{transparenthugepages}).

\subsection{Common path}

\begin{figure}[h!]
        \centering
        \captionbox[General Page Fault flow]{General Page Fault flow \label{fig:pgf_flow_general}}{
            \includegraphics{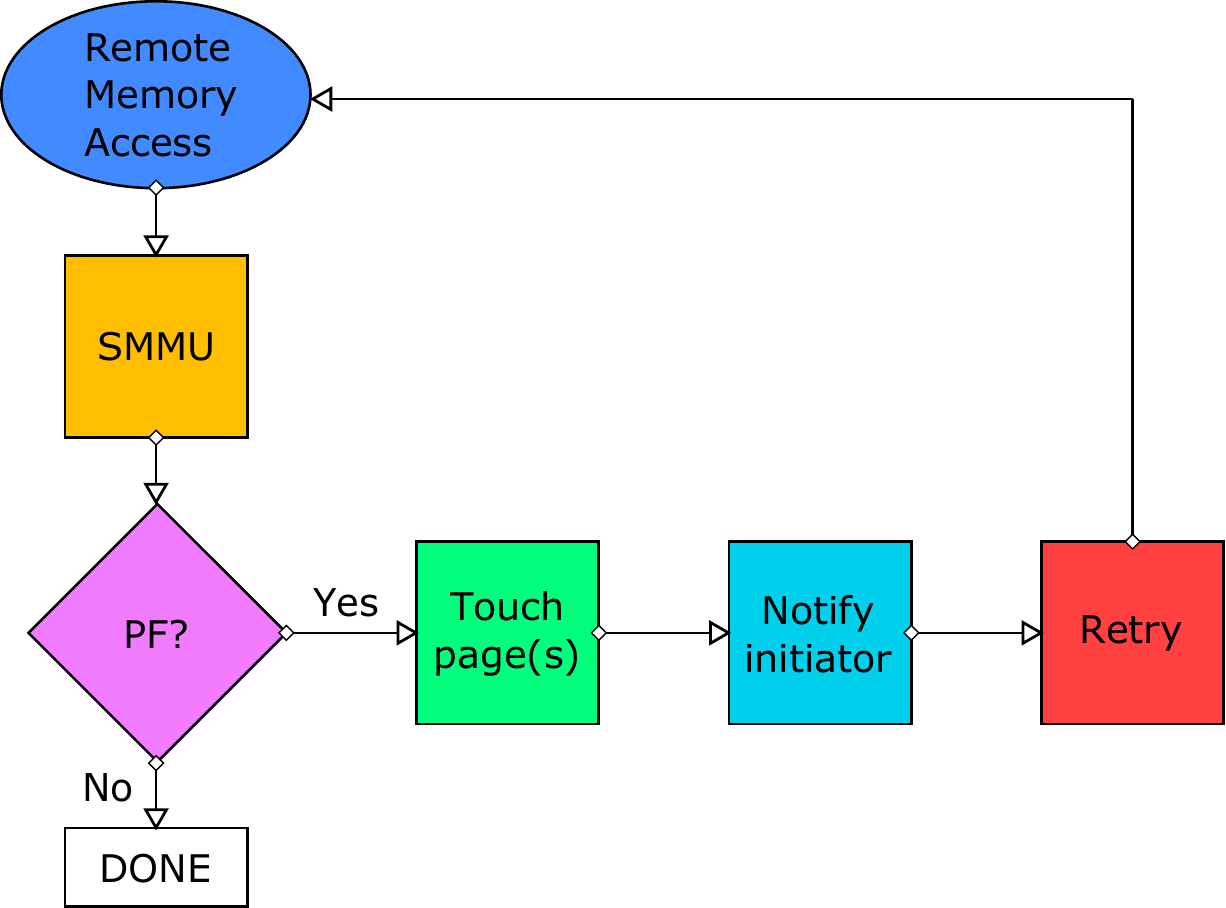}
        }
\end{figure}

The main task of handling the page faults is to make sure that the pages are brought into the main memory, if they are valid, so that after the %requested or timed-out 
retransmission, the transfer will succeed. We designed two implementations to achieve this:
\begin{enumerate}
    \item Netlink Sockets: A user-space thread is responsible to touch the pages that previously triggered a page fault. Kernel (driver) notifies the thread of the corresponding protection domain to touch the pages.
    \item get\_user\_pages(): Kernel-space approach of bringing pages to main memory.
\end{enumerate}
The get\_user\_pages() approach is very recent in our work, which means it might require more investigation in the future about its characteristics and features. It seems interesting and quite beneficial, because using it will allows us to handle page faults completely from kernel-space without any userspace involvement. Also, we will probably be more efficient, since kernel handling will result to less context-switch overheads compared to Netlink/touch pages from user-space approach. Last but not least, we also call this approach \enquote{Touch-Ahead}, because it can page-in more than one pages. Netlink-sockets approach can touch one (1) page per invocation, thus it is called \enquote{Touch-A-Page}.

The current version of the mechanism requires the utilization of Netlink sockets even for get\_user\_pages() approach: the packetizer implemented at ICS-FORTH (from others at FORTH) works when it is configured and utilized by user-space applications. Until this very moment, we can only send messages from the provided packetizer to the mailbox that is polled by the R5 co-processor through user-space. Netlink sockets appear to be the best choice to communicate messages from kernel to userspace and thus to the mailbox, in order for the transactions to be retransmitted when the corresponding page fault has been resolved (this will be covered in detail later). %As we will see later in this thesis, this is only used in the context of a page fault in destination address of a transfer.

We will now describe the common parts of the page fault mechanism both from the sender- and the receiver-side %the common parts here 
and later we will describe the distinctive details of each case (read and write path). 
Common parts can be found both in user-space library and the driver of SMMU. Below we give a description of them.

\paragraph{Page Fault Library}\label{mechanism_user_api}

A page fault user-space library was necessary to fulfill the needs and purposes of our work.

The library includes the following methods:
% \lstset{
% 	language=C,
% 	basicstyle=\footnotesize\color{white}
% }

% \lstset{style=mystyle}
% \lstset{
% 	language=C,
% %	basicstyle=\footnotesize\color{white}
%     basicstyle=\color{white}
% }

%\begin{lstlisting}
% \begin{minted}
% void sig_handler(int signo);
% int pckzer_to_mbox(uint64_t *pckzer_addr, uint64_t dst_coord, int trid, int seqnum, int pdid);
% void* handle_pgfault(void* x_void_ptr);
% void enable_pgfault_mechanism();
% \end{minted}
%\end{lstlisting}

\begin{description}
     \item {\large \textbf{void enable\_pgfault\_mechanism()}}: This method is responsible to create the thread that belongs to the process of a specific protection domain. This thread will be responsible to be woken up and catch the message coming from kernel through Netlink sockets, in order to handle the pafe fault that was triggered.
    \item {\large \textbf{void sig\_handler(int signo)}}:
    This method is responsible to catch and handle a segmentation fault. This method was initially built for testing purposes, since in our environment it is expected that when a segmentation fault occurs (e.g. erroneous not-mapped virtual address), the application will crash, as it happens in most of the systems. 
    
    However, while working on our mechanism and more specifically the \enquote{Netlink sockets} solution (which touches one page, in contrast to our get\_user\_pages() approach), we experienced an interesting phenomenon that can be seen in Figure~\ref{fig:seg_fault_scenario}. In a micro-benchmark that consists of many iterations that each iteration experiences page faults, it is possible to be requested from the userspace library to touch (page in) a page that belongs to a previous iteration and thus no longer belongs to the address space of the process, causing a segmentation fault. With the help of our segmentation fault handler we overcome this obstacle. By using the get\_user\_pages() approach, we do not witness such problem at all, which is why it is the preferred solution. 
    
    \begin{figure}[b!]
        \centering
        \captionbox[Segmentation Fault scenario during Page Fault handling]{Segmentation Fault scenario during Page Fault handling \label{fig:seg_fault_scenario}}{
        \includegraphics[width=1\textwidth]{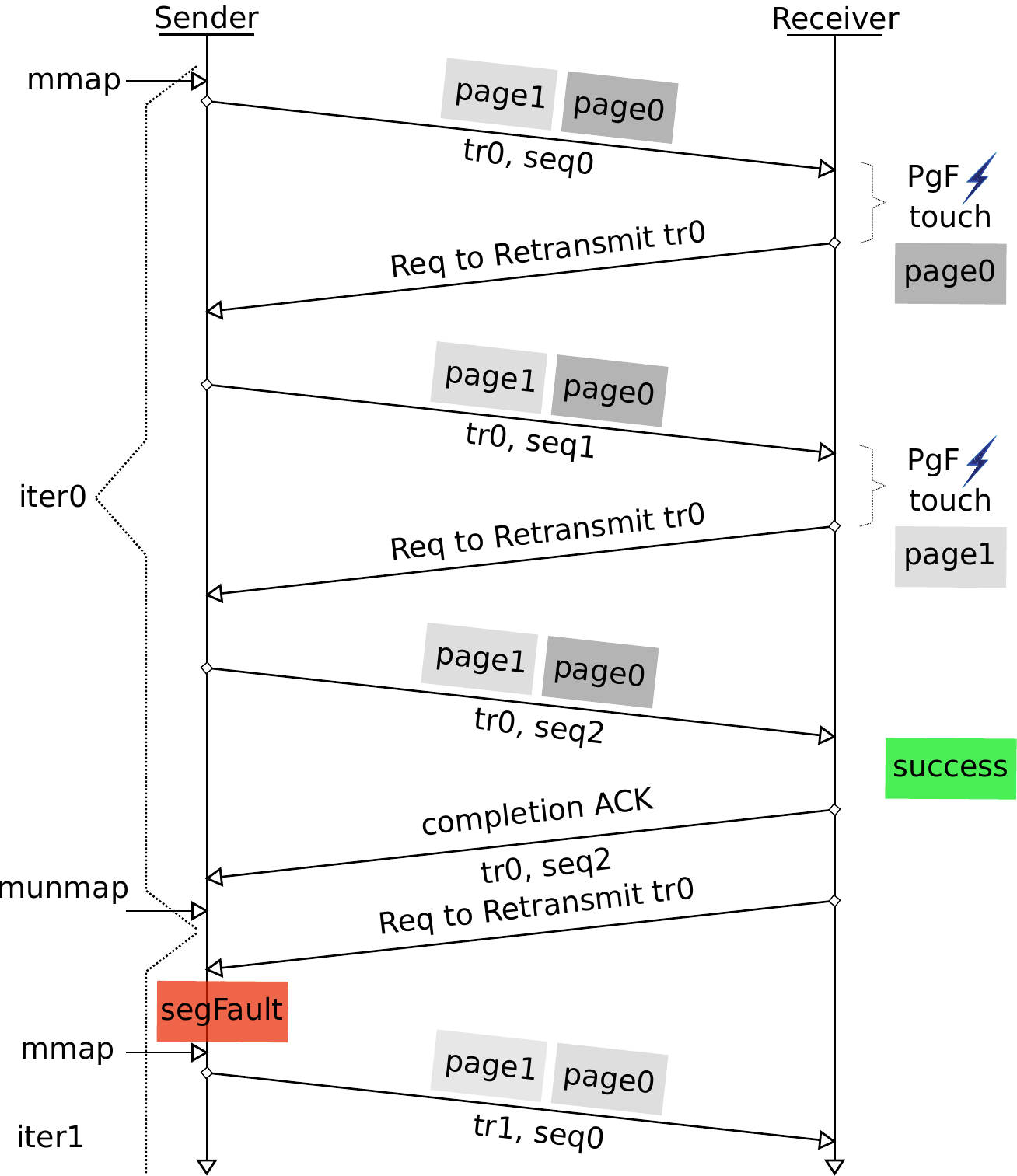}
        %\includesvg{Figures/seg_fault_bw.svg}
        }
    \end{figure}
    
    \item {\large \textbf{int pckzer\_to\_mbox(uint64\_t *pckzer\_addr, uint64\_t dst\_coord, int trid, int seqnum, int pdid)}}:
    This method takes as input arguments the virtual (mmap'ed) address of the packetizer and the useful information for the retransmission of the previously faulty transfer due to a page fault, such as: the coordinates of the computing node we are going to send to its mailbox, that is polled by its local Real-Time R5 co-processor, the transaction identifier (id) that will be retransmitted, the sequence number of this transfer (for debugging purposes mostly), and the protection domain identifier (id), that this transfer belongs to (for safety check, since packetizer will either way have the protection domain wired to the packet -- cannot be changed by user for security reasons). The least significant bits (LSB) of the message to-be-sent consist of the opcode for the message and are given by the driver. The opcode is used by the mailbox to distinguish different types of messages destined to the mailbox, opcode = 2 is for RAPF (Retransmit After Page Fault handled) and other opcodes are used for acknowledgement messages or read requests.
    
    % \begin{minted}{c}
    % int channel = 0; // we only use 1/4 of the packetizer channels
    % uint64_t raw_data[2];
    % uint64_t mbox_addr, data_sz;
    % uint64_t extended_mbox_addr;

    % mbox_addr = MBOX_R5_ADDR;
    % extended_mbox_addr = ((uint64_t) dst_coord<<42 | (uint64_t) mbox_addr);
    % data_sz = MBOX_R5_SZ | _MBOX_WAIT_FOR_PREVIOUS_  | _TIMEOUT_FLAG_;

    % raw_data[0] = (( ( trid ) << 18 ) >> 32);
    % raw_data[0] = ((((0x0FFFFFFFLLU)&((pdid<<12) | (seqnum)))<<32)  | raw_data[0]) ;
    % raw_data[1] = 0;

    % asm volatile ("stp %0, %1, [%2];\n"
    %   "stp %3, %4, [%2, 0x40];\n"::
    %   "r"(raw_data[0]), "r"(raw_data[1]),
    %   "r"(pckzer_addr + 0x10*channel),
    %   "r"(extended_mbox_addr), "r"(data_sz):);

    % return 0;
    % \end{minted}
    
    %\end{lstlisting}
     \item {\large \textbf{void* handle\_pgfault(void* x\_void\_ptr)}}: 
     This method describes the main mechanism for page fault handling. The first thing to do is to \enquote{open} the device (module) that is already generated before, which allows us to allocate and utilize a packetizer by returning its virtual address. This device generation required a module to be implemented and used, utilizing calls such as mmap, which takes as vma\_off the protection domain identifier (PDID). Later, we initialize the source and destination information that is necessary for the header of the message to be sent through Netlink sockets. In our case, since we are now describing the userspace process point of view, the source is the current process (so the src\_addr.nl\_pid takes the pid of the process, using the getpid() method) and the destination is the kernel, which by default has zero (0) as the value for pid, thus dest\_addr.nl\_pid is equal to zero. In our current mechanism, we do not want to send any information to the kernel (only the kernel/driver sends occasionally information to the userspace), but the functionality exists mostly for debugging purposes. %and possible future needs for use. 
     After this, we are entering a while loop of the thread that busy waits until it receives a new message from the kernel, which will include the necessary information for the faulty transaction to be handled. 
     
     When a message is received from kernel, we need to decode it, because originally it was sent as a string of combined information related to the transfer. This string is by design sent as a string of hexadecimal digits and by splitting the incoming message accordingly, we extract the useful information. In Table~\ref{tab:netlinkmsg_format}, we can see the format of the Netlink message as it is sent by kernel (driver) to the process of the related protection domain. Upon arrival of the message, the userspace process checks the least significant bit. If it is equal to 0, we had a page fault in the virtual address of the source (sender) buffer, otherwise (1) the page fault occurred in the destination (receiver) buffer.
    
    At this point, the userspace process touches the faulty virtual address (IOVA), thus triggering an internal page fault that will be handled by the MMU of the CPU, resolving the fault transparently by doing all the necessary actions (mappings, page frames etc). After that, if the fault was triggered by the receiver (destination buffer), based on the type or R/W field (1 bit), we utilize the information we received from the kernel to acknowledge that the fault was resolved so the initiator node can retransmit it. In order to do this, we use the packetizer to send a message to the mailbox that the Real-Time (R5) co-processor polls. R5 uses the information of the transaction id (14 bits), the sequence number and the protection domain identifier (PDID) in order to safely trigger the now-resolved previously-faulty transfer. If the fault was caused by the sender (source buffer), we do not do anything else, because we are restricted by the design, but we expect that after the timeout period (initially it was 200~ms, since the parameter \textit{TIMEOUT\_PERIOD} was 200000000 in \textit{r5\_defines.h}  file, but we changed it --currently the minimum we have used is 1~ms) for the transfer to be-re-transmitted. 
    
    \begin{table}[ht!]
    \caption{Netlink Message format}
    \centering %
   % \resizebox{14cm}{!}{
        %\begin{tabular}{| p{2.5cm} | c | c | c | c | c |}
        \begin{tabular}{| c | c | c | c | c | c |}
        %\toprule \rowstyle %{\bfseries}%
        \hline
            \rowcolor{blue!25} \textbf{Src\_ID} & \textbf{Tr\_ID} & \textbf{Seq\_Num} & \textbf{Faulty IOVA} & \textbf{PDID} & \textbf{R/W} \\\hline %
            6 hex & 4 hex & 4 hex & 8 hex & 4 hex & 1 hex \\\hline
            22 bits & 14 bits & 14 bits & 32 bits & 16 bits & 1 bit \\\hline
            MSB & \multicolumn{4}{c|}{} & LSB \\ \hline
        \end{tabular}
      %  }
        \label{tab:netlinkmsg_format}
    \end{table}
\end{description}

\paragraph{SMMU driver}\label{pgf_driver_common}

% The driver with the name \enquote{arm-smmu.c}, provided by ARM, is responsible to catch the Page Fault exception (interrupt) when it happens in the ARM SMMU. There is already an interrupt handler implemented to catch the translation fault. 

The mainline kernel driver of the ARM SMMU (version 2) includes a context fault handler, that will be triggered on occurrence of new translation (page) faults or permission faults, that might be triggered during a translation targeting an active and valid context bank of the SMMU (we have discussed about context banks in Section~\ref{translation_cb}). This handler is called \textit{arm\_smmu\_context\_fault}. It is important to mention here that the SMMU reports in a different way the translation fault (no valid page frame for the given virtual-address) and the permission fault. While writing this thesis, we aim to support the translation fault, but we believe that permission fault can be handled by the same mechanism -- in the future we expect to extend our tests further to tackle permission faults as well, in order to safely say that they can be supported by our mechanism.

Of course, someone might be wondering what happens if multiple faults occur in one specific context back of SMMU \enquote{simultaneously}, which includes transfers that request translation from a specific context back SMMU, while this context bank is still handling a previous context fault. To answer this question, we first need to describe what are the registers that can provide information about a context fault.

If a fault is encountered when all fields of SMMU\_CB$n$\_FSR, where $n$ is the index number of a corresponding SMMU context bank (\{0...15\} in our system), are zero, the following registers provide full details of the fault:

\begin{itemize}
    \item SMMU\_CB$n$\_FSR
    \item SMMU\_CB$n$\_FAR
    \item SMMU\_CB$n$\_FSYNRm
\end{itemize}

If a fault is encountered when the value of SMMU\_CB$n$\_FSR is non-zero, SMMU\_CB$n$\_FSR.MULTI is set to 1 and no details of the fault are recorded. In other words, we can keep information only on the first fault before it is cleared. SMMU\_CB$n$\_FSR.MULTI indicates that multiple outstanding faults occurred. As we read in Appendix~\ref{Appendix_StreamID}, MMU-500 (the model of our SMMU) supports either 8 or 16 parallel page table walks for a TBU. More details about this topic can be found there.

The Fault Status Register (FSR) of the context bank reveals the type of the fault. More specifically, the translation fault is revealed by the \enquote{TF} bit, that is the second bit (or bit 1, counting from zero) in FSR register (e.g. for context bank $n$, the name of the register is: SMMU\_CB$n$\_FSR) \cite{zynqusplus_registers}.

Both the \enquote{Fault Address Register} (SMMU\_CB$n$\_FAR) register and the \enquote{Fault Address Register - high significant bits} (SMMU\_CB$n$\_FAR\_HIGH) register, hold the input address (i.e. virtual address) bits of the memory access that caused the translation fault. To be more precise, FAR holds the lower input address bits [31:0] and FAR\_HIGH holds the upper input address bits [63:32], that triggered the page fault. In our case, FAR\_HIGH holds only 16 bits, which in total with FAR register can give us: $32+16=40$ bits.

An SMMU handles a context fault, including a translation fault, by either stalling or terminating the transaction that caused the fault:
\begin{itemize}
    \item Terminate the fault: SMMU does not perform the final access. Depending on the value of SMMU\_CB$n$\_SCTLR.CFIE, SMMU reports the fault to the initiator of the transaction that triggered a translation fault.
    \item Stall the fault: Software can either terminate or retry the faulty transaction, by writing to the register SMMU\_CB$n$\_RESUME. It is implementation defined whether SMMU supports the stall mode operation.
\end{itemize}

As the authors of ARM SMMU TRM \cite{arm_smmu_trm} mention, it is not possible to guarantee that a stalled transaction in a context bank will not affect a transaction of another context bank. This is the reason this mode should be used wisely.

In our system we believe that the Stall fault model can be supported, because the corresponding field of all related registers (of all context banks) are read/write, meaning we can have this option enabled to any of them. Another hint is that the SMMU\_SCR0.STALLD is zero (0), which allows SMMU to permit per-context stalling on context faults. However, it is not tested as much as we wanted due to lack of time. Hence, our tests followed the terminate fault mode. But we certainly expect to work and do experiments with Stall fault model in the future. 

\begin{itemize}
    \item If Stall mode:
    \begin{itemize}
        \item If HUPCF==0 and a fault occurs: No more transactions are processed (=any subsequent transaction stalls) until the fault is resolved
        \item If HUPCF==1 and a fault occurs: More transactions can be processed until the fault is resolved for this particular context bank
    \end{itemize}
    According to the manual~\cite{arm_smmu_trm}, the number of transactions processed after the original faulty transaction and the number of subsequent transactions that can raise a fault before no more transactions are processed until the fault in this particular context bank is cleared, is implementation defined.
    
    Another interesting thing with this mode is that if the SMMU is configured to raise an interrupt (which it is, as we mentioned above), one of the following can happen from the supervisory software:
    \begin{itemize}
        \item Fix the fault and resume (retry)
        \item Terminate the fault (no data are returned when read, and no data is affected when write)
    \end{itemize}
    \item If Terminate mode:
    \begin{itemize}
        \item If HUPCF==0 and a fault occurs: if a fault is active for that context, each subsequent transaction (whether it was faulty or not) terminates. FSR records only the original (active) fault.
        \item If HUPCF==1 and a fault occurs: if a fault is active for that context, it terminates the new fault and records multiple faults in FSR.
    \end{itemize}
\end{itemize}

The next thing we need to distinguish is whether the fault was due to a read or a write transaction, since our mechanism supports these cases in a different way due to technical limitations of the current system. In order to do this, we use the \enquote{Write Not Read} (WNR) bit of the \enquote{Fault Syndrome Register} (FSYNR) register of the context back, that experiences the translation fault. This register holds the fault syndrome information about the memory access that caused the fault. The WNR bit (bit 4) indicates whether the fault was part of a write or a read access, which allows us to distinguish whether a fault happened in the source address or the destination address of the RDMA. %A future system, that can provide more information for both types of the translation fault, would allow a general solution.

After this point, we pass the information of the protection domain to the input variable of the tasklet responsible for handling the page fault. In fact, we have implemented two (2) different tasklets, one to handle translation faults in source address and another one to handle the faults in destination address. As it is already mentioned, we have a different handling for each case (source and destination), thus we use two different tasklets. The protection domain is necessary, because it is the only way we can associate the faulty transfer of the context bank with the domain that will handle it. At this point, it is important to remind the readers that each context bank (or page table) is associated with one protection domain. More precisely, there is a one-on-one mapping of a process and a protection domain. Currently the team at FORTH works on having many processes that belong to the same protection domain. Although this is ongoing work, our implementation for the page fault handling takes this into account, which means that in the future it would be easy to adapt to a new system, that supports many processes per rank (or protection domain). As mentioned in Section~\ref{tasklet}, each tasklet will be scheduled and run when it is convenient (time-wise) for the system to run it, which will be after we have exited the interrupt handler. %Tasklets can be preempted, which is the reason we prefer to use them: when we want to leave the interrupt handler quickly.

Another thing we should mention about the driver is the configuration of the context bank. Basically, when we initialize a context bank of the SMMU in order to point to the page table of a process, we also set some settings, including how the faults will be handled for this specific context bank. In the \textit{arm\_smmu\_init\_context\_bank} function in the ARM SMMU driver, we can set the Secure Control register (e.g. SMMU\_CBn\_SCTLR), that provides the top level control of the translation system for the related context bank. The default value of this register (Linux 4.9 version) had the following flags set (in most cases equal to 1) --the rest of the bits for this register were initialized to zero (0).:
\begin{itemize}
    \item SCTLR\_CFIE: Context Fault Interrupt Enable, if set, meaning that when a context fault occurs, an interrupt will be raised.
    \item SCTLR\_CFRE: Context Fault Report Enable, allows the context bank to return an abort when a context fault occurs. 
    \item SCTLR\_AFE: Access Flag Enable, means that in translation table descriptors the AP[0] bit is an access flag.
    \item SCTLR\_TRE: This bit indicates that the TEX Remap Enable is enabled, remapping the TEX[2:1] bits for use as two translation table bits, that can be managed by the operating system. As enabled by default and not having to do with what we were working on, we did not modify it.
    \item SCTLR\_M: MMU (of CPU) behavior for this translation context bank is enabled. Basically, it means that the translation stage (1 or 2) that the context bank belongs to is enabled.
    \item CB\_SCTLR\_SHCFG\_OUTER: Theses two (2) bits indicate the shareable attribute of a transaction where the translation context bank is disabled, which is when SCTLR\_M=1. This is not true in our case --the default value is 2, which means the attribute is Outer Shareable.
\end{itemize}

In order for our mechanism to work when handling page faults, we had to check two (2) of the settings of this register, not enabled by default. The first was the SCTLR\_HUPCF, which means Hit Under Previous Fault context fault. The second was about the SCTLR\_CFCFG, which is the Context Fault Configuration and can have two values, based on the two modes discussed mainly above: 0 (zero) for Terminate mode and 1 (one) for Stall mode -- Terminate mode is the default.

HUPCF setting allows us to process all subsequent transactions independently of any outstanding context fault. This is an interesting setting, since it was one of the reasons we would witness translation faults even in buffers (pages) that were resident in memory. For example, when we had a Remote Write transfer locally to the same node with the source buffer (pages) not being resident in memory, but with the destination pages residing in memory, since they were touched prior the RDMA transfer, we could detect page faults even in the destination pages, because first in time the source pages were \enquote{under a fault}. In other words, translations of destination pages were subsequent translations of source pages, that were experiencing an existing fault, which is the reason why they were terminated in the end. By enabling this mechanism in our example, we could only see page faults that were indeed occurring in the source buffer.

\subsection{Page Fault at Source address}

\begin{figure}[ht!]
        \centering
        \captionbox[Page Fault flow at source address]{Page Fault flow at source address \label{fig:pgf_flow_general}}{ 
            \includegraphics{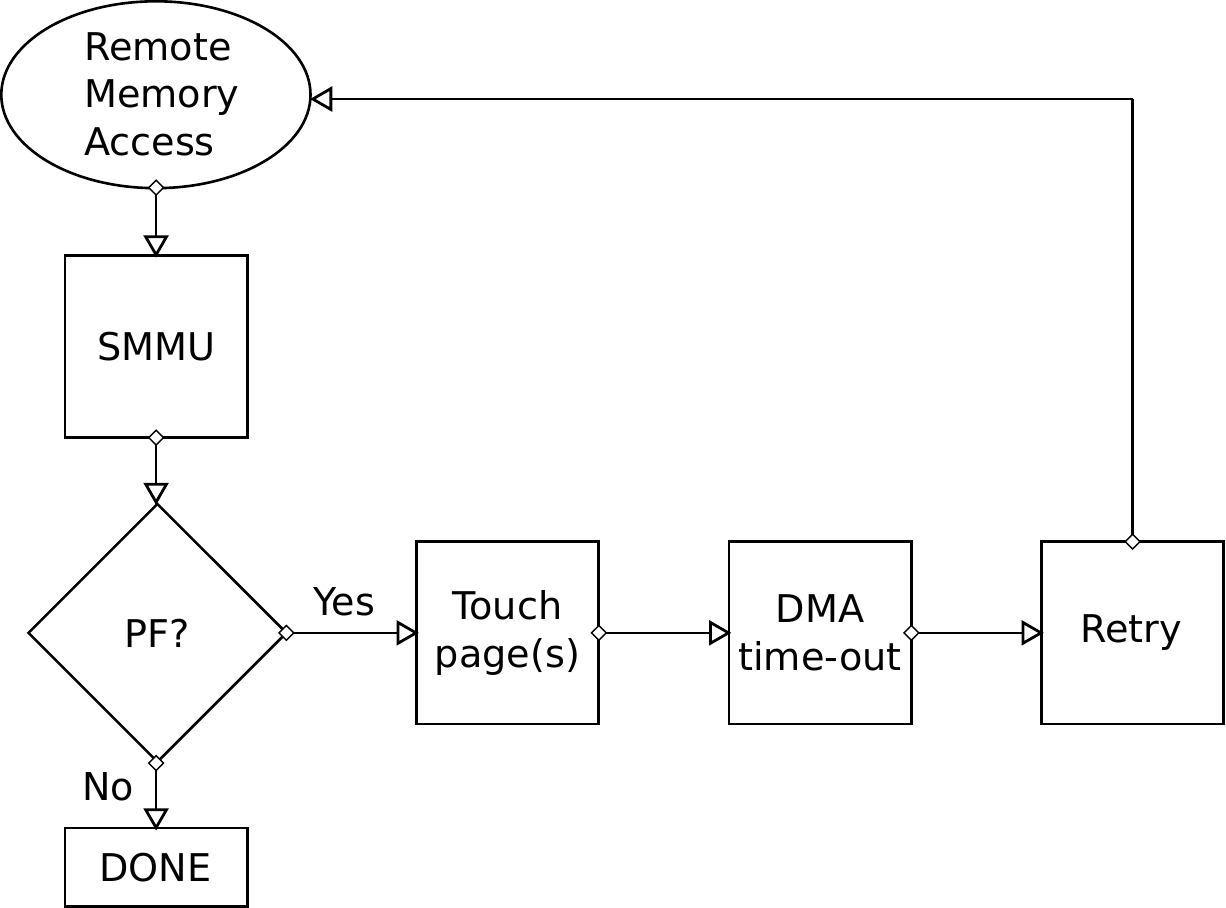}
            %\includesvg{Figures/pgf_read_flow_bw}
        }
\end{figure}

There are two scenarios we can describe for the source buffer: 

\begin{enumerate}
    \item Remote write: In this case, it is expected that the source buffer will be initialized (or written) relatively soon prior to the RDMA transfer, which means the buffer (or pages) usually resides in memory. In other words, we do not expect to have any page faults in this buffer in the common case, except if in the meantime one or more pages have been swapped out to an external drive (major page fault) or the mapping was invalidated for any reason (such as THP optimization).
    \item Remote read: In this case, it is possible that the source buffer is not populated prior to RDMA. At the same time, it will come as no surprise for a source buffer to-be written with an initial value first, just to make sure whether the data has been sent (read) correctly or not (e.g. when polling for a specific value). %We assume that prior to a remote read DMA, the source buffer is initialized by either writing an initial value, or reading the buffer and writing back to it its data.
\end{enumerate}

Initially, translation faults in source buffer of the custom DMA were not supported. This means, that a when a page fault occurred, the transfer would be completed with no information (feedback) indicating that a fault was triggered and \enquote{garbage} data would be sent.

During the last one or two months of this thesis, modifications were made by the original designers of the FORTH PLDMA, in order to at least not complete the transfer in a case of a fault in the source buffer. This would give our page fault handling mechanism a time window to solve the page fault by bringing the pages to memory and then a time-out would trigger a re-transmission. Time-out retransmissions is a resiliency feature of the FORTH PLDMA implemented by others at FORTH. By the time a time-out occurs, we expect the pages that previously experienced a page fault to reside in memory --after that, the transfer will be completed.

The main modifications we had to do in order to handle this translation fault case, were in the driver of the SMMU (\textit{arm-smmu.c}) and more specifically in the context fault handler, as described above.

\subsubsection{Driver}\label{pgf_src_driver}

%At the same time, because it is possible that at the same time we have a page fault at the destination buffer (right after the fault at the source buffer) we also schedule the corresponding tasklet, responsible for faults in destination addresses. In theory, everytime 

The tasklet, called \textit{pf\_send\_handler}, %, is scheduled and will run when the system is ready to run it, probably after we have exited the interrupt handler. This tasklet 
will only need to make sure that the pages are brought to main memory so after the time-out of R5 co-processor, the transfer will be completed.

The only thing necessary for this to work is to pass both the protection domain, the process index and the faulty virtual address to the tasklet, that as we know will run independently. We had to configure iommu\_domain struct (iommu.h) to hold information for both the protection domain and the process index in order to be able to associate a context bank with a protection domain and a process index. Reading both SMMU\_CB$n$\_FAR (Fault Address Register) and the SMMU\_CB$n$\_FAR\_HIGH (Fault Address Register - high significant bits) while inside the handler and before calling the tasklet was sufficient. Then, we utilized some bits of the 64-bit unsigned integer input, that is given to the tasklet of the sender, in order for it to know what is the protection domain and process that would handle the page fault. Using the Netlink sockets approach, we send a Netlink message to the corresponding user-space process to request a touch of a specific 4KB page. Using the get\_user\_pages() approach, as an optimization, we request up to four (4) pages (the one that was faulty and the next three after it) and the kernel call will return the number of pages that it was allowed to bring to memory -- it is possible that a next page does not belong to the user application, or in other words it was not mapped, and in that case we are not allowed to bring it to memory.

Currently, after scheduling the \textit{pf\_send\_handler} tasklet, we also schedule the tasklet responsible for page faults in the destination buffer, just in case there is a new fault to-be-handled from that side, as well. This works as an optimization, thus it does not seem necessary. %It is not clear yet why it is necessary, but while working on our experiments, it was the only way to work. %\todo[inline]{future work (rcv tasklet call after sender): More research into this is expected in the future.}

\subsection{Page Fault at Destination address}

\begin{figure}[H]
        \centering
        \captionbox[Page Fault flow at destination address]{Page Fault flow at destination address \label{fig:pgf_flow_dst}}{
             \includegraphics{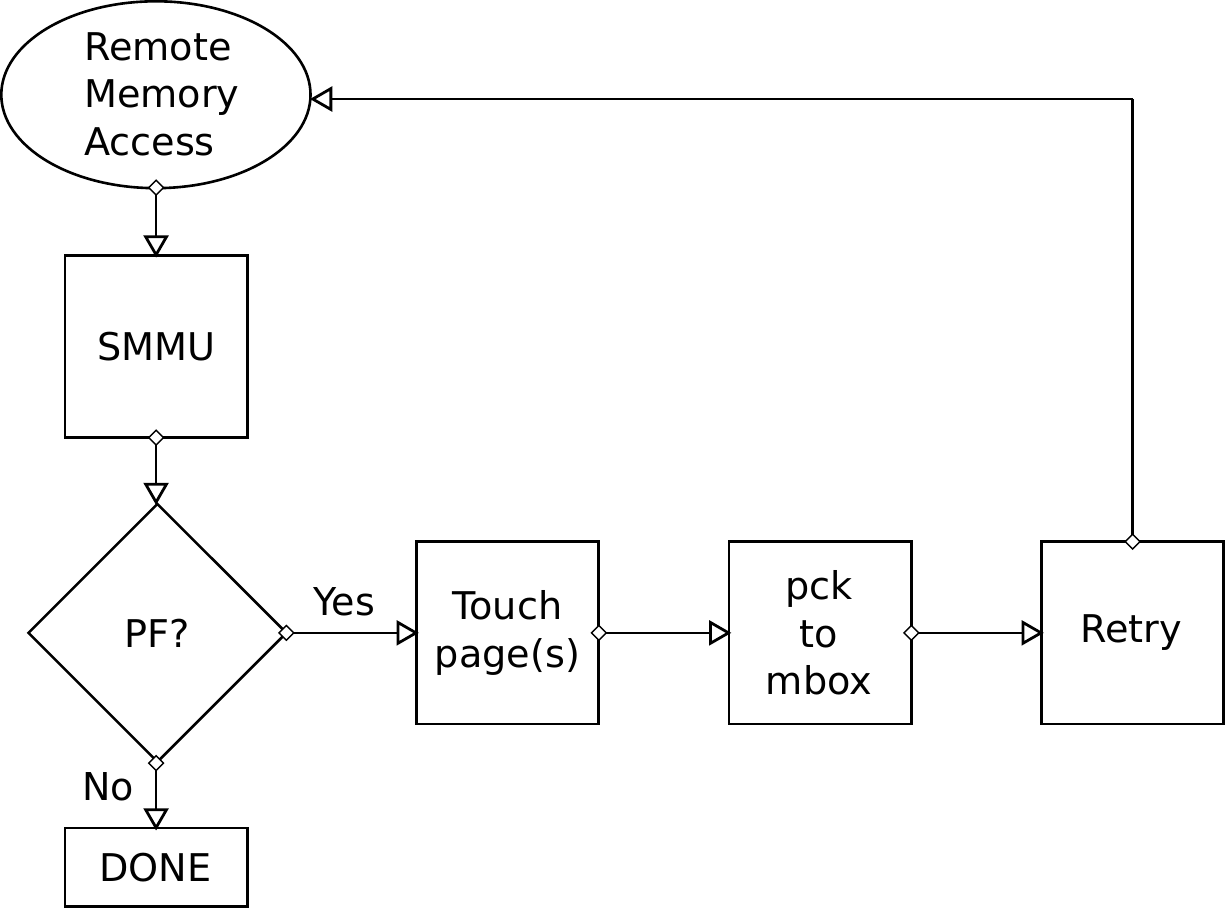}
        }
\end{figure}

%While writing this document for the thesis the only way to support a Page Fault is when it happens at the destination address.
The main mechanism includes modifications to the mainline kernel driver for the SMMU (ARM's IOMMU), a user-space library and additions to the hardware design and the firmware of the R5 processor. The common parts of the mechanism were described previously, so here we will describe the different characteristics of the receive (destination) path of the page fault handling mechanism.

\subsubsection{Hardware}

In order to serve all page faults caused when the receiver part of the custom DMA, described in Section~\ref{custom_PLDMA}, tries to access the memory, we implemented a FIFO, responsible to log all necessary information of the faulty transfers. We will use this information to solve the page fault and to also request from the sender to retransmit the previous faulty transfer --which mainly works as an optimization, because the system, as already mentioned before, has the functionality of a time-out, that would eventually trigger a re-transmission. %Currently time-out is 200 ms, but in general it could be either too early or late after the page fault was handled.

The FIFO originally was created to be read through an interface (AXI-lite) that when working in our Trenz experimental prototype allowed read requests of 32-bits. Later, when moving to new designs we could have read requests of 64-bits. So this is what our design has in mind, although the current interface in-use allows us to read up to 128-bits. The 64-bit mechanism includes a Finite State Machine (FSM), that makes sure that when a user has consumed/read the related data, only then it pops this information (FIFO entry). But since we need two (2) read requests to pop the entry from the FIFO, the FSM ensures that this happens in a safe order and way (e.g. if someone tries to read the second 64-bits of data, the entry will not be popped out from the FIFO). The depth of our FIFO is currently 512 and the width is 128 bits.

Our FIFO basically logs the information of all incoming packets that experienced a slave error (AXI NACK), which by the way could happen in different scenarios and not only in page faults. That is why we expect our mechanism to be used in the future by embedding solutions for other problems/type of faults as well, including invalid addresses. Currently, as an optimization, when a new negative acknowledgement (slave error) arrives in the receiver part of the FORTH PLDMA, we check it and if it has the same source node identification, transaction identification, sequence number and virtual page (excluding the page offset) with the entry we added (pushed) last time in our FIFO, we do not add it again. In the future, using get\_user\_pages(), it would be okay to check only the source node id, the transaction id and the sequence number, provided that pages arrive in order and when we request to get the user pages of a virtual address, the first time we add it to our FIFO, it will belong to the first page, in order for all pages to be touched.

The author of the design of this thesis had two (2) things in mind: first was to be able to serve 32-bit read requests if necessary (currently we do two (2) 64-bit read requests) and also to be convenient for the user and the developer to check these entries, while debugging the mechanism. In the future, we expect this FIFO to be optimized space-wise, which was in our plans but due to lack of time it did not happen before submitting this thesis. %\todo[inline]{Check column lines: In the Table~\ref{tab:fifoentry_format} below}
In the Table~\ref{tab:fifoentry_format} below we can see the format of each entry in our FIFO.

\begin{figure}[h!]
        \centering
        \captionbox[The detailed flow of handling a Page Fault at the destination buffer using NL]{The detailed flow of handling a Page Fault at the destination buffer using Netlink sockets \label{fig:remwrite_pfatdst}}{%
        %\fontsize{10pt}{10pt}\selectfont\includesvg[width=1\textwidth]{Figures/remwrite_pfatdst}
    %   \includesvg{Figures/remwrite_pfatdst_cp}
       \includegraphics{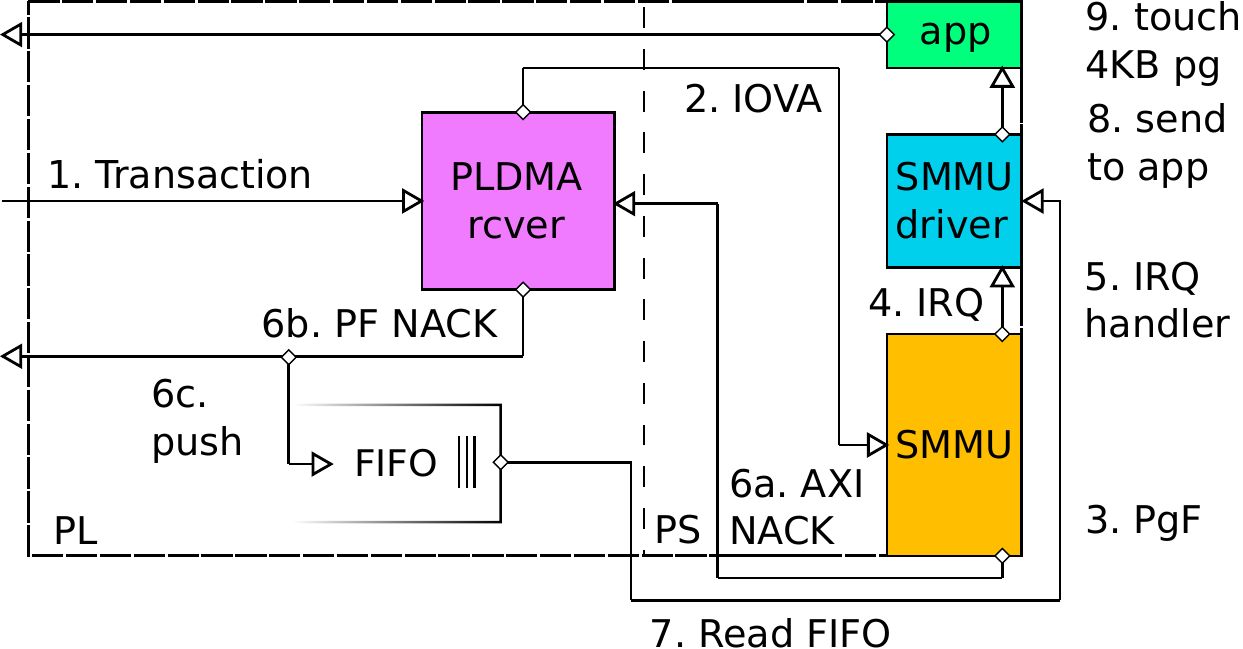}
        }
        %\caption{SVG in superior directory}
\end{figure}

\begin{table}[ht!]	
%\begin{table}[ht!]\Huge
    \caption{FIFO entry in Receiver side of custom PLDMA}
    \centering %
   \resizebox{13.5cm}{!}{
        \begin{tabular}{| c | c | c | c | c | c | c |}
        %\toprule \rowstyle %{\bfseries}%
        \hline
            % // First Read  (32 bits): 0s (2 bits) + src_ID (22 bits) + 0s (2 bits) + tr_ID (2 / 14 bits) +  0s (3 bits) + Valid (1 bit)
            \rowcolor{blue!25} \textbf{00} & \textbf{src\_ID} & \textbf{00} & \textbf{tr\_ID MSB} & \textbf{000} & \textbf{Valid} \\\hline
            2 bits & 22 bits & 2 bits & 2 bits & 3 bits & 1 bit \\\hline
            \multicolumn{2}{|c|}{6 hex} & \multicolumn{2}{c|}{1 hex} & \multicolumn{2}{c|}{1 hex} \\\hline
            %  // Second Read (32 bits): tr_ID (12 / 14 bits) + 0s (2 bits) + seq_num (14 bits) + 0s (3 bits) + Valid (1 bit)
            \rowcolor{blue!25} \textbf{tr\_ID LSB} & \textbf{00} & \multicolumn{2}{c|}{\textbf{seq\_num}} & \textbf{000} & \textbf{Valid} \\\hline
            12 bits & 2 bits & \multicolumn{2}{c|}{14 bits} & 3 bits & 1 bit \\\hline
            3 hex & \multicolumn{3}{c|}{4 hex} & \multicolumn{2}{c|}{1 hex} \\\hline
            % // Third Read  (32 bits): PDID (16 bits) + IOVA (12 / 32 bits - MSB, i_IOVA[AXI_ADDR_WIDTH-1:AXI_ADDR_WIDTH-1-12]) + 0 (1 bit) + EXA_ACK (2 bits) + Valid (1 bit)
            \rowcolor{blue!25} \multicolumn{2}{|c}{\textbf{PDID}} & \multicolumn{1}{|c|}{\textbf{Faulty IOVA MSB}} & \textbf{0} & \textbf{EXA\_ACK}  & \textbf{Valid} \\\hline
            \multicolumn{2}{|c}{16 bits} & \multicolumn{1}{|c|}{12 bits} & 1 bit & 2 bits & 1 bit \\\hline
            \multicolumn{2}{|c}{4 hex} & \multicolumn{1}{|c|}{3 hex} & \multicolumn{3}{c|}{1 hex} \\\hline
            % // Fourth Read (32 bits): IOVA (20 / 32 bits) + (some 0s: 32 - (20+1) = 11 ) + Valid (1 bit)
            \rowcolor{blue!25} \multicolumn{3}{|c}{\textbf{Faulty IOVA LSB}} & \multicolumn{2}{|c|}{\textbf{0000\_0000\_000}} & \textbf{Valid} \\\hline
            \multicolumn{3}{|c}{20 bits} & \multicolumn{2}{|c|}{11 bits} & 1 bit \\\hline
            \multicolumn{3}{|c}{5 hex} & \multicolumn{3}{|c|}{3 hex} \\\hline
        \end{tabular}
        }
        \label{tab:fifoentry_format}
    \end{table}

% \begin{figure}[htb]
%         \centering
%         \includesvg[width=0.1\textwidth]{Figures/fifo_inkscape}
%         \caption{SVG in superior directory}
% \end{figure}

% not sure if it's working...
% \begin{figure}[htb]
%         \centering
%       \input{Figures/fifo_inkscape.pdf_tex}
%         \caption{SVG in superior directory}
% \end{figure}

\subsubsection{Driver}\label{pgf_dst_driver}

% \todo[inline]{Add figure/schematic}

When the interrupt handler, \textit{arm\_smmu\_context\_fault}, catches the related exception and distinguishes that it was due to a translation fault of a local write request to the memory, it schedules a \textit{tasklet}, called \textit{pf\_rcv\_tasklet}. This tasklet is responsible to do a couple of things, that will be described below:

% \begin{figure}[h!]
%         \centering
%       \input{Figures/remwrite_pfatdst_cp.pdf_tex}
%         \caption{SVG in superior directory}
% \end{figure}

%\the\textwidth

\begin{itemize}
    \item Read from the 128-bit FIFO added to the hardware part of the receiver all the related information to the transaction (block) that failed, such as: 
    \begin{enumerate}
        \item the source node ID (src\_ID) - 22~bits: corresponds to the initiator computing node of the transfer
        \item transaction ID (tr\_ID) - 14~bits
        \item sequence number (seq\_num) - 14~bits
        \item protection domain ID (PDID) - 16~bits
        \item faulty virtual address (iova) - 32~bits: 4 most significant bits correspond to the process index in a rank/protection domain and the remaining 28 bits are the most significant bits of a 39-bit virtual address, which is currently the width of the address our system supports. The 12 less significant bits of the virtual address correspond to the page offset which is not needed in our mechanism, since we handle requests per page. 39-12 = 27 bits. The 28th bit is wired to zero.
        \item other kind of bits (such as Read/Write, Valid bits etc)
    \end{enumerate}
    This information is crucial and necessary in order to be able to successfully initiate a re-transmission of the faulty transfer -- when the time comes and we know for sure that the page is already resident in memory. This seems rational when we do not rely on time-out re-transmissions, which cannot be effective when we cannot be really sure when exactly the page is paged in.
    \item Now that we know the information of the transaction that failed due to a page fault, we can initiate the handling mechanism - a mechanism responsible to page in the page that is not resident yet. As it is already mentioned, we mainly used Netlink sockets in order to transfer the information of the faulty transaction to the corresponding process. This process belongs to the same protection domain, that the faulty virtual address belongs to. Our initial approach causes an internal page fault, that will be handled by the Memory Management Unit (MMU) of the CPU. In simple words, we inform the corresponding process to touch the faulty page through the Netlink sockets, triggering an internal fault. The other approach, is using get\_user\_pages(), which allows us to be benefited by some optimizations as described in the page fault in the source buffer approach. %\todo[inline]{future work but already mentioned before.}
    \item After this step, we need to let the initiator node know that the page fault is handled, so the initiator node should re-transmit it. For the purposes of this sub-task, we utilize a packetizer that sends the necessary information of the transfer to the mailbox that the R5 polls, in order to initiate the re-transmission. We have mentioned this in Section~\ref{mechanism_user_api}, because currently this kind of re-transmission is only supported from user-space. In the future, we expect to provide such service from kernel-space, that will probably lead to less context-switches when we do all the work completely in the kernel (using get\_user\_pages()).
\end{itemize}

% In order for this part (receive path) of the mechanism to work, we also modified the firmware of the R5 Real-Time processor of the Zynq UltraScale+ MPSoC, to be able to work in accordance with our needs. %In Appendix~\ref{Appendix_R5}, we have a detailed description of all the tweaks we developed in already existing firmware for R5.

As already mentioned in Section~\ref{mechanism_user_api}, we also developed a module that, when inserted, it is detected as a device. When a user calls mmap() with the file descriptor of the new device, the call-back of mmap() allocates channel 0 out of {0..3} (4 channels in total) of the first available packetizer. Also, in the end, when the user no longer needs this mapping (e.g. exiting the program), the related call-back frees the previously allocated packetizer. We mention this module in this part of the document, since modules and drivers are somehow connected as terms (see Section~\ref{driversvsmodules}).

\subsubsection{Real-Time co-processor modifications}\label{R5} % For referencing this appendix elsewhere, use \ref{AppendixD}

%\section{Real-Time co-processor modifications}

In order for this part (receive path) of the mechanism to work, we modified the firmware of Real-Time co-processor R5 of Zynq UltraScale+.

At this point, it is important to clarify that the development of the R5 framework was part of the work conducted by other colleagues of ours at FORTH and mostly as part of \cite{TzanakisMScThesis2019}.

The modifications for this work were made in the r5\_dma\_controller.c source file that when built with other files results to the firmware file (.elf).

When receiving a Page Fault NACK and more specifically, any AXI Slave Error response (2'b10), instead of triggering an instant retransmission (if the transfer was valid), R5 pauses (for this transaction). In that case, there are only two (2) ways for this transfer to be re-transmitted:

\begin{enumerate}
    \item Either by the expiration of the time-out period of the corresponding transaction
    \item Or by receiving a specific message in its mailbox, that explicitly requests a retransmission for a transaction identification (trID), with a specific sequence number that might be valid and the protection domain
\end{enumerate}

Explicit or on-demand requests to retransmit using the packetizer to send to the mailbox polled either way by the R5 co-processor works probably, as already mentioned before, as an optimization. Initially we did not know how much time it takes for a faulty, due to page faults, transfer to be resolved, meaning we could not decide what is the optimum time-out period for retransmission -- something still uncertain, because it is not easy to generalize a solution based on our current environment, that behaves differently when a fault occurs in the sending or the receiving part of the PLDMA.

Below we see the part of the code that was added when R5 decodes the Page Fault Negative acknowledgement that is sent from the receiver part of the FORTH PLDMA. The previous (default) behavior was to instantly retransmit. Now we have modified this behavior and we expect a retransmission to be triggered either by a time-out or by a corresponding message that will be received in the mailbox polled by R5.

\begin{lstlisting}[language=C]
// [...]
// PF Negative ACK: Message for R5 to NOT retransmit, which was the default behavior
else if(((mbox_value >> 16) & 0x07) == 1){ 
	transaction_id = ((mbox_value >> 2) & 0x03FFF);
    #if PF_DEBUG
    	xil_printf("PAGE FAULT NACK for trans_id %d with errorcode %d @ set\n\r",transaction_id,(mbox_value >> 16));
    #endif
	return;
}
else{
// [...]
}
// [...]
\end{lstlisting}

The code below, decodes the message that was received in R5's mailbox. If this message has as opcode the value 2, it means that is a \enquote{Page Fault: Ready to retransmit} message. 

It is important to acknowledge here that the first (least significant) twelve (12) bits of this message after the two (2) bits of the opcode are wired by the packetizer (kernel-space) and not by the user who sent the message. This way we can avoid malicious users who may pretend they legitimately request an access to the local memory of a protection domain that does not exist. We need to elaborate further on this. One approach would be to check the wired protection domain and the one sent/given by the user to see if they match. 

We expect that when moving all the mechanism to kernel space, we will not have to deal with such issues.

\begin{lstlisting}[language=C]
// [...]
// Request-to-retransmit sent from packetizer as part of page fault mechanism (receiver side)
case 2:			
    #if PF_DEBUG
        xil_printf("NEW PF msg received! \n\r");
    #endif
    
    word0 = mbox_value; // consume first 32 of 64 bits from msg received
    wired_opcode = word0 & 0x03; // wired opcode (2 bits -- if 2, means PF msg)
    wired_pdid = ((word0 >> 2) & 0x0FFFF); // wired protection domain ID (4x4 = 16 bits)
    transaction_id = ((word0 >> (2+16)) & 0x3FFF); // received transaction id (2+3x4 = 14 bits)
    
    word1 = *(ack_mbox_b); // consume second 32 of 64 bits from msg received
    seq_num = (word1 & 0XFFF); // received sequence number (3x4 = 12 bits)
    rcved_pdid = ((word1 >> 12) & 0x0FFFF); // received protection domain ID (4x4 = 16 bits)
    
    #if PF_DEBUG
        xil_printf("TID is: %d, SEQ_NUM is: %d\n\r", transaction_id, seq_num);
        xil_printf("Wired PDID is: %d, received PDID is: %d\n\r", wired_pdid, rcved_pdid);
    #endif
    
    // Message with out-dated seq. number was received -- it will be ignored
    if(seq_num != pending_transactions[transaction_id].seq_num){
    #if PF_DEBUG
    	xil_printf("PF msg: expected seq was: %d -----BUT------  seq received was: %d and tid %d \n\r",pending_transactions[transaction_id].seq_num,seq_num,transaction_id);
    #endif
    	return;
    }
    
    // call timeout_drop_transaction function in order to drop out the transaction that received NACKED
    timeout_drop_transaction(&timeout,transaction_id);
    // find the virtual channel of transfer
    int virtual_channel_num = pending_transactions[ transaction_id ].transfer_id;
    // retransmit this transaction and send completion notification again if this is the last block
    transaction_retransmission(virtual_channel_num, transaction_id);
    
    return;
}
// [...]
\end{lstlisting}

%\cleardoublepage

%----------------------------------------------------------------------------------------

% \vfill
% \footnoterule
% Chapter 4
\chapter{Evaluation}\label{chap:evaluation}

In order to evaluate our work, we conducted some experiments to measure mostly the latency of RDMA transfers when using our mechanism.

Ideally, we would like to have results from measurements on real applications but due to lack of time, we can only evaluate our mechanism based on custom-made micro-benchmarks. In other words, we are able to only run an application that triggers RDMA transfers using the FORTH PLDMA, that experience minor page faults. We expect our mechanism to also work with major page faults. %\todo[inline]{future work: major faults, measurements on real applications}.

When performing latency measurements, it is common to exclude the overhead of pinning or touching the pages before they are DMA'ed. As a result, we do not expect our approach to have better measurements compared to these cases, because our mechanism causes pages to be paged in on-demand, which adds an extra overhead during the transfer.

On the other hand, we propose that page faults are not common and this type of overhead will not be paid often; page faults are considered a rare phenomenon.

Another measurement we would like to conduct, that was not possible due to lack of time, was the memory utilization. Our mechanism ensures better utilization of the physical memory, because in the common case we will have pages that belong to a process (application) reside in memory only when they have to be there (i.e.~they are actually used). %\todo[inline]{future work: measure memory utilization}.

During the time we had on our hands to evaluate our work, we were interested in measuring:

\begin{enumerate}
    \item The latency / execution time of an RDMA transfer (Remote Write) with: 
    \begin{itemize}
        \item our mechanism enabled
        \item pinned buffers prior to the RDMA transfer %(optionally: unpin them after the transfer)
        \item touched buffers prior to the RDMA transfer
    \end{itemize}
    \item The cost (time) of pinning pages
    \item The cost (time) of unpinning pages
    \item The cost (time) of touch
    \item The latency / execution time of the page fault handler (driver)
\end{enumerate}

In order to perform measurements in user-space, we made use of the method named \enquote{clock\_gettime()}, defined in the Linux header file \textit{time.h}. The resolution of the result from this method is in nanoseconds. Although this method is not a system call (it is a VDSO, which in theory is used to minimize the overheads of system calls), it might add a delay - as we will see later in our results. In kernel-space, we made use of the call/method named \enquote{ktime\_to\_ns}, which also produces a result in nanoseconds resolution.

% \section{Remote Write}

% For the measurements during a Remote Write DMA transfer, we used the corresponding PLDMA library implemented at FORTH, that supports remote write DMA requests.

We have conducted our measurements in one FPGA of a QFDB (intra-QFDB), but we do not expect significant differences when moving to more hops (FPGAs) or even other computing nodes (QFDBs) -- the hop-to-hop latency is about 100 nanoseconds.

In our experiments the transfer sizes are: 16 Bytes, 64 Bytes, 256 Bytes, 1024 Bytes, 4K Bytes, 16K Bytes, 32K Bytes and 64K Bytes. We consider 64K Bytes as the maximum transfer size in our experiments, because in Linux kernel, as mentioned in \cite{DBLP:conf/asplos/LesokhinERSGLBA17}, %\todo[inline]{add reference to mlock upper limit}
the upper limit of Bytes that a user can lock in memory (using the \enquote{mlock} method), is 64K Bytes.

Because of this upper limit in our PLDMA transfers, it would not make any difference whether we have Transparent Huge Pages (THP) mechanism (see Section~\ref{transparenthugepages}) enabled or disabled -- since we do not work with Huge Pages in this set of experiments, we do not expect any THP to occur. Moreover, we hope that our mechanism will be used by the main ExaNeSt prototype soon, which is when we expect the system to be able to recover from translation faults caused during THP activity. %which means it will be tested then whether it is able to support translation faults in SMMU, that might occur during THP.

\begin{table}[ht!]
    \caption{Overhead (time in μsec) of mmap, munmap, pin, unpin and touch}
    \centering %
    \resizebox{13.5cm}{!}{
        \begin{tabular}{|>{\columncolor[gray]{0.8}} l | c | c | c | c | c | c | c | c |}
        \hline
        \rowcolor{blue!25}
        \multicolumn{1}{|c|}{} & \textbf{16 B} & \textbf{64 B} & \textbf{256 B} & \textbf{1 KB} & \textbf{4 KB} & \textbf{16 KB} & \textbf{32 KB} & \textbf{64 KB} \\\hline
        mmap & 2 & 2 & 2 & 2 & 2 & 2 & 2 & 2 \\\hline
        munmap & 6 & 6 & 6 & 6 & 7 & 10 & 12 & 19 \\\hline
        pin & 6 & 6 & 6 & 6 & 6 & 15 & 27 & 49 \\\hline
        unpin & 2 & 2 & 2 & 2 & 2 & 5 & 8 & 14 \\\hline
        touch & 3 & 3 & 3 & 3 & 3 & 10 & 19 & 40 \\\hline
            
        \end{tabular}
    }
    \label{tab:ovrheads}
\end{table}

Table~\ref{tab:ovrheads} describes the overhead of each different method for one buffer (e.g. the source address of a DMA transfer) --in our measurements someone would have to consider each overhead twice (two buffers: source and destination). The way we measured the overhead is described in Listing~\ref{lst:real_measurements}, where instead of the call appeared as \enquote{PLDMA\_transfer} we had the call of the corresponding method (e.g. \enquote{mmap}). This means that an overhead for \enquote{get\_time()} is probably included in our %\enquote{Real} (Listing~\ref{lst:real_measurements}) type of 
measurements. Initially, our RDMA measurements for page faults included the overheads of these methods (Table~\ref{tab:ovrheads}), which is the reason we evaluated them.

In the case of the \enquote{pin}, we also expect a call of \enquote{unpin} to follow for two reasons:
\begin{itemize}
    \item This is the only way for the \enquote{pin} system call to be effective in each next iteration of the loop, when using the same address (page). Also, the approach of pinning/unpinning of the buffers before/after they are DMAed is used by many~\cite{DBLP:conf/asplos/LesokhinERSGLBA17}. 
    \item It is expected that when someone pins a page, they will eventually have to unpin it (e.g. at the end of the PLDMA transfer), in order to make more space available for the current or other users (memory utilization).
\end{itemize}

While running experiments that were triggering page faults, we faced many challenges because the fault path of the custom FORTH PLDMA was not stressed enough until that moment. This is the main reason that all measurements related to page faults were a result of only 500 iterations for each transfer size.

\begin{lstlisting}[language=C, caption={Pseudo-code for \enquote{ideal} measurements}, label=lst:ideal_measurements, captionpos=b]
start = get_time();
for(i=0; i<iterations; i++){
    PLDMA_transfer();
}
end = get_time();
time = (end-start)/iterations;
\end{lstlisting}

\begin{lstlisting}[language=C, caption={Pseudo-code for \enquote{real} measurements}, label=lst:real_measurements, captionpos=b]
for(i=0; i<iterations; i++){
    start = get_time();
    PLDMA_transfer();
    end = get_time();
    time += (end-start);   
}
avg_time = time/iterations;
\end{lstlisting}

\begin{figure}[h!]
        \centering
        \captionbox[Remote Write: All buffers pre-touched, Transfer-Only Latency]{Remote Write Transfer (One FPGA), All buffers pre-touched, Transfer-Only Latency \label{fig:RemWr_1FPGA_AllTouched_OnlyTransfer}}{
           \includegraphics[keepaspectratio,width=1\textwidth]{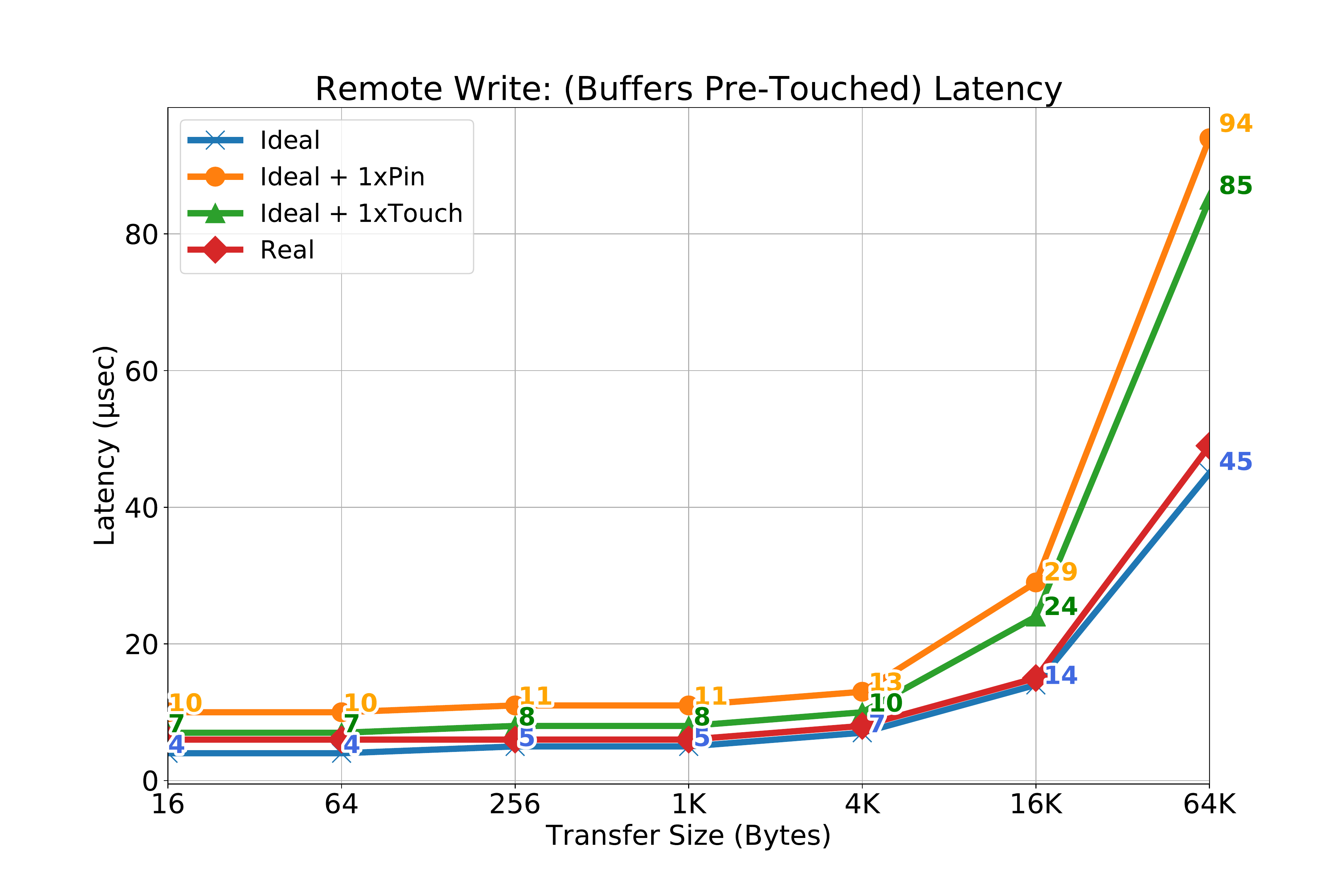}
        }
\end{figure}

The round-trip latency of a remote DMA write transfer that experiences zero (0) page faults during the RDMA is low, i.e. 4 μsec for 16 Bytes, when using the \enquote{Ideal} (Listing~\ref{lst:ideal_measurements}) type of measurements. 

In the \enquote{Ideal} type of measurements we do all the memory mapping/unmapping work before/after the RDMA transfer. This measurement includes only the method, that triggers the transfer and the method that is polling the status register of the PLDMA, that signals the completion of a transfer. For this experiment, we conducted 10000 iterations for each transfer size and we calculated the average. We use the name \enquote{Ideal}, because we expect no overhead from any kind of system call during the measurement.

Initially our case study involved measurements that used a \enquote{for} loop, which included some system calls such as memory mapping/unmapping of buffers, mainly for two (2) reasons:
\begin{enumerate}
    \item In order for a page fault to be triggered in each iteration, a call that unmaps the buffer from the address space was necessary
    \item The measurement of the latency (time) was outside of the loop (as in Listing~\ref{lst:ideal_measurements})
\end{enumerate}
Another reason behind this experiment choice is that we wanted to evaluate transfers that cause page faults and compare the results to the measurements of transfers that do not cause page faults during the RDMA, because they were either pinned or touched prior to the DMA transfer. Since touching pages is part of our page fault handling mechanism, it would only make sense to compare it with a mechanism that basically pre-faults the pages prior to the RDMA transfer.

Although this approach seemed rational we thought that the results would probably cause confusion to the readers due to the cost/overhead of the other methods and system calls involved, so we decided to follow the methodology that is depicted in Listing~\ref{lst:real_measurements} for the measurements that involve page faults. In other words, in measurements where we evaluate the page fault mechanism we compare against the no-page fault during RDMA evaluated using the \enquote{Real} type of measurements.

In Figure~\ref{fig:RemWr_1FPGA_AllTouched_OnlyTransfer}, we can see the measurements conducted using the average of the execution time of the transfer, that experiences no page faults and thus no handling from our mechanism is involved. In order to achieve this, we touch one (1) Byte for each page (4K Bytes) prior to the DMA transfer. 

In the same Figure (Fig.~\ref{fig:RemWr_1FPGA_AllTouched_OnlyTransfer}) we can see that the \enquote{Ideal} latency increases when we add the overhead of one \enquote{pin} or \enquote{touch} operation of a buffer. %As we will discuss later in the experiments we conducted in order to evaluate our mechanism, overheads such as these were also included, so someone should have this in mind.
We also see a line that describes the results of the \enquote{Real} type of measurements, as described above. %This is the type of measurement we used in order to be able to evaluate many iterations of page faults, that require the buffers to be unmapped before the next iteration in order to also trigger a minor page fault.
Due to lack of space we could not show the value of \enquote{Real}, but it can be found in the next Figures, whenever we show results of no involvement of our mechanism (no page faults).%, since when we had page faults our measurements were conducted the same way.

Since we did not want our process under-test to be preempted by other tasks in a random way that could affect our measurements, we made sure that all interrupts of the system were moved to the CPU 0 and we conducted our experiments on CPU 2. In order to do this, a script was written that would overwrite the smp\_affinity value of each IRQ in /proc/irq/. Of course this cannot guarantee that all IRQs will be moved to CPU 0, but it seemed sufficient to move the majority of them. For example, the interrupt with the name \enquote{arch\_timer} could not be moved, but it appears to be a timer necessary for the system and each CPU, which is the reason it runs periodically on each one of them. At the same time, we were calling \enquote{taskset 0x4}, before running the executable of our application, in order for it to run on CPU 2 --0x4, or 0100 in binary, is a mask that indicates the CPUs that an executable will run. This would allow us to get as much reliable as possible measurements for our mechanism. Although using this approach allowed us to eliminate the outliers, we noticed that it did not affect significantly the average values of our measurements since page faults are costly in any case. We also noticed that sometimes the process would fall asleep and never wake up, which means it requires further investigation before actually using the move of IRQs.

In the next Figures we evaluate the latency of a Remote DMA Write inside one MPSoC (FPGA) of one QFDB (node). We have three (3) different types of bars. The first bar (blue color), indicates the latency of a Remote DMA Write where no page fault occurs thus no invocation of our page fault handling mechanism, since the buffers were pre-touched. The second bar (orange color) indicates the latency of a Remote DMA Write, which experiences a page fault in the source/destination address and the paging-in mechanism used is touch-ing one page (\enquote{Touch-A-Page}) per invocation of the handler (tasklet), by sending the corresponding information to the user-space and the related process through Netlink sockets. The third bar (green color) indicates the latency of a Remote DMA Write, which experiences a page fault in the source/destination address and the paging-in mechanism used is touch-ing \enquote{ahead} (\enquote{Touch-Ahead}), which means paging-in up to four (4) pages per invocation of the handler (tasklet), using the get\_user\_pages method. As already mentioned before, the reason we chose to \enquote{touch} up to four (4) pages is that each transaction (or block) is at most four (4) 4KB pages and we expect that if a page fault is caused on the first page, it is very likely that the same will happen to the next three (3) pages (locality).

\begin{figure}[h!]
        \centering
        \captionbox[Remote Write: Page Fault at Destination - Latency]{Remote Write Transfers (One FPGA) with Page Fault at Destination - Latency\label{fig:RemWr_1FPGA_PgF_Dst}}{
            \includegraphics[keepaspectratio,width=1\textwidth]{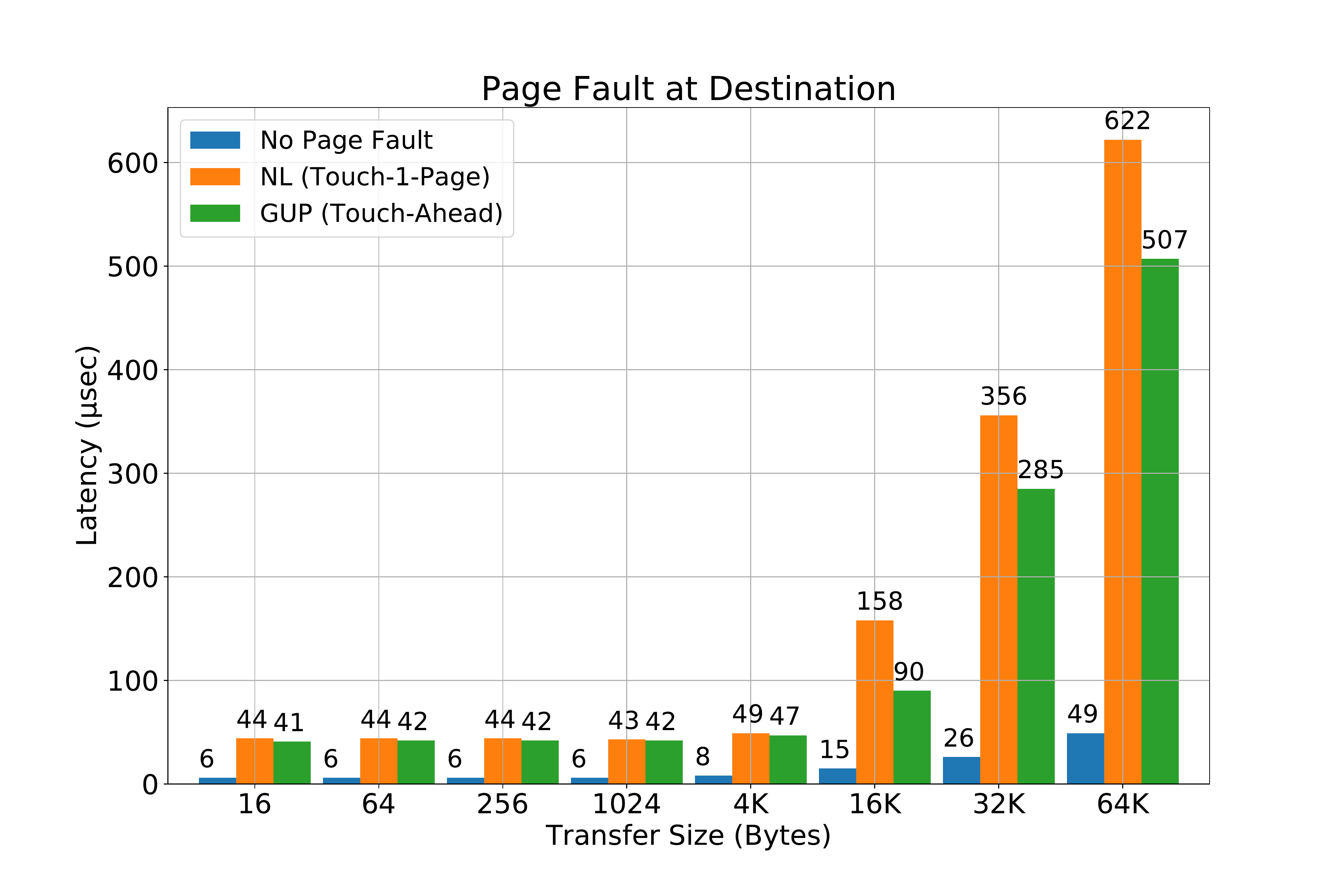}
        }
\end{figure}

In Figure~\ref{fig:RemWr_1FPGA_PgF_Dst} we see the latency of a Remote DMA Write in one MPSoC (FPGA), where a (minor) page fault occurs in the destination address during the Remote DMA Write transfer. As expected, the results seem similar up to 4KB, which is the size of one page -- the handling up to the size of one page is the same cost-wise. For the sizes of 16~KB, 32~KB and 64~KB we see a benefit of 1.7x, 1.2x and 1.2x when using \enquote{Touch-Ahead} instead of \enquote{Touch-A-Page}. We believe that the decrease of benefit in 32~KB and 64~KB is due to the interleaving effect in FIFO: we noticed that we might have duplicate packets enqueued when the transfer size is more than 32~KB (two transactions or in other words, blocks). Because of this interleaving effect, it takes more time to find a new page / set of pages to page-in during the handling. As an optimization, we keep track of the last two (2) entries (transactions) that triggered a page fault, on the driver, and we do not repeat the handling for a page we have already handled. By design, only two (2) different blocks might come interleaved, so checking only the last two (2) entries of one source node, seems sufficient. In the future we expect to add this check on hardware, in order to eliminate the duplicates that can appear in our current design of the FIFO.

\begin{figure}[h!]
        \centering
        \captionbox[Remote Write: Page Fault at Source - Latency]{Remote Write Transfers (One FPGA) with Page Fault at Source - Latency\label{fig:RemWr_1FPGA_PgFSrc}}{
            \includegraphics[keepaspectratio,width=1\textwidth]{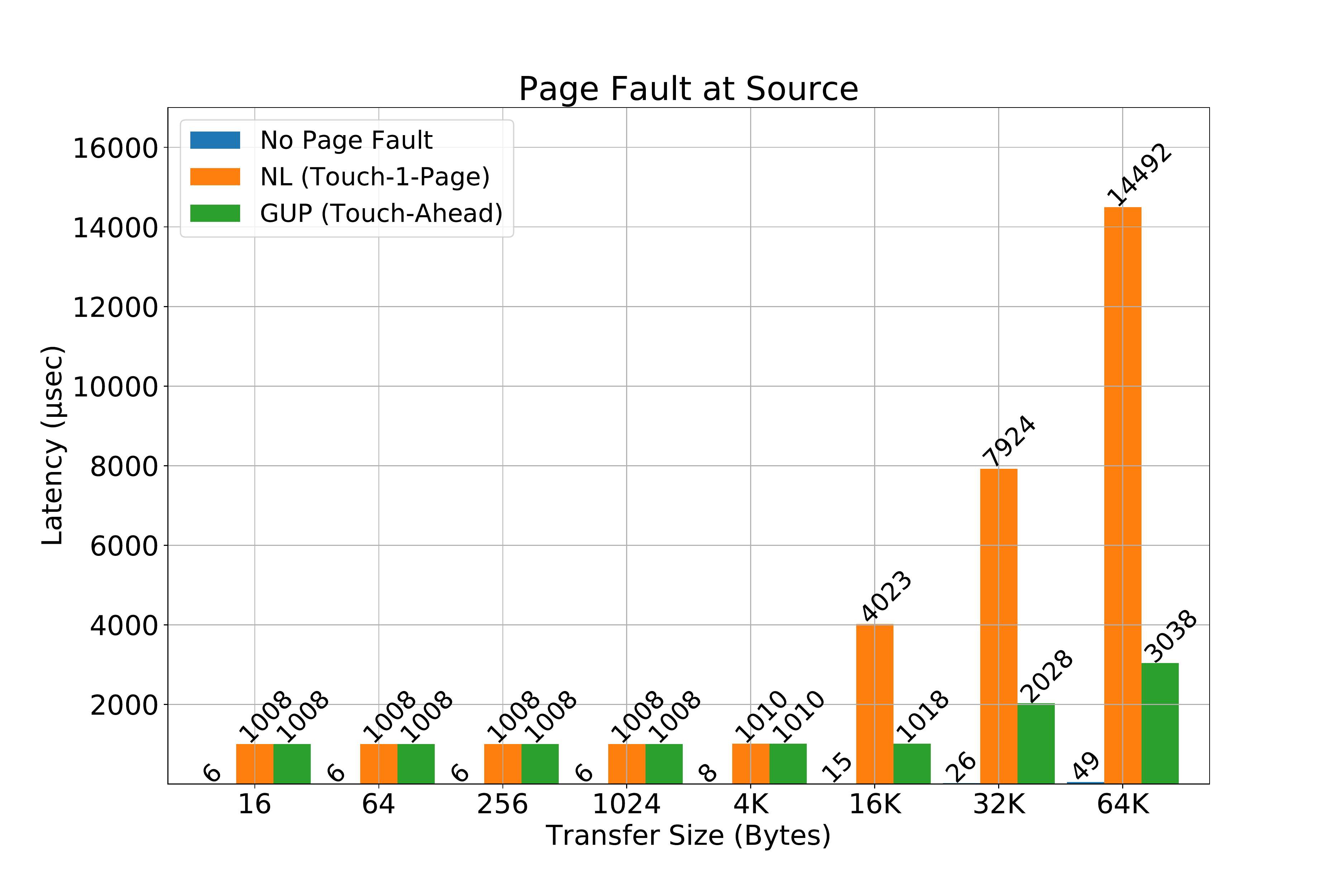}
        }
\end{figure}

We tried three (3) different time-out periods: 25~ms, 2.5~ms and 1~ms. As expected our best scenario was with time-out period set to 1~ms, which is reasonable due to the fact that we can complete the transfers that encountered a page fault sooner. At this point, it is important to remind to the readers that our mechanism expects time-out retransmissions when a page fault is encountered in the source addresses. This means that in all cases we expect the corresponding latency to be dominated by the time-out delay, which currently is 1~ms.

In Figure~\ref{fig:RemWr_1FPGA_PgFSrc}, we see a similar behavior for transfer sizes up to 4~KB, that we saw in Figure~\ref{fig:RemWr_1FPGA_PgF_Dst}. For transfer sizes of 16~KB, 32~KB and 64~KB we see a benefit of 3.9x, 3.9x and 4.7x when using \enquote{Touch-Ahead} instead of \enquote{Touch-A-Page}. 3.9x seems a rational benefit since \enquote{Touch-Ahead} pre-touches up to four (4) pages per transaction (block). It is uncertain why the 64~KB transfer size results to a better benefit (4.7x), but we believe it might be an effect due to the interleaving and the FIFO effect, already mentioned before. More investigation on this is expected in the future. 

\begin{figure}[h!]
        \centering
        \captionbox[Remote Write: Page Fault at Source and Destination - Latency]{Remote Write Transfers (One FPGA) with Page Fault at Source and Destination - Latency \label{fig:RemWr_1FPGA_PgFSrcnDst}}{
            \includegraphics[keepaspectratio,width=1\textwidth]{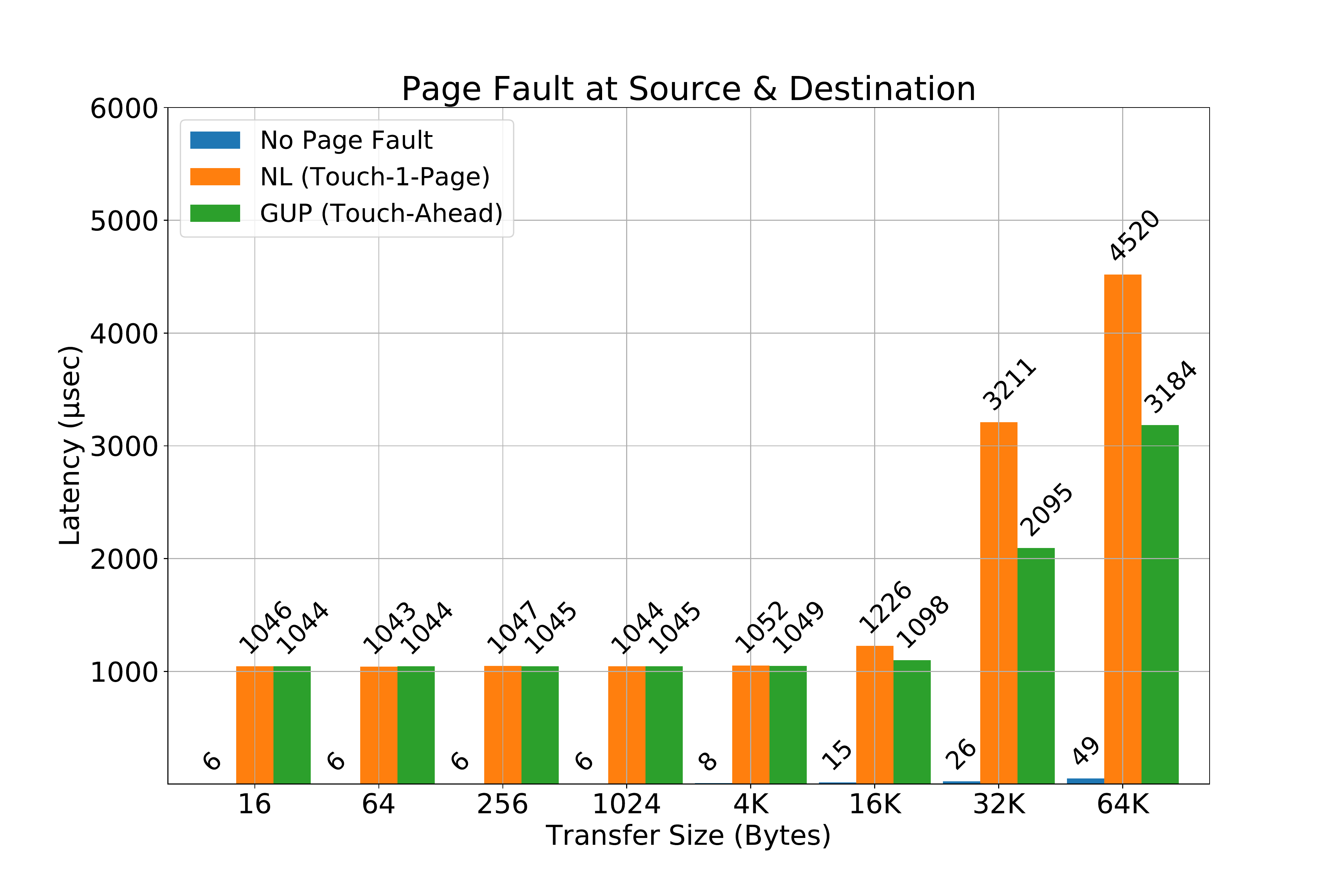}
        }
\end{figure}

In Figure~\ref{fig:RemWr_1FPGA_PgFSrcnDst}, we see a similar behavior for transfer sizes up to 4~KB, that we saw in Figure~\ref{fig:RemWr_1FPGA_PgF_Dst}. It is obvious that transfers with smaller transfer sizes are dominated by the time-out in the source buffer. For the sizes of 32~KB and 64~KB we see a benefit of 1.5x and 1.4x when using \enquote{Touch-Ahead} instead of \enquote{Touch-A-Page}. For the 16~KB transfer size the benefit when using \enquote{Touch-Ahead} is way smaller and we believe that is due to the fact that even when we utilize \enquote{Touch-A-Page}, the mechanism manages to touch all pages (4) before the time-out retransmission. This can explain why \enquote{Touch-Ahead} and \enquote{Touch-A-Page} have similar results.

\begin{figure}[h!]
        \centering
        \captionbox[Remote Write: Page Fault at Source Vs Source and Destination - Latency]{Remote Write Transfers (One FPGA) with Page Fault at Source Vs Source and Destination - Latency\label{fig:RemWr_PgFSrcvsPgFSrcnDst}}{
        \includegraphics[keepaspectratio,width=1\textwidth]{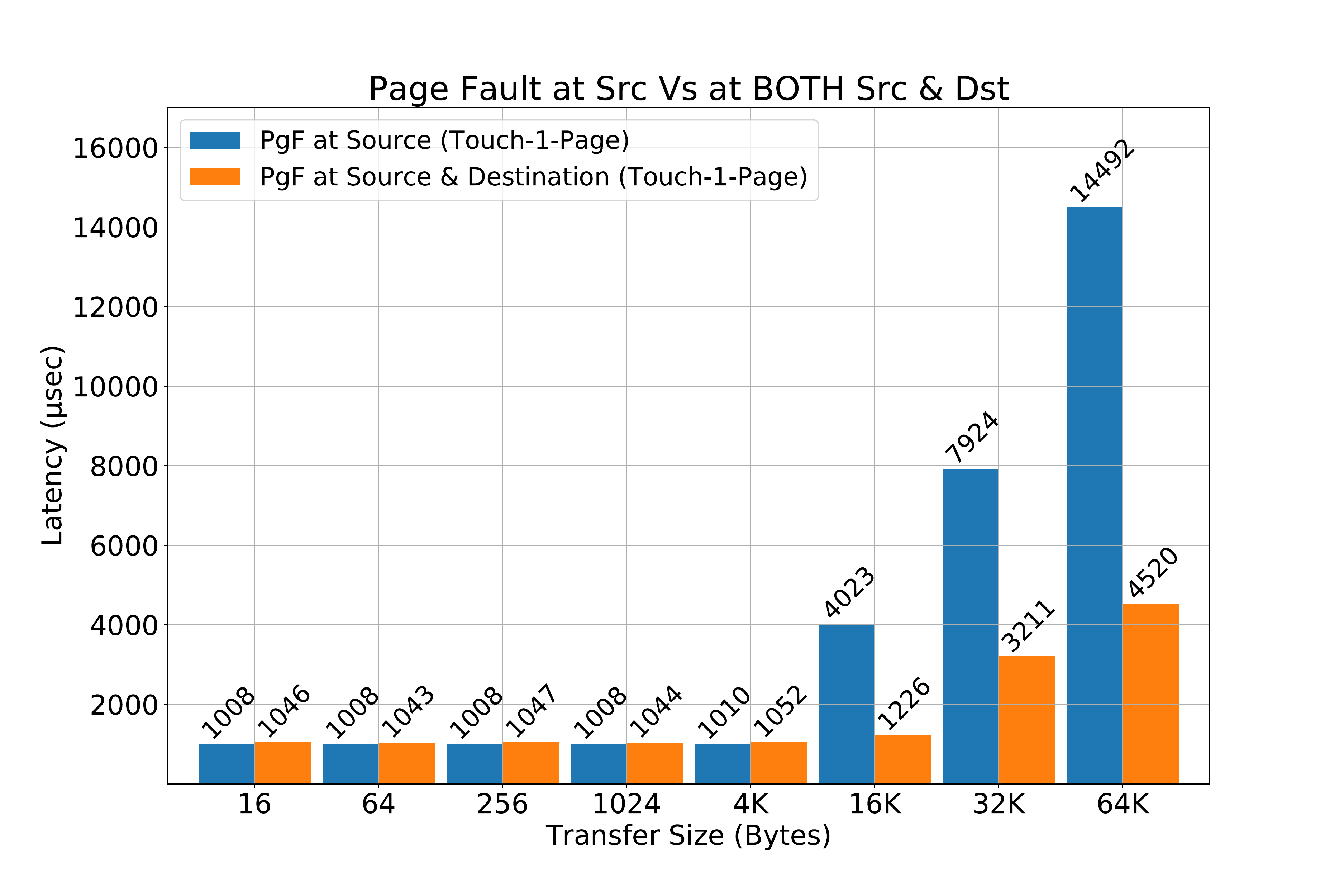}
        }
\end{figure}

\begin{figure}[h!]
        \centering
        \captionbox[Page Fault at source and destination addresses simultaneously leads to less time-outs than Page Fault only at source address]{Page Fault at source and destination addresses simultaneously leads to less time-outs than Page Fault only at source address. In this example, we have one (1) transaction (tr0), with page 0 and 1 (light grey) constituting the source address (buffer) and page 0 and 1 (grey) constituting the destination address (buffer). \label{fig:pgf_both_at_src_and_dst}}{
        \includegraphics[keepaspectratio,width=0.9\textwidth]{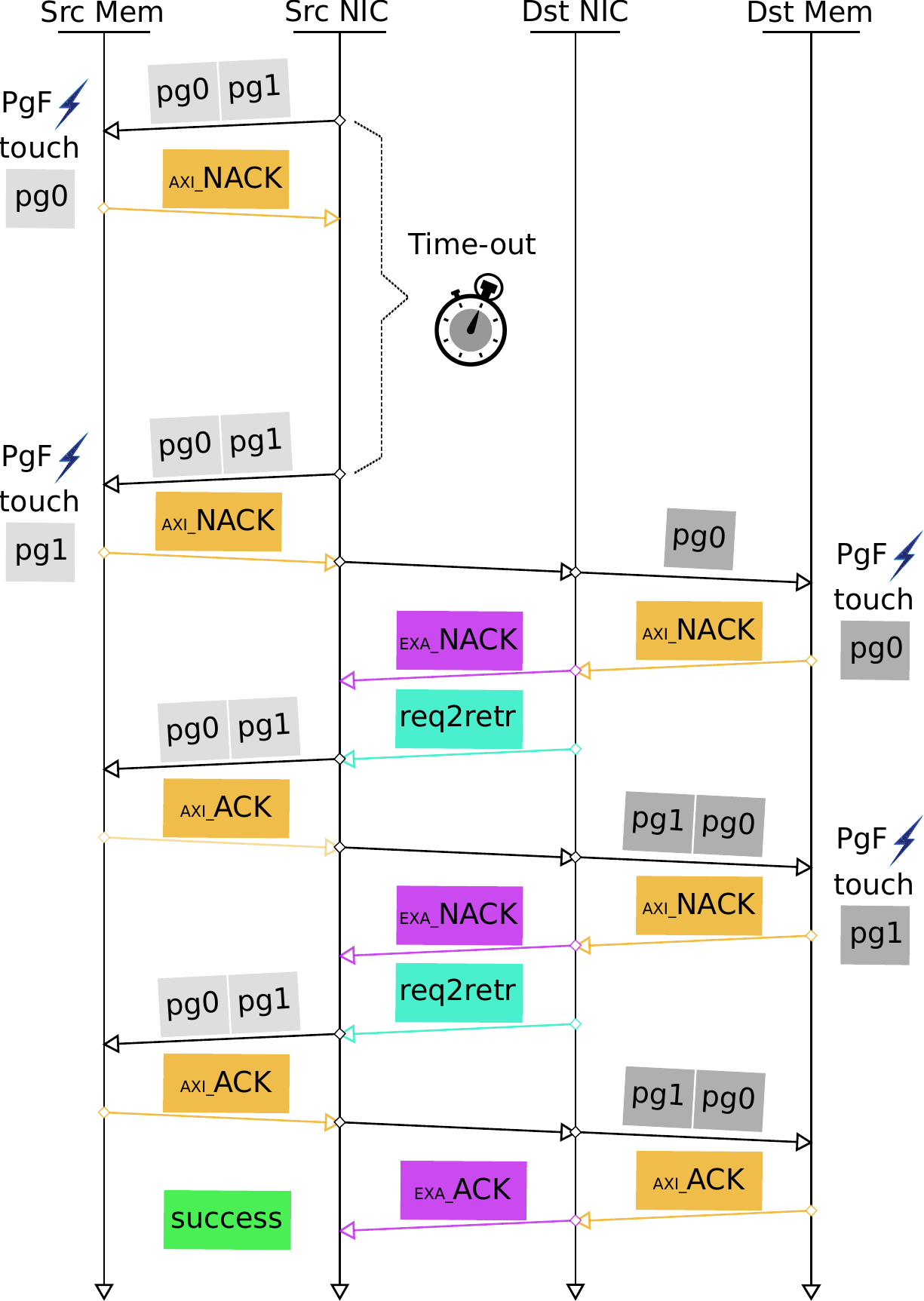}
        }
\end{figure}

In Figure~\ref{fig:RemWr_PgFSrcvsPgFSrcnDst} it is noticeable that when a page fault occurs both in the source and destination buffers, the overall latency is significantly less than when we handle a (minor) page fault at the source, when moving to bigger transfer sizes. In Figure~\ref{fig:pgf_both_at_src_and_dst}, we argue that this can probably be explained due to the less time-outs due to page faults at source buffer, which are considered the most costly ones in our mechanism. While a page fault at the source address is not resolved yet, requests for the faulty destination address begin to arrive and our mechanism can request explicit re-transmissions sooner than the time-out. In other words, when we have page faults only from the side of the source address and while using no optimizations (\enquote{Touch-A-Page} and not \enquote{Touch-Ahead}), we expect as many time-outs as the number of pages, since the handling takes place per page.

%In Figure~\ref{fig:RemWr_NLvsGUP} and Figure~\ref{fig:RemWr_NLvsGUP_SrcDst}, we see the results of both of the option-techniques that we utilized in our mechanism. One of the benefits that comes with the use of get\_user\_pages() is that it can request (\enquote{touch}) more than one pages, as long as they belong to the corresponding user process. Pre-faulting more than one pages can lead to less interrupt invocations of the handler, which overall can provide better performance. In our mechanism we only request four (4) pages per faulty page, since four (4) is the number of pages that constitute a transaction in terms of the custom DMA of FORTH. As we can see from the results in these Figures, using get\_user\_pages() is indeed better in performance than the Netlink approach, mainly in cases where the transfer size is larger than 32~KB. We would expect better results in the case of 16~KB as well, which is why it is currently under investigation, since as we have mentioned already in this thesis, unexpected errors and problems might occur while working in a prototype. 

Another type of measurement we conducted in order to evaluate our mechanism was from the side of the kernel -how much time the page fault handling spends in kernel space, since a part of our mechanism is based on the driver of ARM's IOMMU (SMMU). 

First, we calculate the number of invocations of SMMU's context bank fault handler due to translation faults, for different transfer sizes. Then, as already mentioned before, judging from the type of fault (read/write), we schedule accordingly a tasklet, that of course will run when it is possible for the system, offloading this burden from the interrupt handler and allowing it to exit sooner. So, apart from measuring the time inside the interrupt handler, we also evaluate the time spent in the corresponding tasklet - excluding the time it took the tasklet to actually run. We are currently not sure if it is reasonable for someone to calculate it, since it will probably depend on the load a system has each time.

\begin{figure}[h!]
        \centering
        \captionbox[Remote Write: Driver Latency]{Remote Write: Driver Latency\label{fig:RemWr_driver}}{
        \includegraphics[keepaspectratio,width=1\textwidth]{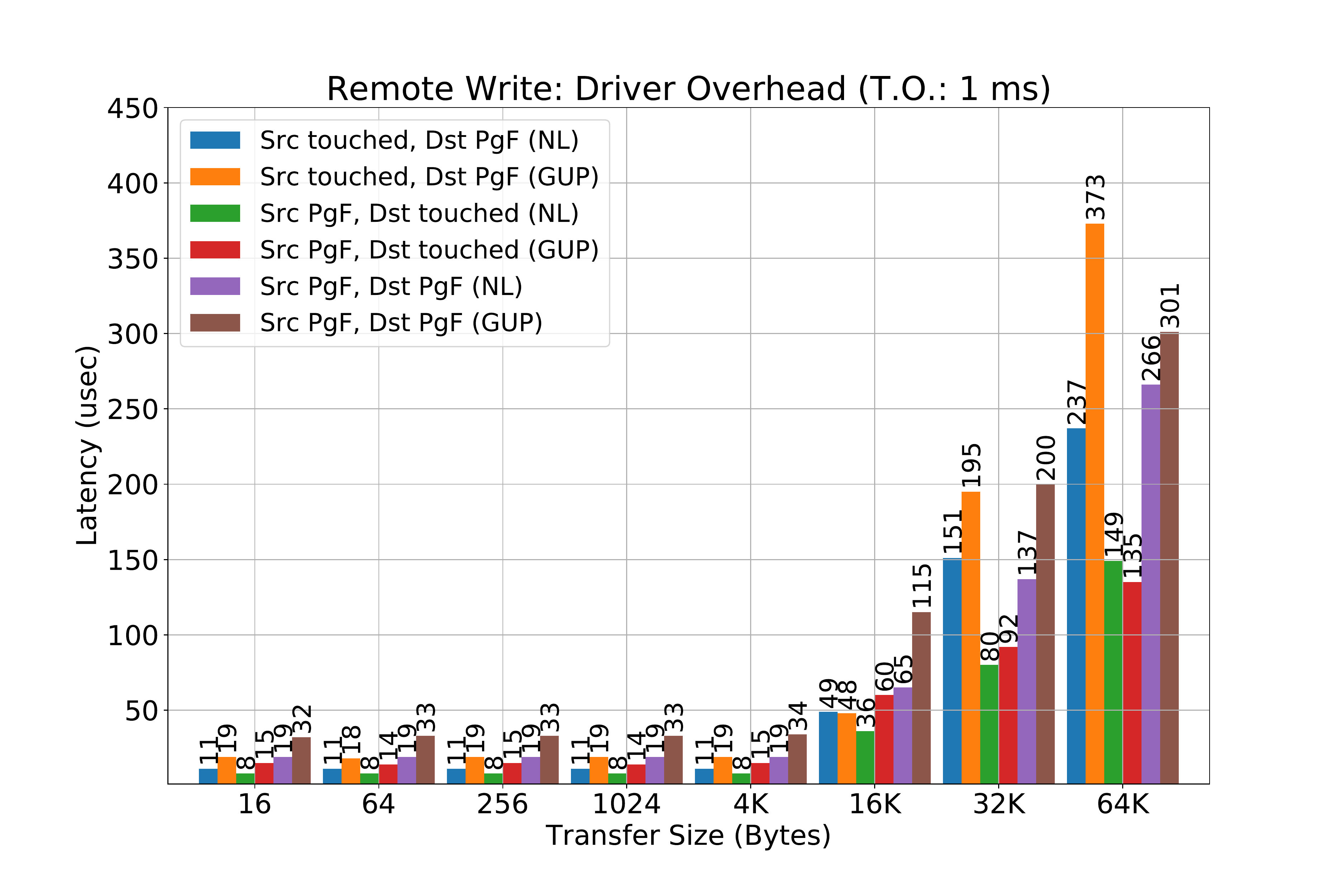}
        }
\end{figure}

In the kernel driver measurements that we present in Figure~\ref{fig:RemWr_driver}, we show numbers only with 1~ms time-out, but having measured also scenarios with time-out as 25ms and 2.5ms, we know that even in this type of measurements, 1~ms time-out outperforms the other two scenarios.

In Figure~\ref{fig:RemWr_driver} we see the driver latency of each approach. In order to measure accurately the latency, we conducted a \enquote{cold run} (e.g. cold cashes) prior to our \enquote{for} loop experiment, that we excluded from our results. To be more precise, the \enquote{cold run} was an 8-byte transfer that we knew in advance that will invoke the translation fault handler only once in case the page fault was caused in the source or the destination buffer. When a page fault occurs in both buffers, we expect two (2) invocations of the translation fault handler, so in these scenarios we excluded the first two (2) \enquote{cold run} measurements. 

 %First of all, we notice high overheads when a page fault occurs only in the destination address of a transfer size $\geq 32K$ Bytes. In fact, since we know that the time-out is set to 1~ms, we suspect that a time-out is triggered even though our mechanism supports explicit requests for retransmissions in order to avoid the high overhead of time-outs. Initially we thought that there was a problem with the FIFO, that will log most of the incoming entries: after $\geq 32K$ Bytes of transfers, we have two (2) or more transactions, that will probably come interleaved. Because of the way that our FIFO was implemented we only check the head value before we push something into it, which is not sufficient in that case, meaning we might have more than one duplicates for the same page. After further investigation, we noticed that there is a problem with the packetizer that cannot send on time a request to the mailbox of R5, in order to initiate the retransmission and as a result a time-out retransmission is triggered. We expect to investigate this behavior further in the future.

There are some insights we can extract from Figure~\ref{fig:RemWr_driver}. It is evident that \enquote{get\_user\_pages} (GUP) or \enquote{Touching-Ahead} costs more time in driver, because the burden of \enquote{touching} the pages is moved to it. In fact, our mechanism tries to touch up to four (4) pages each time when using GUP, as long as it is possible and permitted (e.g. it might be the case that a user application does not own more than one pages). This is a speculative approach, that comes with some costs that can be seen in Figure~\ref{fig:RemWr_driver}.

One thing we find interesting is that in 64~KB when a page fault occurs in the source buffer, the Touch1Page through Netlink seems to be more expensive than the Touch-Ahead through \enquote{get\_user\_pages} (GUP), which requires further investigation and possible re-run of our experiments in order to elaborate on the reasons behind this. Due to the prototype and the on-going work on it, there are many factors that could affect this measurement, including unexpected time-outs or timing problems.

\chapter{Conclusion}\label{chap:conclusion}

\section{Summary}

As part of the ExaNeSt project, we worked on a mechanism that provides support on page faults that might be triggered during remote direct memory accesses (RDMAs).

This mechanism supports translation faults in SMMU, that are expected to occur under many circumstances, including when having optimization mechanisms enabled like the Transparent Huge Pages (THP). Another reason to provide this kind of mechanism is to overcome the consequences of the alternatives such as pinning/unpinning or touch of the buffers before/after they are DMA'ed, which do not not seem good and viable solutions memory-utilization-wise in the long run.

In the end, we evaluated our work in terms of latency overhead and also measured the cost of the alternatives.

\section{Future Work}

There are many areas related to our work that, due to lack of time, we could not cover in this work. Some of the areas that we would like to investigate in the future are listed below:

\begin{enumerate}
    \item Extend our mechanism to fully support get\_user\_pages() and packetizer-to-mailbox exchange of messages in kernel (full kernel implementation).
    \item Finalize and evaluate the handling of page faults during Remote Read RDMA requests.
    \item Build a smarter mechanism of logging information of faulty transfers (considering the interleaved packets), including a NACKer, that would offload the network from the multiple NACKs per faulty packet.
    \item Add support for other types of faults such as permission faults (as mentioned before, more testing is required) and possible segmentation faults from invalid addresses.
    \item Evaluate our mechanism on major page faults that require I/O (e.g. disk access).
    \item Perform measurements on memory utilization, to see what is the fraction of the benefit we gain with our mechanism.
    \item Elaborate more on the SMMU setting of stalling the faults and try to resume them from the driver after the handling.
\end{enumerate}

 %\the\columnwidth 

%CHAPTER_1
%\chapter{Introduction}
%\label{chapter:introduction}

%There is an increasing interest on recommending to the user instantly

%BIB
%\bibliographystyle{plain}
\bibliographystyle{unsrt}
\cleardoublepage
\addcontentsline{toc}{chapter}{Bibliography}
\bibliography{bib/thesis}

%\addcontentsline{toc}{chapter}{Appendix}
% \appendix{} % Cue to tell LaTeX that the following "chapters" are Appendices	
% \include{Appendices/Appendix_0.tex}
% \include{Appendices/AppendixA.tex}

\appendix
%\chapter{Appendix}
% Appendix 

\chapter{StreamID explained}\label{Appendix_StreamID} % For referencing this appendix elsewhere, use \ref{AppendixD}

%\section{StreamID explained}

System Memory Management Unit (SMMU) implements a TBU (Translation Buffer Unit) for sets of masters (devices) on the PS port. The Master ID for each AXI master and each port of the PS for the PL are listed in a Table titled \enquote{Master IDs List} in Chapter~16 on the Zynq MPSoC Technical Reference Manual \cite{zynq_usplus}.

Each PS port to the PL has a fixed master ID that is used together with the TBU number and an AXI ID to construct the StreamID. The AXI protocol includes two kinds of AXI transaction identifiers, AXI IDs and Master IDs. The Master ID is used to uniquely identify the master that initiated a transaction. AXI IDs are used to identify separate transactions that must be processed in order, for example when having multiple contexts or threads within a single master.

\begin{table}[ht!]
    \caption{StreamID format (15 bits)}
    \centering %
    %\resizebox{14cm}{!}{
        \begin{tabular}{| c | c | c |}
        \hline
            \rowcolor{blue!25} \textbf{TBU Number} & \textbf{Master ID} & \textbf{AXI ID} \\\hline
            [14:10] & [9:6] & [5:0] \\\hline
            5 bits & 4 bits & 6 bits \\\hline
        \end{tabular}
    %    }
        \label{tab:streamid_format}
\end{table}

\begin{table}[ht!]
    \caption{PS port connections to TBUs}
    \centering %
    %\resizebox{14cm}{!}{
        \begin{tabular}{| c | c |}
        \hline
            \rowcolor{blue!25} \textbf{TBU Number} & \textbf{PS port} \\\hline
            0 & S\_AXI\_HPC[0]\_FPD \\\hline
             0 & S\_AXI\_HPC[1]\_FPD \\\hline
              1 &  \\\hline
              2 &  \\\hline
              3 & S\_AXI\_HP[0]\_FPD
 \\\hline
            4 & S\_AXI\_HP[1]\_FPD  \\\hline
             5 & S\_AXI\_HP[2]\_FPD  \\\hline
              6 & S\_AXI\_HP[3]\_FPD  \\\hline
        \end{tabular}
    %    }
        \label{tab:tbu_psports_connections}
\end{table}

In Table~\ref{tab:tbu_psports_connections} we can see where each port of the Processing System connects to the TBU (essentially to SMMU). This information is extracted from the top-level diagram of the Zynq Ultrascale Plus MPSoC (Figure~\ref{fig:zynqusplus_toplvldiagram}).

\section{TBU (Translation Buffer Unit)}

The Translation Buffer Unit (TBU) contains a Translation Look-Aside buffer (TLB) that caches page tables maintained by the Translation Control Unit (TCU). The SMMU implements a TBU for system masters. Each TBU has the following characteristics:
\begin{itemize}
    \item designed to be local to the master
    \item 256 outstanding transactions for each TBU
    \item 16-deep TBU queue support
    \item 32-deep write buffer support (for each TBU) : 0, 4, 8 or 16  bursts
    \item Best-case hit latency
    \begin{itemize}
        \item 2 clocks (when TBU addr. slave register slices are NOT implemented)
        \item 3 clocks (when TBU add. slave register slices are implemented)
    \end{itemize}
    \item Micro-TLB
    \begin{itemize}
        \item caches PTW (page table walk) results returned by the TCU
        \item fully associative
        \item configurable depth (based on our requirements)
    \end{itemize}
\end{itemize}

\subsection{Outstanding transactions per TBU}

Outstanding transactions are defined as transactions for which, the physical address access is generated and accepted by the slave, or write/read responses are stalled.

For every TBU, the MMU-500 supports 256 outstanding transactions each for write and read accesses. The MMU-500 generates a PTW when accesses from the master result in a TLB miss. However, based on the configuration, the MMU-500 supports either 8 or 16 such parallel PTWs for a TBU. If more than 8 or 16 PTWs are pending, a TLB miss on a channel indicates that the MMU-500 cannot accept additional transactions on the write or read channels.

% The SMMU\_CBn\_FSYNR0.S1PTWF bit indicates a fault on the stage 1 translation table walk.
% If an external fault is reported to the SMMU in response to a fetch issued as part of a translation table walk, the fault is recorded synchronously in the translation context bank. For an external fault on a translation table walk, SMMU\_CBn\_FSYNR0.PTWF is set to 1. 

% An external fault encountered in an instruction read, a data read or a data write can be reported to the SMMU, and the SMMU might record the fault synchronously or asynchronously. External faults in an SMMU implementation are generated in response to:

% \begin{itemize}
%     \item A read issued as part of a translation table walk
%     \item Optionally: 
%     \begin{itemize}
%         \item A translated client transaction where the downstream system returns an abort
%         \item n SMMUv1 only, an unsupported upstream transaction.
%     \end{itemize}
% \end{itemize}

% SMMUv2 supports a dedicated unsupported upstream transaction fault.

\section{TCU (Translation Control Unit)}

This unit is responsible to control and manage the address translations (translation tables for the TBUs). It uses a private AXI stream interface to update the translation tables in the TBUs. TCU core can run at half the clock speed of TCU external interfaces.

It consists of four (4) main components:
\begin{enumerate}
    \item Macro-TLB: responsible to cache PTW results in the TCU
    \item PTW cache: MMU-500 caches partial PTWs to reduce the number of PTWs on a TLB miss. It caches both stage 1 and stage 2 level 2 PTWs.
    \item Prefetch buffer: MMU-500 fetches in advance 4KB and 64KB sized pages into the prefetch buffer, which reduces the latency for future PTWs. It has configurable width. It is a single four-way set associative cache, that can be enabled/disabled depending on the context. It also shares RAMs with the TLB cache.
    \item IPA-to-PA cache: MMU-500 implements an IPA to PA single four-way set associative cache for stage 1 followed by stage 2 translations. It can be enabled/disabled depending on the context. It shares RAMs with the PTW cache.
\end{enumerate}
% Appendix 

\chapter{Boot-up environment}\label{Appendix_Environment} % For referencing this appendix elsewhere, use \ref{AppendixD}

%\section{Boot-up environment}

In order to effectively work and run our experiments, we had to initialize our Linux environment. This included many settings, libraries, development and insertion of kernel modules, necessary for our work. In this Appendix, we will give a fully detailed description of all modules and libraries developed as part of this work and some other utilities developed by other members of the Lab at FORTH.

Assuming we are working on a QFDB, one had to run the following three calls before having an environment to work:
\begin{enumerate}
    \item ./load\_boot\_package.sh $<$number of FPGA (0, .., 3)$>$ (load boot package of a specific design)
    \item ./kexec\_script.sh (load kernel Image modified to serve our purposes - mostly for the arm-smmu.c driver modifications)
    \item ./new\_remoteproc.sh (initialize environment)
\end{enumerate}

% \lstset{style=mystyle}
% \lstset{
% 	language=bash,
% %	basicstyle=\footnotesize\color{white}
%     basicstyle=\color{white}
% }

\section{Boot package loading}

% #!/bin/bash
% echo the $# parameter did not destroy pygments syntax highlighting

% \begin{lstlisting}[language=bash,caption={bash version}]
% #!/bin/bash
% echo "Hello, world!" $1;
% \end{lstlisting}

\begin{lstlisting}[language=bash]
    #!/bin/bash
    DESIGN_NAME="2.2.4";
    FPGA_NUM=$1;
    range_of_num='^[0-9]+$';
    range_of_fpga='[0-3]'
    
    if [ -z "$FPGA_NUM" ] ; then
            echo "ERROR: FPGA number is REQUIRED in order to load boot package." >&2;
            exit 1;
    elif ! [[ "$FPGA_NUM" =~ $range_of_num ]] ; then
            echo "ERROR: Input parameter is NOT a number." >&2;
            exit 1;
    elif ! [[ "$FPGA_NUM" =~ $range_of_fpga ]] ; then
            echo "ERROR: Input parameter is OUT of range." >&2;
            exit 1;
    fi
    
    echo "Loading boot package of $DESIGN_NAME design for F$((FPGA_NUM+1)).....";
    
    if [ "$FPGA_NUM" = "0" ] ; then
            /root/qfdb-stuff/boot_package /home/pxirouch/bps/$DESIGN_NAME/output.bp.F1
    elif [ "$FPGA_NUM" = "1" ] ; then
            /root/qfdb-stuff/boot_package /home/pxirouch/bps/$DESIGN_NAME/output.bp.F2
    elif [ "$FPGA_NUM" = "2" ] ; then
            /root/qfdb-stuff/boot_package /home/pxirouch/bps/$DESIGN_NAME/output.bp.F3
    elif [ "$FPGA_NUM" = "3" ] ; then
            /root/qfdb-stuff/boot_package /home/pxirouch/bps/$DESIGN_NAME/output.bp.F4
    fi
\end{lstlisting}

\section{Loading and booting a different kernel Image}

\begin{lstlisting}[language=bash]
    #!/bin/bash
    kexec -l /home/psistakis/Image --reuse-cmdline
    kexec -e
\end{lstlisting}
    
The first command (line 2) is responsible to link the new kernel Image file. We used the Linux kernel 4.9 that was taken from Xilinx repository and is part of \enquote{xilinx-v2017.2} tag. After having a copy (clone) of this repository we only had to cross-compile it, having in mind the target architecture, which in our case is AArch64. The \enquote{--reuse-cmdline}, basically says the new Linux Image to boot in the current console (command line).

The second command (line 3) reboots (executes) the system with the new Linux kernel Image linked before.

An example of the command to build the Linux kernel and generate the Linux kernel Image is:

\begin{lstlisting}[language=bash]
    make ARCH=arm64 CROSS_COMPILE=aarch64-linux-gnu- -j8
\end{lstlisting}

\section{Environment initialization}

We use the script below to initialize our environment, before we run any application. This script makes sure that all necessary modules are inserted along with the R5 firmware.

\begin{lstlisting}[language=bash]
    #!/bin/sh
    FPGA_NUM=$1;
    range_of_num='^[0-9]+$';
    range_of_fpga='[0-3]'
    
    if [ -z "$FPGA_NUM" ] ; then
            echo "ERROR: FPGA number is REQUIRED in order to insert unimem_coord module." >&2;
            exit 1;
    elif ! [[ "$FPGA_NUM" =~ $range_of_num ]] ; then
            echo "ERROR: Input parameter is NOT a number." >&2;
            exit 1;
    elif ! [[ "$FPGA_NUM" =~ $range_of_fpga ]] ; then
            echo "ERROR: Input parameter is OUT of range." >&2;
            exit 1;
    fi
    
    if [ "$FPGA_NUM" = "0" ] ; then
            U_COORDS=0x0;
    elif [ "$FPGA_NUM" = "1" ] ; then
            U_COORDS=0x1000;
    elif [ "$FPGA_NUM" = "2" ] ; then
            U_COORDS=0x2000;
    elif [ "$FPGA_NUM" = "3" ] ; then
            U_COORDS=0x3000;
    fi
    
    echo "Loading all necessary modules (exanest etc) and R5 firmware in F$((FPGA_NUM+1)).....";
    
    /root/qfdb-stuff/rwphys/write32 0x80070004 0x1
    /root/qfdb-stuff/rwphys/write32 0x80070000 0xffffffff
    
    # Path of drivers
    DRIVERS_PATH=/home/psistakis/unimem-exanet-drivers
    
    insmod -f $DRIVERS_PATH/remoteproc.ko  # A generic framework with which AMP remote processors can be controlled (powered up/down), provided by Linux
    insmod -f $DRIVERS_PATH/zynqmp_r5_remoteproc.ko    # A Xilinx ZynqMP R5 remoteproc driver to enable Linux kernel to bringup R5, and enable communication between Linux kernel and R5 (provided by Xilinx)
    insmod -f $DRIVERS_PATH/unimem_coord.ko unimem_coords="$U_COORDS"
    insmod -f $DRIVERS_PATH/mbox_back.ko   # Mailbox Driver developed by others at FORTH
    insmod -f $DRIVERS_PATH/pack_back.ko   # Packetizer Driver developed by others at FORTH
    insmod -f $DRIVERS_PATH/pldma.ko # PLDMA Module developed by others at FORTH
    
    # Scratchpad module developed by others at FORTH
    insmod -f $DRIVERS_PATH/scratchpad/scratchpad_alloc.ko
    
    # Load R5 firmware
    cp /home/psistakis/pf_r5.elf /lib/firmware/
    cd /home/psistakis/
    echo pf_r5.elf > /sys/class/remoteproc/remoteproc0/firmware
    echo start > /sys/class/remoteproc/remoteproc0/state 
    
    # Insert module developed as part of our page-fault mechanism for packetizer
    insmod -f $DRIVERS_PATH/pf_pckzer/pf_pckzer.ko
    
    # Disable Transparent Huge Pages (THP) (if/when needed)
    echo never > /sys/kernel/mm/transparent_hugepage/enabled
\end{lstlisting}

\end{document}